\documentclass[aps,prb,reprint,
twocolumn,amsmath,superscriptaddress,amssymb,longbibliography,nofootinbib]{revtex4-2}
\usepackage{amsmath,braket,amssymb}
\usepackage[final]{graphicx}
\usepackage{subfigure}
\usepackage[english]{babel}
\usepackage[utf8]{inputenc}
\usepackage{mathtools}
\usepackage{empheq}
\usepackage{tensor}
\usepackage{cancel}

\DeclareRobustCommand{\stirling}{\genfrac\{\}{0pt}{}}

\usepackage{bbm}

\usepackage{simplewick}
\usepackage{xfrac}

\usepackage[usenames,dvipsnames]{color}

\newcommand{\blue}{\color{blue}}

\usepackage{enumitem}
\usepackage{slashed}
\usepackage{graphicx}
\usepackage[dvipsnames]{xcolor}
\usepackage{tikz}
\usetikzlibrary{arrows.meta}
\usepackage{bm}
\usepackage[colorlinks=true,citecolor=blue,linkcolor=blue,urlcolor=blue]{hyperref}
\usepackage{dsfont}
\usepackage{changepage}
\usepackage{array}
\usepackage{soul}
\usepackage[capitalise]{cleveref}
\crefname{section}{Sec.}{Secs.}
\Crefname{section}{Sec.}{Secs.}
\usepackage{xifthen}
\usepackage{xargs}

\newcommand\eq[1]{\begin{align}#1\end{align}}

\newcommand{\partitions}[1]{\Pi_{#1}}

\usepackage{bm}

\newcommand{\bit}{\begin{itemize}}
\newcommand{\eit}{\end{itemize}}

\newcommand{\f}{\frac}
\renewcommand{\>}{\right\rangle}
\newcommand{\<}{\left\langle}
\newcommand{\ba}{\begin{align}}
\newcommand{\ea}{\end{align}}
\newcommand{\be}{\begin{equation}}
\newcommand{\ee}{\end{equation}}
\newcommand{\bi}{\begin{itemize}}
\newcommand{\ei}{\end{itemize}}
\newcommand{\lf}{\left(}
\newcommand{\ri}{\right)}
\newcommand{\dd}{\mathrm{d}}

\DeclareMathAlphabet{\mymathbb}{U}{BOONDOX-ds}{m}{n}

\usepackage{comment}

\usepackage{braket}
\usepackage[normalem]{ulem}

\definecolor{colorforA}{rgb}{1, 0.249, 0.02745}
\definecolor{colorforB}{rgb}{0.08625, 0.813725, 0.08625}
\definecolor{colorforC}{rgb}{0.06075, 0.23333, 0.851}

\def\colZ{white}
\def\colA{colorforA!75}
\def\colB{colorforB!75}
\def\colC{colorforC!75}

   \newcommand{\emptysite}{\protect\tikz[baseline=-2.5pt]
 \protect \draw[thick] (-0.2,-0.2) -- (0.2,-0.2);}

 \newcommand{\ballZ}{\protect\tikz[baseline=-2.5pt]
  {\protect\node[minimum width=0,inner sep=0.5] (u) at (0,0) [draw, circle, fill=\colZ] {$\phantom{1}$}; \draw[thick] (-0.2,-0.2) -- (0.2,-0.2);}
  }
 
 \newcommand{\ballA}{\protect\tikz[baseline=-2.5pt]
  {\protect\node[minimum width=0,inner sep=0.5] (u) at (0,0) [draw, circle, fill=\colA] {$1$}; \draw[thick] (-0.2,-0.2) -- (0.2,-0.2);}
  }
 \newcommand{\ballB}{\protect\tikz[baseline=-2.5pt]
  {\protect\node[minimum width=0,inner sep=0.5] (u) at (0,0) [draw, circle, fill=\colB] { $2$}; \draw[thick] (-0.2,-0.2) -- (0.2,-0.2);}}
   \newcommand{\ballC}{\protect\tikz[baseline=-2.5pt]
{\protect\node[minimum width=0,inner sep=0.5] (u) at (0,0) [draw, circle, fill=\colC] { $3$};   \protect \draw[thick] (-0.2,-0.2) -- (0.2,-0.2);}}

\begin{document}

\title{The damage spreading transition: a hierarchy of renormalization group fixed points}
\author{Adam Nahum}
\affiliation{Laboratoire de Physique de l’\'Ecole Normale Sup\'erieure, CNRS, ENS \& Universit\'e PSL, Sorbonne Universit\'e, Universit\'e Paris Cit\'e, 75005 Paris, France}

\author{Sthitadhi Roy}
\affiliation{International Centre for Theoretical Sciences, Tata Institute of Fundamental Research, Bengaluru 560089, India}

\date{\today}

\begin{abstract}
Deterministic classical cellular automata can be in two phases, depending on how irreversible the dynamical rules are. In the strongly irreversible phase, trajectories with different initial conditions coalesce quickly, while in the weakly irreversible phase, trajectories with different initial conditions can remain different for a time exponential in the system volume. The transition between these phases is referred to as the damage-spreading transition (the ``damaged'' sites are those that differ between the trajectories). We develop a theory for this transition. In the simplest and most generic setting, the transition is known to be related to directed percolation, one of the best-studied nonequilibrium phase transitions. However, we show that full theory of the damage-spreading critical point is richer than directed percolation, and contains an infinite hierarchy of sectors of local observables. Directed percolation describes the first level of the hierarchy. The higher observables include ``overlaps'' for multiple trajectories, and may be labeled by set partitions. (These higher observables arise naturally if, for example, we consider decay of entropy under the irreversible dynamics.) The full hierarchy yields a hierarchy of nonequilibrium fixed points for reaction-diffusion-type processes, all of which contain directed percolation as a subsector, but which possess additional universal critical exponents. We analyze these higher fixed points using a field theory formulation and renormalization group arguments, and using simulations in 1+1 dimensions.
\end{abstract}

\maketitle


\section{Introduction}

Dynamical systems for discrete spins, 
such as cellular automata \cite{walker1966temporal,kauffman1969metabolic,wolfram1983statistical,kauffman1984emergent,games1970fantastic,derrida1989dynamical}, 
can be classified according to 
the effect of a localized change to the initial conditions \cite{kauffman1969metabolic,creutz1986deterministic,stanley1987dynamics,derrida1987dynamical}.
The difference induced between the two initial configurations, referred to as ``damage'', may die out rapidly under the dynamics, so that the two distinct trajectories converge.
Alternatively the damage may spread throughout the system \cite{derrida1989dynamical,derrida1986phase,weisbuch1987phase,dearcangelis1987period, grassberger1995damage,bagnoli1996damage,willsher2022measurement}.

The least structured settings for studying these phenomena are random cellular automata such as the Kauffman model and its variants, which may be defined either in a mean-field-like regime or in finite dimensions \cite{kauffman1969metabolic,derrida1989dynamical,derrida1986phase,derrida1986multivalley,derrida1987distribution,derrida1987random,derrida1988statistical,kurten1988correspondence}. In these models, the rules which define the spins' (deterministic) updates are selected randomly from the set of all possible rules.
By varying the  prevalence of irreversible local updates, it is possible to tune to a continuous  damage spreading transition. 

In the simplest case, the update rules for the automaton are random not only in space 
(if the model is local) but also in time  \cite{derrida1986random}.
Simple random-in-time automata were initially studied as 
an approximation for the  Kauffman model 
without temporal randomness \cite{derrida1986random}, 
but in fact temporally random automata are  interesting in their own  right as null models for complex dynamical processes.
In the spatially local case these automata may be formulated as random classical circuits (e.g. random Boolean circuits).
Once a random realization of the circuit has been fixed, it defines a \textit{deterministic} evolution of the  classical bits or spins. 
However, this evolution is in general \textit{irreversible}.

Defining the damage using \textit{two} trajectories that are evolved using the same update rules, but with distinct initial conditions, these temporally random automata show a  robust damage-spreading phase transition whose basic observables are governed by the directed percolation universality class  \cite{derrida1986random, grassberger1995damage, bagnoli1996damage}. In the language of absorbing state transitions \cite{hinrichsen2000non,odor2004universality,janssen2005field}, the 
 ``active'' sites are those where damage is present.
 
This paper revisits the damage spreading transition in order to argue that it is in fact much richer than directed percolation.
The full set of observables for damage spreading includes observables that involve $n$ trajectories for arbitrary ${n\geq 2}$ \cite{derrida1986evolution}. Directed percolation captures only a small sector of the full theory, namely observables that can be expressed using two copies (${n=2}$). 
We argue that, once ${n>2}$, the natural observables are labelled by set partitions \cite{stanley2011enumerative} of the set of replicas, i.e. divisions of the replicas $\{1,2,\ldots, n\}$ into distinct blocks. 
For example, the partition
\be
\pi = (135)(2)(46),
\ee
for the case ${n=6}$, labels a spatial region 
(a site or a group of adjacent sites)
 on which replicas 1, 3 and 5 share the same state, 
replicas 4 and 6 share another distinct state, and replica 2 has yet a different state.
Local activity  (damage) variables can be defined for each partition $\pi\in \Pi_n$.
Loosely speaking, $\rho_\pi(x,t)$ is the density of sites (or bonds, etc.) with damage of type $\pi$.

A universal scaling theory of the damage-spreading critical point must account for 
this hierarchy of observables. 
We describe such a theory and find that the critical point is unusual from the point of view of the renormalization group (RG).

After formulating a mean field theory for the transition, we derive finite-dimensional 
Langevin equations, or equivalently finite-dimensional field theories.
The transition is described by a hierarchy of such theories, for $n\in \{2,3,4,5,\ldots\}$, in which the ``elementary'' fields are a subset of the damage densities ${\rho_\pi(x,t)}$.
The stochastic equations take the schematic form (repeated indices summed) 
\ba \label{eq:langevingeneralintro}
\partial_t \rho_\pi
& = 
D \nabla^2 \rho_\pi
+ r \rho_\pi 
 - {\mathcal{K}}_{\sigma,\sigma'}^\pi
\rho_\sigma\rho_{\sigma'}
+ \eta_\pi.
\end{align}
where   the tensor $\mathcal{K}$ will be specified in Sec.~\ref{sec:continuummeanfield}, and $\eta_\pi$ is spacetime noise of variance  $\propto\rho_\pi$.
At first sight,  $\pi, \sigma, \sigma'$ should run over all possible partitions, 
but in fact we find that the critical continuum theory can be expressed using only fields corresponding to  partitions with two blocks: the other damage densities may be expressed as composite fields.

The above field theory is most useful near the upper critical dimension.
However, we also find that many features of the critical point 
can be fixed without reference to a continuum formulation, 
on general combinatorial grounds --- 
that is, using the partial ordering on   set partitions together with real-space RG logic.

In particular we show that there are distinct RG scaling operators for every partition $\pi$, but that the set of distinct scaling dimensions $\Delta_\pi$ is much smaller.\footnote{The basic scaling dimensions for damage densities are labelled by the number of blocks in the partition. (In addition to the damage densities there are also  ``dual'' response fields that we discuss in the following.)}
In general, the mathematical structure of the critical theory seems interesting and may be worth studying further: for example, operators are classified by data that go beyond the symmetry ``quantum numbers'' which are sufficient at a conventional critical point.

Before describing these analytical approaches,
we show numerical simulations for a 1D model. These demonstrate concretely that observables with ${n\geq 2}$ trajectories lead to new critical exponents. 
We give numerical values for the new exponents that appear at ${n=3}$ and ${n=4}$.
The simulations also confirm a surprising time-reversal symmetry of the critical theory.

Finally, let us make a brief clarification about the role of randomness, to avoid any possible confusion.

The classical circuits we consider are 
defined by local update functions $F_{i,t}$ associated with spacetime points $(i,t)$, 
which specify the local rule by which the degree of freedom $s_i$ is to be updated in the $t^\text{th}$ timestep.
These functions will be chosen randomly.
But we emphasize that, once the  $F_{i,t}$ are chosen, they define a \textit{deterministic} 
evolution 
(a Boolean circuit) for the spins ${s_i}(t)$.
We refer to these deterministic processes as the ``physical'' dynamics.
When we consider damage observables involving multiple copies of the system,  the copies are evolved with the \textit{same} deterministic Boolean circuit \cite{derrida1989dynamical}.

Nevertheless, we ultimately want to consider observables averaged over the ensemble of Boolean circuits. 
This averaging leads to a Markovian evolution of both single and multi-replica observables.
We call these Markovian dynamics ``effective'' to distinguish them from the ``physical'' deterministic dynamics.\footnote{Our terminology in which the deterministic dynamics is ``physical'' and the Markovian dynamics is ``effective'' is  natural if we think of the random circuit as a toy model for some complex dynamical system, in which the update rules are not random, but perhaps spatiotemporally inhomogeneous. In other contexts \cite{hinrichsen2000non,odor2004universality} a different terminology may be more natural. (Of course the choice of terminology does not affect calculations.)} 
For observables that only involve a single trajectory (a single copy of the system) this effective Markovian dynamics is trivial for the present models, but, for observables involving multiple copies, it gives a nontrivial Markov process for the overlaps between trajectories, i.e. for the damage variables.

\addtocontents{toc}{\protect\setcounter{tocdepth}{2}}
\tableofcontents

\section{Overview of models}
\label{sec:overviewofmodels}

We consider temporally random \cite{derrida1986random} Kauffman networks \cite{kauffman1969metabolic}
describing dynamics of classical bits or, more generally, classical $q$-state spins ${{s_i}\in \{0,\ldots,q-1\}}$.
Here we describe a 1D model and a mean field model.
These are essentially equivalent to models in the literature (see e.g. \cite{derrida1989dynamical}), 
modulo microscopic variations that are unimportant for the universal physics.

The simplifying features of these models are  that 
(1) the ``update rules'' for spins are given by random functions, 
and (2) the probability distribution for these random functions satisfies an invariance property.
This property allows the random functions to be averaged over exactly, giving an effective dynamics for the damage variables
(or ``activity'' variables)
alone \cite{derrida1986random}.
This effective dynamics is a Markov process for the activity variables. 
In the simplest case of only ${n=2}$ replicas 
this Markov process is directed percolation.

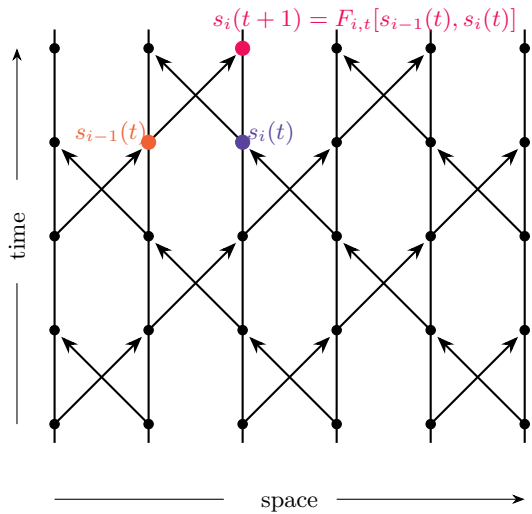
\begin{figure}
\begin{tikzpicture}[scale=1.25]
\foreach \x in {0,...,5}{
    \draw[thick] (\x,-0.2) -- (\x,4.2);
    \foreach \y in {0,...,4}{
        \filldraw (\x,\y) circle (0.05);   
    }
}
\foreach \x in {0,2,4}{
\foreach \y in {0,2}{
\draw[thick, -Stealth] (\x,\y) -- (\x+1-0.07,\y+1-0.07);
\draw[thick, -Stealth] (\x+1,\y) -- (\x+0.07,\y+1-0.07);
}
}
\foreach \x in {1,3}{
\foreach \y in {1,3}{
\draw[thick, -Stealth] (\x,\y) -- (\x+1-0.07,\y+1-0.07);
\draw[thick, -Stealth] (\x+1,\y) -- (\x+0.07,\y+1-0.07);
}
}
\node at (0.6,3.1) {{\color{RedOrange}{$s_{i-1}(t)$}}};
\node at (2.3,3.1) {{\color{Violet}{$s_{i}(t)$}}};
\node at (3.3,4.3) {{\color{OrangeRed}{$s_{i}(t+1)=F_{i,t}[s_{i-1}(t),s_{i}(t)]$}}};
\filldraw[OrangeRed] (2,4) circle (0.075);
\filldraw[Violet] (2,3) circle (0.075);
\filldraw[RedOrange] (1,3) circle (0.075);
\node at (2.5,-0.85) {space};
\draw[ -Stealth] (3,-0.8) -- (5,-0.8);
\draw[] (0,-0.8) -- (1.8,-0.8);

\node[rotate=90] at (-0.4,2) {time};
\draw[ -Stealth] (-0.4,2.6) -- (-0.4,4);
\draw[] (-0.4,0) -- (-0.4,1.5);
\end{tikzpicture}
\caption{Structure of the 1+1D classical circuit. Each dot represents the value $s_i(t)$ of a spin at a certain time. 
Arrows show which neighbor is involved in the update of a given spin at a given time, Eqs.~\ref{eq1plus1update},~\ref{eq1plus1update2}.}
\label{fig:circuit}
\end{figure}

\emph{Model in one spatial dimension.}
In the 1D model the physical degrees of freedom  are bits ${s_i\in\{0,1\}}$ and they are arranged on the line:  ${i=1,\ldots, L}$.
Dynamics is in discrete time, with an even-odd pattern that is shown in Fig.~\ref{fig:circuit}.
The updates are of the form 
\ba\label{eq1plus1update}
s_i(t + 1) & = F_{i,t}\big(
s_i(t), \, s_{i-1}(t) \big)
& 
& \text{for $i+t$ even,}
\\ \label{eq1plus1update2}
s_i(t + 1) & = F_{i,t}\big(
s_i(t), \, s_{i+1}(t) \big)
& 
& \text{for $i+t$ odd.}
\end{align}
The dynamics is given by a ``random circuit'' in the sense that 
each function $F_{i,t}$ is independently random, and 
picking these functions for all $(i,t)$
yields a definite Boolean circuit.
But once a given circuit is picked, it defines a   \textit{deterministic}
dynamics for the spins $s_i(t)$.

The functions are
drawn from a distribution specified in Sec.~\ref{sec:choiceof1Dmodel}.
The distribution contains a parameter $p_s$ 
(with $0\leq p_s<1$) 
which controls the degree of irreversibility: ${p_s=0}$ is the maximally irreversible limit, where the output of the function is independent of the inputs.  (In this limit, the spins forget their initial conditions after a single timestep.)

Note that each function $F$ has $4$ possible input values and 2 possible output values. 
The key invariance property of the ensemble is that a function $F$ is equally likely as any other function $\widetilde F$ that can be obtained from $F$ by permuting the output values and/or the input values.
For more general models with $q$-state spins, like that below, the corresponding property is invariance under permutations in $S_q \times S_{q^2}$ (where $S_k$ is the symmetric group on $k$ elements).\footnote{This invariance is a convenient simplification for analytical work, but the universal behavior we will discuss is robust even if the invariance is broken.}

\emph{Mean-field model.}
In order to formulate a field theory it is convenient also to examine a ``mean-field'' model similar to Ref.~\cite{derrida1986random}. In the mean field model
(more precisely, a model without spatial locality)
we allow for an arbitrary number $q$ of local states,  $s_i\in \{0,\ldots, q-1\}$.
It is also slightly more convenient to work in continuous time.

The random circuit is then constructed in the following way.
In an infinitesimal time interval $(t,t+\dd t)$, each site $i\in \{1,\ldots, N\}$ has a probability $\dd t$ of being updated.
If some site $i$ is updated, its new value is 
of the form
\be\label{eq:MFupdate}
s_i(t+\dd t) = F_{i,t}\big( 
s_{j_1}(t), s_{j_2}(t) \big).
\ee
The randomness on the right-hand side consists in 
(1) a choice of random function $F_{i,t}$;
and 
(2) a random choice of two inputs $j_1$ and $j_2$, which are chosen independently and uniformly from the $N$ sites. Neither $j_1$ nor $j_2$ need be equal to $i$. 
The update functions are similar to those in the 1D case:
a simple  distribution for $F$ is specified in Sec.~\ref{sec:meanfieldeqnsgeneralform}
(and the effect of varying the distribution is discussed in Sec.~\ref{sec:moreoninteractiontensor}).
Again the most important 
property of the distribution of updates is encoded in a parameter  $p_s$;  
when ${p_s=0}$ the dynamics is maximally irreversible (the outputs of the functions $F$ are independent of their inputs) while when ${p_s>0}$ the updates are not completely irreversible.

\section{Damage variables}
\label{sec:damagevars}

Consider two or more generally $n$ initial conditions 
which evolve under same random circuit.
The resulting trajectories are
\ba
& s_i^{{\blue a}}(t),
& & {\blue a} = 1,\ldots, n,
\end{align}
where ${\blue a}$ labels the copies (replicas) of the system. 
Rather than studying the dynamics of the spins themselves,  we wish to study the dynamics of the pattern of similarity/difference between the $n$ copies.
The case $n=2$ has been extensively studied and maps to directed percolation. There appears to be little work on ${n>2}$ replicas, though see Ref.~\cite{derrida1986evolution}.

\subsection{Two replicas (review)}
\label{sec:tworeplicareview}

Let us first recall the well-studied case with $n=2$  \cite{derrida1989dynamical}.
The damage variable  is
(suppressing the time argument for clarity) 
\be\label{eq:n2damagedefn}
\rho(i) = 1 - \delta_{s_i^{{\blue 1}}, s_i^{{\blue 2}}},
\ee
and is 1 where the replicas differ and 0 where they agree.
(For damage variables, we will write  the spatial site as an argument rather than a subscript.)

A convenient feature of the present models is that, after  averaging over the circuit randomness,
we obtain an effective Markovian evolution for the damage variables ${\rho(i)=0,1}$ alone.\footnote{More formally: 
imagine fixing some initial conditions $s_i^1(0)$ and $s_i^2(0)$ that yield some initial damage pattern $\rho(i,0)$. We can then write (e.g.) the expectation value of the damage at some later time as (the initial conditions are left implicit):
\be
\mathbb{E}_\text{circuits} \, \rho(i,t) = \< \rho(i,t)\>.
\ee
On the left-hand side, we think of $\rho_i$ as the  observable ${1 - \delta_{s_i^{{ 1}}, s_i^{{ 2}}}}$ defined on two copies of the system, 
and the averaging is over the functions $F$.
On the right-hand side, the angle brackets represent an expectation value in a stochastic process in which the only degrees of freedom are the  variables $\rho(i)=0,1$.}
(That is, we can forget about the underlying spin configurations $s_i^a$.)

This simplification is due to the invariance property of the $F$ ensemble (Sec.~\ref{sec:overviewofmodels}).
However, it may be argued that the resulting universal behavior is robust \cite{grassberger1995damage, bagnoli1996damage} and also applies for more general models with temporal randomness.
The damage variables discussed here and in Sec.~\ref{sec:introducengt2damage} can of course be defined in a broader setting, for example in models without temporal randomness \cite{kauffman1969metabolic}.

Explicit derviations for general $n$ will be discussed in the following Sections. For now we just review the result for $n=2$, focussing on the 1D model for concreteness.
 
Recall that  each time a spin is updated,  two sites are used as inputs and and a single site is updated (Eq.~\ref{eq1plus1update}).
Correspondingly there are input damage variables $(\rho_\text{in}, \rho_\text{in}')$ and an output damage variable $\rho_\text{out}$.
Averaging over the random function that is applied, one finds that
\begin{align}\label{eq:n2update1D}
& \text{If $(\rho_\text{in}, \rho_\text{in}') = (0,0)$},
& 
& \text{then $\rho_\text{out} = 0$}
\\
& \text{If $(\rho_\text{in}, \rho_\text{in}') \neq (0,0)$},
& 
& \text{then } \rho_\text{out} = 
\left\{
\begin{array}{lll}
1  &  \text{with prob. $p_s$}    \\
0  &   \text{with prob. $1-p_s$.}  
\end{array}
\right.
\notag
\end{align}
The first line just says that if the two replicas agree on both of the inputs, they must necessarily agree on the output.
As a result the state where ${\rho(i)=0}$ for every site $i$ is an absorbing (``dead'') state that is unchanged by the stochastic dynamics.

Representing an active site by a particle, the rules above are 
\ba\label{eq:dpvisual1}
 (\emptysite, \emptysite)  \quad\,
 & \longrightarrow \quad
\emptysite 
\\  \label{eq:dpvisual2}
\left.\begin{array}{rl}
 (\emptysite, \ballZ)  \\
\text{or } (\ballZ ,\emptysite) \\
\text{or } (\ballZ ,\ballZ)
\end{array}\right\}
& \longrightarrow
\left\{
\begin{array}{ll}
\ballZ &  \text{with prob. $p_s$}    \\
\emptysite &   \text{with prob. $1-p_s$.}  
\end{array}\right. 
\end{align}
Here the left shows the initial state of the  two sites used for the input (for example, sites $i$ and $i+1$) and the right shows the updated state of one of these two sites (say, site $i$); the other of the two sites is also updated, independently, with the same probabilities.

The Markov process for $\rho_i$ defined by these rates maps to a standard model \cite{hinrichsen2000non, cardy1996scaling}  of  directed \textit{bond percolation} on the tilted square lattice.
Spacetime points  $(i,t)$, indicated by dots in  Fig.~\ref{fig:circuit},
map to \textit{bonds} of the tilted square lattice. (The bonds of the tilted lattice lie along the diagonal arrows of Fig.~\ref{fig:circuit}.)
A damaged site $(i,t)$ maps to a \textit{wetted} bond in the bond percolation problem.
There is a directed percolation \cite{hinrichsen2000non,odor2004universality,janssen2005field}  transition at \cite{jensen1999low}
\ba
p_s^*  & = 0.644700185(5).
\end{align}
For $p_s< p_s^*$ only the dead state is stable, and there is a finite characteristic timescale for activity to die out locally.
For ${p>p_s^*}$ there is an active state, with a nonzero average damage density $\<\rho\>$, which is stable in the thermodynamic limit.\footnote{In a large finite system activity will eventually be eliminated  even when $p>p_s^*$ because of  exponentially rare system-wide extinction events. However the timescale for this is exponential in $L$.}
At the critical point, 
the average damage density decays as $\< \rho \> \sim t^{-\alpha_2}$
if damage is initially present homogeneously throughout the system, where $\alpha_2 \simeq 0.16$ is the standard directed percolation  exponent  ``$\alpha$'' \cite{hinrichsen2000non,jensen1999low,wang2013high}.
(This decay is the upper curve in Fig.~\ref{fig:rho}, which is discussed below. See Sec.~\ref{sec:numerics} for more general initial conditions and observables.)

\subsection{Damage for $n>2$ replicas: Set partitions}
\label{sec:introducengt2damage}

Once we have ${n>2}$ replicas there are many more possibilities  for how sites or regions agree or disagree.
In general, the pattern of agreement/disagreement can be defined using a partition $\pi\in \Pi_n$, 
i.e. a splitting of the replicas ${1, \ldots, n}$ into some number of ``blocks'', 
with replicas that share the same state grouped into the same block of $\pi$.
We argue that, for general $n$, partitions give a useful labelling for the local damage state.

For ${n=2}$ the only possibilities are $\pi=(12)$, meaning that the replicas agree, and $\pi=(1)(2)$, meaning that they disagree.
For general $n$ the number of partitions is the Bell number $B_n$, which grows superexponentially with $n$:
$B_1 = 1$, $B_2=2$, $B_3=5$, $B_4=15$, $B_5=52\ldots$.

We define damage variables for all the nontrivial partitions,
\ba
& \rho_{\pi}(i), &  \pi & \in \Pi_n \setminus \{\mathbbm{1} \}.
\end{align}
Here $\mathbbm{1}$ denotes the coarsest partition, 
with all elements in the same block: this is the ``dead state'' where all replicas agree. $\rho_{\pi}(i)$ is 1 if site $i$ hosts partition type $\pi$ and 0 otherwise.

For a fixed number $q$ of local spin states (e.g. $q=2$ for a model of bits),
not all of the partitions $\pi$ can arise at a single site $i$. 
For example, when $q=2$ the possibilites at a single site are the
single-block partition $\mathbbm{1}$
and all the \textit{two-block} partitions 
(there are  ${2^{n-1}-1}$ of these).
For general $q$ we can have at most $q$ blocks in the partition, i.e. ${|\pi|\leq q}$.

Nevertheless, we can  define nontrivial densities $\rho_\pi$  for any partition $\pi$, 
simply by using more than one spatial site.
For example, for $q=2$, we can define nontrivial operators $\rho_\pi(x)$ on a \textit{bond} $x$ for $|\pi|\leq 4$.
(See Sec.~\ref{sec:simulationsnreplicaobservables} for a more explicit definition.)
Similarly we can define more complex operators using larger groups of sites.

For this reason, when we consider the RG, we will need to consider scaling operators associated with all possible $\pi$. 
It is also worth emphasizing that the universality class of the damage spreading transition does not depend on $q$ (this will be shown explicitly in the continuum treatment). Unless a given model suffers from further fine-tuning,\footnote{An example of a fine-tuned model is discussed in App.~\ref{app:finetuning}.} the value of $q$ is simply a ``microscopic detail''.

\begin{figure}
\includegraphics[width=\linewidth]{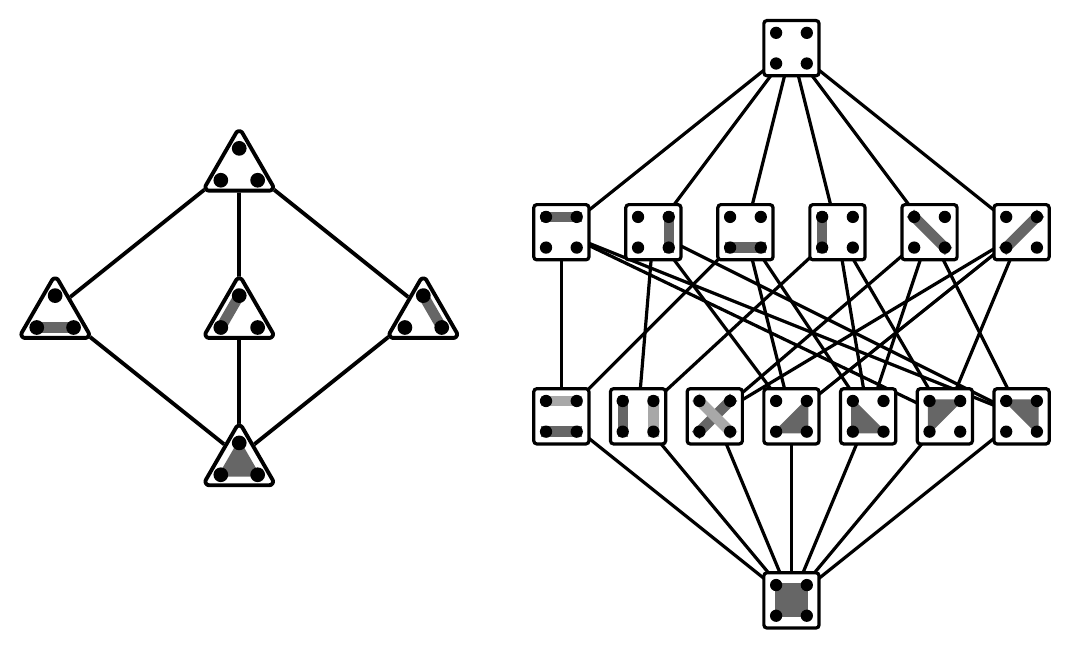}
\caption{Hasse diagrams for $\Pi_3$ and $\Pi_4$ \cite{stanley_enumerative_1999}:
${\pi< \sigma}$ iff $\pi$ can be reached from $\sigma$ by following the edges of the graph upward. The node at the bottom of a given diagram represents the partition ${\mathbbm{1}\in \Pi_n}$, i.e. an undamaged state where all replicas agree. The node at the top represents the partition $(1)(2)\ldots(n)$, i.e. a state where every replica differs from every other.}
\label{fig:hassediagrams}
\end{figure}

As for the case of $n=2$ replicas, the simple models we consider allow the dynamics to be reduced to an effective Markovian dynamics for the on-site damage variables (see below for an example).
In fact, when we perform simulations of the 1+1D model we will work directly with the physical spins $s_i$, 
so we will not need to use these effective  Markov rules explictly in 1+1D. 
But the mean-field version of the effective Markov processes (for $n>2$) will be our starting point for deriving continuum descriptions in Sec.~\ref{sec:continuummeanfield}.

The set $\Pi_n$ of partitions admits a partial order which will be important for the RG (Sec.~\ref{sec:rgstructure}).
Fig.~\ref{fig:hassediagrams} shows the ``Hasse diagrams''  that illustrate this partial order for
for $n=3$ and $n=4$ \cite{stanley_enumerative_1999}.
The coarsest partition, $\mathbbm{1}$, is at the bottom.
We write ${\pi< \sigma}$ if $\pi$ can be reached from $\sigma$ by splitting blocks: in this case,  $\pi$ can be reached from $\sigma$ by following the edges in the Hasse diagram upwards.
The second-lowest level of each diagram contains the two-block partitions, which will play a special role in the field theory treatment (Sec.~\ref{sec:continuummeanfield}).

\subsection{Example of damage dynamics: $n=3$} \label{sec:effectivedynamicsn3}

For illustrative purposes only,  we summarize the   rates for this Markov process in the case ${n=3}$,
for the 1D model with $q=2$.
In this model the only possible partition states at a given site are 
\ba\label{eq:n3states}
& \mathbbm{1} = \emptysite , & 
& (1)(23) = \ballA, 
& 
& (2)(13)  = \ballB,
& 
& (3)(12) =  \ballC,
\end{align}
with ${\mathbbm{1} =(123)}$ representing the completely dead/passive state.
We have represented the other three states as three types of active particle.
In a model with ${q>2}$, we would have an additional state ${(1)(2)(3)}$ at each site; however we will show  that the simple ${q=2}$ model captures the generic universality class of the damage spreading transition (Sec.~\ref{sec:fieldtheories}). 
The fact that all the 
damage types in Eq.~\ref{eq:n3states} 
(the three colors of particle)
are related to each other by symmetry is a special feature of the ${n=3}$ case; for ${n>3}$ we would have nonequivalent damage types even for ${q=2}$, as a result of the nonequivalent shapes of two-block partitions in Fig.~\ref{fig:hassediagrams}.

If, in a given 
update of a site,
the two ``input'' sites involve at most a single type of active particle
(a single color in Eq.~\ref{eq:n3states})
then the update rule reduces to Eqs.~\ref{eq:dpvisual1},~\ref{eq:dpvisual2}
but for particles with the given color.\footnote{This is because, in this situation, at least two replicas are identical,
and may effectively be considered a single replica. This idea will be useful in 
Sec.~\ref{sec:fieldtheoryfinitedgeneraln} and Sec.~\ref{sec:rgstructure}.}
But when the input involves two distinct types of activity (it does not matter which two so long as they are distinct) then 
\ba\label{eq:3siteupdatepictures}
\left.
\begin{array}{r}
 (\ballA, \ballB)  \\
\text{or }  (\ballA ,\ballC) \\
\text{or }  (\ballB ,\ballA) \\
\text{or }  (\ballB ,\ballC) \\
\text{or }  (\ballC ,\ballA) \\
\text{or }  (\ballC ,\ballB)
\end{array}
\right\}
\longrightarrow
\left\{
\begin{array}{ll}
\ballA &  \text{with prob. $p_s/2$}    \\
\ballB &  \text{with prob. $p_s/2$}   \\
\ballC & \text{with prob. $p_s/2$}     \\
\emptysite & \text{with prob. $1-3p_s/2$.}
\end{array}\right. 
\end{align}

\begin{figure}
\includegraphics[width=\linewidth]{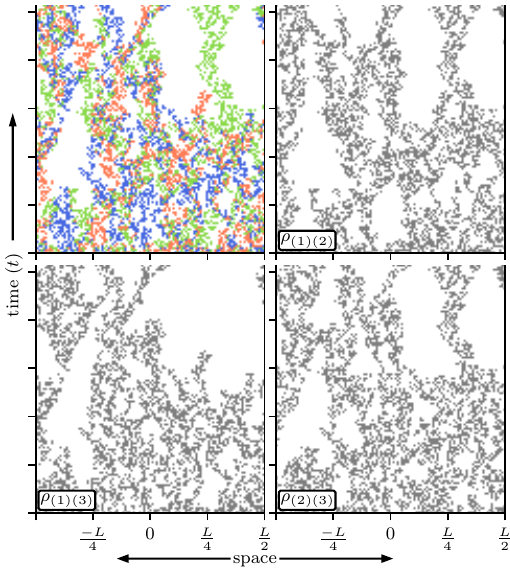}
\caption{Top left: a damage trajectory for three replicas with the three nontrivial partitions of the form ${(a)(bc)}$ represented by three colors (Eq.~\ref{eq:n3states}). Other panels: neglecting any one of the three replicas gives a two-replica damage trajectory, as explained below Eq.~\ref{eq:3siteupdatepictures}. Here $L=256$ and $t=256$.}
\label{fig:traj}
\end{figure}

An example of a spacetime trajectory for this dynamics, at its critical point, is shown in the upper left panel of Fig.~\ref{fig:traj}.
The other panels show the two-replica damages, for the same configuration. 
These are obtained by ``forgetting about'' one of the replicas. 
For example, since  $\rho_{(1)(2)}$ is nonzero if replica 1 is different from replica 2 (irrespective of the state of 3), 
$\rho_{(1)(2)}$ is equal to the \textit{sum} of the densities of the $\ballA$ particles and the $\ballB$ particles.
(The possibility of ``forgetting'' one of the replicas will be important in the structure of the continuum description and the RG in Secs.~\ref{sec:continuummeanfield}--\ref{sec:rgstructure}.)

If we only consider the upper right or one of the lower panels on its own, then it has directed percolation statistics. So the upper left panel of Fig.~\ref{fig:traj} determines three different ``directed percolation configurations'', but they are nontrivially correlated with each other (see the brief comment in Sec.~\ref{sec:conclusions}).

In this ${q=2}$ model the three-block partition ${(1)(2)(3)}$ cannot appear at a site, but it appear on a bond (Sec.~\ref{sec:introducengt2damage}): a given bond has partition type ${(1)(2)(3)}$ if it hosts two particles with different colors.

The analogous effective dynamics for ${n=4}$  replicas of the present 1+1D model  has eight states per site, 
corresponding to the undamaged state and the seven possible two-block partitions depicted  in the second level of Fig.~\ref{fig:hassediagrams}, Right.

\section{Overview of approaches}
\label{sec:overview}

Since the theoretical approaches in the 
following Sections are independent of each other to a large extent, we give an overview here.

{\bf Simulations.} In Sec.~\ref{sec:numerics} we simulate the 1+1D model,  demonstrating the existence of new critical exponents for the decay of damage with time, starting from a globally damaged state, and for the survival probability of localized initial damage. 

{\bf Mean field theory.} In Sec.~\ref{sec:continuummeanfield} we derive the mean field theory of damage spreading for an arbitary number $n$ of replicas. 
This is applicable to all-to-all-coupled models or in sufficiently high dimensions.
An example of a mean-field result is the spectrum of mean-field density decay exponents
$\rho_\pi(t) \sim t^{-\alpha_\pi}$, where ${\alpha_\pi=\lceil\log_2(|\pi|)\rceil}$, and $|\pi|$ is the number of blocks in partition~$\pi$.

{\bf Finite-dimensional Lagrangians.} In Sec.~\ref{sec:fieldtheories} we write field theories valid in finite dimensions $d$,
showing for example that the mean field exponents receive corrections in ${d<4}$. This Section builds on the mean-field results.

{\bf Structure of the damage spreading fixed point.}
In conventional critical theories, 
observables (``scaling operators'') are classified by global symmetries. 
Interestingly, the damage-spreading problem  permits a different kind of classification of observables, based on the partial order on the set of partitions (recall Fig.~\ref{fig:hassediagrams}). 
This property is demonstrated, and shown to be consistent with simulations, 
in Sec.~\ref{sec:rgstructure}.
 It allows some exact statements about the spectrum of critical exponents (for example that the decay exponents $\alpha_\pi$ depend only on $|\pi|$).
The treatment  does not rely on  continuum field theory, so is independent of the two preceding Sections.

In Sec.~\ref{sec:trs} we use the field theory formalism to discuss an unexpected time-reversal symmetry which imposes a relation between scaling dimensions for observables involving $n=3$ replicas. Sec.~\ref{sec:conclusions} outlines possible future directions.

\section{Simulation results in 1+1D}
\label{sec:numerics}

Here we describe  simulations of the 1+1D model sketched in Sec.~\ref{sec:overviewofmodels}, for $n\leq 4$ replicas.  
We define observables in general terms and then specify some details of the lattice setup.
Recall that the updates involve random functions applied to pairs of bits: the  probabilities with which we choose the various possible random functions for each update are detailed in Sec.~\ref{sec:choiceof1Dmodel}.
This gives a one-parameter phase diagram. 
We denote the distance to the critical point in this phase diagram by $\varepsilon$; the  survival probability $p_s$ appearing in
in Eqs.~\ref{eq:dpvisual2},~\ref{eq:3siteupdatepictures} is given  by  ${p_s = p_s^* - \varepsilon/6}$. Simulation results are in Sec.~\ref{subsec:numresults}.

\subsection{Review: scaling in directed percolation}\label{sec:dpreview}

Let us first  recall
some standard observables in directed percolation \cite{hinrichsen2000non}, which describes the two-replica problem.

The late-time limit of the density $\varrho(t) = \<\rho(x,t)\>$ of active (i.e. damaged) sites provides one natural order parameter for the transition.
Consider  an infinitely large system that at ${t=0}$ has activity everywhere in space.
Then $\varrho(t)$ decays to zero 
in the ``healed'' phase,  
whereas it saturates to a finite stationary density 
$\varrho_\infty=\lim_{t\to\infty} \varrho(t)$
in the active (damage-spreading) phase. 
The order parameter vanishes with an exponent $\beta$ as the transition is approached,
\eq{\label{eq:orparam}
\varrho_\infty\sim (-\varepsilon)^\beta\,,
}
where $\varepsilon$ denotes the distance from criticality
in the parameter space,
with ${\varepsilon<0}$ corresponding to the active  phase.
The characteristic
lengthscale $\xi_\perp$ and timescale 
 $\xi_\parallel$ for the dynamics 
diverge near the critical point as 
\ba
\xi_\perp & \sim |\varepsilon|^{-\nu_\perp}
\,,
& \xi_\parallel\sim |\varepsilon|^{-\nu_\parallel}\, ,
\end{align}
and the dynamical exponent 
$z$ is defined as~${z=\nu_\parallel/\nu_\perp}$.
Standard scaling arguments give the dynamic scaling form 
\eq{
\varrho(t)\sim t^{-\alpha} f\lf \varepsilon t^{1/\nu_\parallel}\ri
\label{eq:dyn-sca1}
}
for some scaling function $f$,
with ${\alpha = \beta/\nu_\parallel}$ in order to match (\ref{eq:orparam}).
At the critical point we have ${\varrho(t)\sim t^{-\alpha}}$.\footnote{In field theory language, $\varrho(t)$ is a one-point function of the field $\rho(x,t)$,
taken with a certain initial-time boundary condition. 
Denoting the scaling dimension (inverse length dimension) of $\rho(x,t)$ by $\Delta$, we have ${\alpha = \Delta/z}$ and ${\beta = \Delta \nu_\perp}$ (so long as we are below the upper critical dimension, where hyperscaling  holds \cite{cardy1980directed};  mean field exponents above the upper critical dimension are discussed in Sec.~\ref{sec:continuummeanfield}).} 

An alternative order parameter is the probability ${\cal S}(t)$ that activity/damage which is seeded \textit{locally} in an otherwise empty state
survives  until time $t$ \cite{hinrichsen2000non}. 
The limit ${{\cal S}_\infty \equiv \lim_{t\to\infty}{\cal S}(t)}$
is nonzero in the active phase, 
with 
${\cal S}_\infty\sim (-\varepsilon)^{\beta^\prime}$.
We have the scaling form
\eq{
{\cal S}(t)\sim t^{-\delta}\tilde f\lf \varepsilon t^{1/\nu_\parallel}\ri,
\label{eq:dyn-sca2}
}
(for ${\delta = \beta'/\nu_\parallel}$)  and the critical decay ${{\cal S}(t)\sim t^{-\delta}}$.

This gives four (a priori) independent exponents in the ${n=2}$ problem, 
which we can take to be 
the time-decay exponents
$\alpha$ and  $\delta$,   the correlation length exponent 
$\nu_\perp$, and the dynamical exponent $z$. 
In fact, a time-reversal symmetry of directed percolation \cite{hinrichsen2000non} imposes the relation $\alpha=\delta$.\footnote{(And the equality of the scaling functions $f$ and $\tilde f$.)}
In field theory language, the exponents $\alpha$ and $\delta$ are 
proportional to the scaling dimension of the density field and of the dual ``response'' field, respectively, and they are forced to be equal because a time-reversal symmetry exchanges the two fields (see  Sec.~\ref{sec:trs}).

\subsection{Lattice observables for ${n>2}$}
\label{sec:simulationsnreplicaobservables}

Let us describe the generalizations of the abovementioned activity density and survival probability that we will use in our simulations.

In Sec.~\ref{sec:introducengt2damage} we have defined damage observables $\rho_\pi$ for ${n>2}$ replicas in general terms. 
Let us be slightly more concrete.
Our physical state variables  bits, ${s_i\in\{0,1\}}$, on the lattice sites. This means that only  partitions with two blocks are possible for single-site damage operators. 
To allow for non-trivial partitions 
for ${n>2}$, 
we  define variables $S_i^{(k)}$,
which run over $2^k$ possible values, 
and which encode  the state of $k$ consecutive sites. For example, a possible explicit labelling is 
\eq{
S_i^{(k)} = \sum_{l=0}^{r-1}2^{l}s_{i+l}\,.
\label{eq:tau-i-r}}
We may then compute the damage densities 
$\rho_\pi^{(k)}(i)$ from the equality/inequality of the $S^{(k)}_i$ values for the various replicas.
In fact, as we will restrict to ${n\leq 4}$ replicas, it is sufficent to consider damage densities defined on \textit{bonds} $(i,i+1)$ of the 1D chain, which we denote by $\rho_\pi(x,t)$.

A given range $k$ allows all possible damage types (all possible partitions)  for up to ${n=2^k}$ replicas to have nontrivial densities.
However, in this Section we will present results only for the finest partition
for each $n$, viz. $\rho_{(1)(2)\cdots(n)}$, where there are as many blocks as the number of replicas, each block containing only one replica.  This is sufficient to see the emergence of new exponents $\alpha_n$ for each ${n\in \{2,3,4,5,\ldots\}}$.

We defer numerical results for more general (coarser) partitions $\pi$ until after we have discussed how to construct scaling fields in  Sec.~\ref{sec:rgstructure}.
This is because the damage operators for coarser partitions are not scaling operators on their own; instead, scaling operators $\mathcal{O}_\pi$ are formed by  taking appropriate linear combinations of the form (Sec.~\ref{sec:rgstructure})
\be
\mathcal{O}_\pi = \rho_\pi + (\text{densities of finer partitions}).
\ee
We will see in Sec.~\ref{sec:rgstructure} that this structure is fixed by general RG arguments and dictates power law forms that are in agreement with simulations.
For now we restrict to the $\rho_{(1)(2)\cdots(n)}$.

We denote the density of this finest damage type, starting from a state in which each replica has an independent uniformly random configuration, 
by $\varrho_n(t)$, and denote the corresponding decay exponent (see Eq.~\ref{eq:dyn-sca1})  by $\alpha_n$. That is, at criticality,
\eq{
\varrho_n(t) = \Braket{\rho_{(1)(2)\cdots(n)}(x,t)}\sim t^{-\alpha_n}\,.
\label{eq:rho-n-t}
}
and more generally 
\eq{
\varrho_n(t) \sim t^{-\alpha_n}
f_n\lf \varepsilon t^{1/\nu_\parallel}\ri
\,.
\label{eq:rho-n-t-scaling}
}
$\{\alpha_n\}$ is the first set of exponents that we will examine below.
We expect on theoretical grounds\footnote{Since $\varrho_n$ and $\varrho_{n'}$ can be regarded simply as different observables in the $m$-replica theory, for any ${m\geq \operatorname{max}\{ n,n'\}}$~(Sec.~\ref{sec:rgstructure}).} and will confirm  numerically that  the exponents $\nu_\parallel$ and $\nu_\perp$ are independent of the number of replicas.

We next define the survival probability 
$\mathcal{S}_n(t)$
for multiple replicas, starting from a state with  damage on only a single bond.
For $n=2$,
${\cal S}_2(t)$ is the probability 
that there is  at least one damaged site at time $t$.
Alternately we can say that 
${\cal S}_2(t)$ is the probability that the ``global'' (system-wide) state of the two  replicas differs  at time $t$.  

For $n$ replicas, we define $\mathcal{S}_n$ as the  probability that
all of the $n$ replicas have distinct  global states at time $t$. Explicitly, if we write
\eq{
\mu^{ab}(t) = \begin{cases}
        1\,;&\exists~i~{\rm s.t.}~s_i^a (t)\neq s_i^b(t)\\
        0\,;&{\rm otherwise}
        \end{cases}\,,
}
then (the initial condition is implicit)
\eq{
{\cal S}_n(t) = \Braket{\prod_{1\leq a < b\leq n} \mu^{ab}(t)}\,.
\label{eq:surv-prob}
}
Note that this is does not require 
that we are able to distinguish all $n$ replicas by looking at any \textit{individual}  site or bond.
As an example, the configurations
\eq{
{\rm replica }~1\,:~~ 001000000\nonumber\\
{\rm replica }~2\,:~~ 001000010\nonumber\\
{\rm replica }~3\,:~~ 000000010\nonumber
}
are not all distinguishable on any single bond, but are  globally different.\footnote{For $q>2$ (e.g. trits ${s_i=0,1,2}$), it is possible to have  three-replica damage at a single site, whereas for ${q=2}$ at least 2 sites are needed. Global 3-replica damage can be due either to configurations similar to the one above, where no individual site/bond has three replica damage, or to configurations which do have 3-replica damage at some single site/bond. In fact scaling arguments like that in App.~\ref{app:latticetrs} suggest that $\mathcal{S}_3(t)$ decays with  the same exponent as $\mathcal{S}_3^\text{loc}(t)$, defined as the probability of configurations only of the second kind.}
[An alternative to (\ref{eq:surv-prob}) is to write ${\cal S}_n(t)$ as 
an average damage density 
$\varrho^{(L)}_n(t)$ 
similar to Eq.~\ref{eq:rho-n-t}, but for a global damage operator  defined via $S_1^{(r)}$ with $r=L$.]

We will denote the decay exponents for ${\cal S}_n(t)$ by $\delta_n$.

\subsection{Choice of 1+1D model}
\label{sec:choiceof1Dmodel}

The basic structure of the model has been described above around Eqs.~\ref{eq1plus1update},~\ref{eq1plus1update2}. It remains to specify the probability distribution $P(F)$ from which each update function $F_{i,t}$ is drawn.
We will first describe a model with two parameters, and then specify to a line in this two-parameter space.

Functions ${F:\,\{0,1\}^2 \rightarrow \{0,1\}}$ may be divided into three types, according to their degree of reversibility.
(Note that there are four possible values $\in \{0,1\}^2$ for the input of a function.)
``Type 0'' functions are maximally irreversible: they give the same output regardless of their input.
``Type 1'' functions give the same output for three  inputs but a different output for the fourth,
and 
``Type 2'' functions take the value $F=0$ on two inputs and $F=1$ on the other two.\footnote{An example of a type 0 function is one where ${F(0,0)=F(0,1)=}{F(1,0)=F(1,1)=0}$.
A type 1 example is ${F(0,0)=F(1,1)=F(0,1)=0}$, $\,F(1,0)=1$.
A type 2 example is ${F(0,0)=F(1,1)=0}$, ${\, F(1,0)=F(0,1)=1}$.}

We take functions of the same type to be equally likely. 
Then the model is fully defined by specifying the probabilities $p_0$, $p_1$, $p_2$ (with ${p_0+p_1+p_2=1}$) 
of the three types.
Considering the survival of damage in the 2-replica problem gives,\footnote{In more detail: if we assume that damage is initially present on the pair of sites used as input for the update, i.e. that  ${(s^1,{s^1}')\neq (s^2,{s^2}')}$, then the probability of damage surviving on the updated site is easily checked to equal the RHS of Eq.~\ref{eq:pschoiceofmodelsection}.
In turn, a standard mapping relates 
the spacetime configuration of damaged sites to a  directed percolation configuration on the tilted square lattice 
(the tilted square lattice formed by the arrows in Fig.~\ref{fig:circuit}).} in the notation introduced in Eq.~\ref{eq:n2update1D},
 \be\label{eq:pschoiceofmodelsection}
p_s = \f{1}{2} p_1 + \f{2}{3} p_2 .
 \ee 

The parameters $(p_1, p_2)$ define a two-dimensional parameter space of models. 
The line  
\be\label{eq:criticalline}
\f{1}{2} p_1 + \f{2}{3} p_2 = 0.644700185(5)
\ee
 separates the active phase, which 
 is at larger values of the LHS, from
 the inactive phase.
 (For the present models, the active phase occupies only a small fraction of the phase diagram.)
 Eq.~\ref{eq:criticalline} follows from   the mapping of the 2-replica problem to directed percolation \cite{derrida1986random}, together with Jensen's estimate of the  directed percolation threshold \cite{jensen1999low}.

At first we might have guessed that the universal properties would be equivalent for any point on the critical line. 
This is the case for 2-replica observables, as shown by the mapping to directed percolation, 
and also  for 3-replica observables, as shown by explicitly computing the rates in the effective Markov process, which depend only on $p_s$ (Sec.~\ref{sec:effectivedynamicsn3}).

However, when we consider observables with ${n>3}$ replicas, we notice that the ${p_1=0}$ boundary of the phase diagram is fine-tuned.
The circuits with $p_1=0$ are not sufficiently generic: 
there are constraints on the dynamics which force certain damage correlation functions to vanish (these are described in App.~\ref{app:finetuning}).

\begin{figure}[t]
\includegraphics[width=\linewidth]{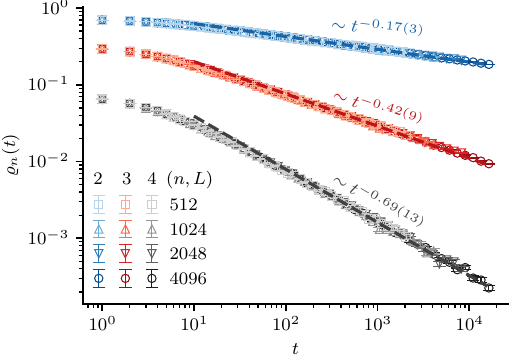}
\caption{The density of damages, $\varrho_{n}(t)$, for $n=2$, $3$, and $4$ replicas,  as a function of time $t$ at the critical point. Different colours correspond to the different numbers of replicas, $n$, whereas different system sizes, $L$, are denoted by different markers (and colour intensities). The dashed lines show the fits used to extract the $\alpha_n$ exponents, values of which are also mentioned in the figures. All data are averaged over 500 realisations of the dynamics.}
\label{fig:rho}
\end{figure}

Therefore we expect that the \textit{endpoint} of the critical line with ${p_1=0}$ is a fine-tuned multicritical point, while the rest of the critical line, with ${p_1>0}$,
is all described by the same generic universality class for damage spreading. 

Our aim is to study the universal properties of the generic damage-spreading transition. 
We therefore perform all simulations at and around the point on the critical line which is as far away from the fine-tuned boundary of the  phase diagram as possible:
\be
(p_{1,c}, p_{2,c}) \simeq (0.131799,0.868201)\,.
\label{eq:crit-1+1D}
\ee
At this point we have $p_0=0$, but we can argue that this does not represent fine-tuning in the RG sense (App.~\ref{app:finetuning}).
To tune away from the critical point, we will parametrise our problem as
\ba
p_1 = p_{1,c}+\varepsilon\,,~~p_2 = p_{2,c}-\varepsilon\,,
\end{align}
with $(p_{1,c}, p_{2,c})$ given in Eq.~\ref{eq:crit-1+1D}, so that ${p_s = p_s^* -\varepsilon/6}$.

\begin{figure}[t]
\includegraphics[width=\linewidth]{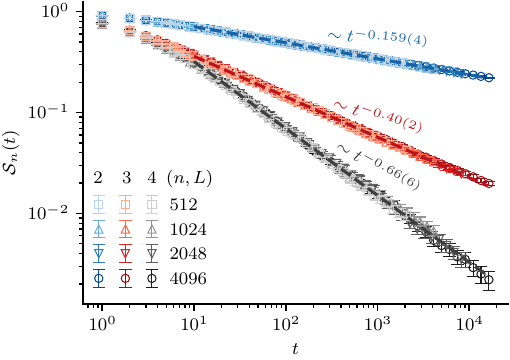}
\caption{The survival probability, ${\cal S}_n(t)$, defined in Eq.~\ref{eq:surv-prob}, as a function of time $t$, for different numbers of replicas $n$ (different colours), and different system sizes $L$ (different markers and colour intensities) at the critical point. The dashed lines denote fits to extract the $\delta_n$ exponent as ${\cal S}_n(t)\sim t^{-\delta_n}$ with the values mentioned in the figure. All data are averaged over 1000 realisations of the dynamics.}
\label{fig:surv-prob}
\end{figure}

\subsection{Numerical results}
\label{subsec:numresults}

Let us now present the numerical results for the 1+1D model defined above, for  ${n\leq 4}$. 
We begin with simulations at the critical point, ${\varepsilon=0}$, for initial conditions where each replica is initialised uniformly randomly and independently.
This  leads to a macroscopic fraction of damaged sites in the initial state.
We define $\varrho_n(t)$ with respect to two-site variables, i.e. $S_i^{(r)}$ with $r=2$ (see Eq.~\ref{eq:tau-i-r}).

The results for $\varrho_n(t)$  are shown in Fig.~\ref{fig:rho} for $n=2$, $3$, and $4$.
Note that the data is well converged with $L$ (until the largest $t$ considered) and falls onto clear power-law decay, 
$\varrho_n(t)\sim t^{-\alpha_n}$,
consistent with the scaling ansatz in Eq.~\ref{eq:rho-n-t}.
The exponents can be extracted from linear fits
to the data on logarithmic axes.
As expected, $\alpha_2$ is consistent with the usual directed percolation value~${\alpha\simeq 0.159}$~\cite{hinrichsen2000non}.
Importantly, the exponents $\alpha_n$  show a clear  hierarchy with $n$.

Fig.~\ref{fig:rho} is our first concrete evidence of the emergence of new universality classes in damage spreading upon the introduction of more replicas. 

Later in Sec.~\ref{sec:generalpartitionssimulations} we will show data for densities associated with more general partitions, such as $\rho_{(12)(3)(4)}$ in the ${n=4}$ theory, for which the number $k$ of blocks is smaller than $n$. 
We wil show that these more general densities are governed by the same decay exponents $\alpha_k$ found above, but have additional subleading contributions.

\begin{table}[b]
\begin{tabular}{c||c|c|c|c|c|}
$n$ & $\alpha_n$ & $\delta_n$ &$\nu_\perp$ & $\nu_\parallel$\\
\hline
2 &0.17(3) & 0.159(4) & 1.734(1) & 1.097(1)\\
3 & 0.42(9) & 0.40(2)& 1.734(1)& 1.097(1)\\
4 & 0.69(13) & 0.66(6)& 1.734(1)& 1.097(1)\\
\end{tabular}
\caption{Table summarising the numerically obtained critical exponents for the 1+1D lattice model. The exponents for $n=2$ are standard directed percolation exponents.}
\label{tab:1+1D-exp}
\end{table}

\begin{figure}
\includegraphics[width=\linewidth]{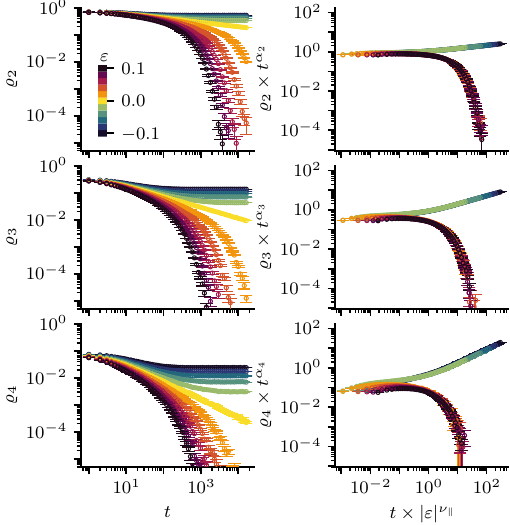}
\caption{Off-critical scaling of the damages $\varrho_n$ for $n=2$, $3$, and $4$ (rows). The left panels shows bare data for different values of $\varepsilon$, indicated by different colours (colourscale in  upper left panel).
The right panels show that the data for different $\varepsilon$ collapse onto two common curves (for $\varepsilon\gtrless 0$) when rescaled with $t^{\alpha_n}$ and plotted against $t|\varepsilon|^{\nu_\parallel}$; see scaling ansatz in Eq.~\ref{eq:rho-n-t-scaling}. The best collapse is obtained for $\alpha_n$ values given in Fig.~\ref{fig:rho} and ${\nu_\parallel}=1.734(1)$ for all $n$ which is also the exponent for conventional directed percolation. All data are for $L=4096$ and averaged over 500 realisations of the dynamics.}
\label{fig:rho-dyn-sca}
\end{figure}

\begin{figure}
\includegraphics[width=\linewidth]{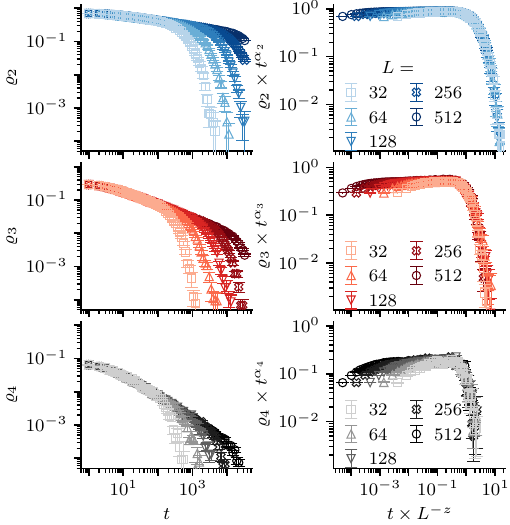}
\caption{Finite-size scaling for $\varrho_n(t,L)$ at the critical point. The different rows correspond to different $n$. The left panels show the bare data at the critical point for different $L$ (different markers and colour intensities, see legends). The data for different $L$ collapses onto each other according to the finite-size scaling ansatz ${\varrho_n(t,L)= t^{-\alpha_n} \hat f_n(t L^{-z})}$. The dynamical exponent $z$ giving the best collapse is consistent with the  standard directed percolation value. }
\label{fig:fss-crit}
\end{figure}

Next we extract the $\delta_n$ exponents (or equivalently the exponents ${\beta^\prime_n=  \delta_n \nu_\parallel}$). 
For this, we analyze the   survival probabilities 
defined in Eq.~\ref{eq:surv-prob}, which decay as  ${{\cal S}_n(t)\sim t^{-\delta_n}}$ at the critical point.

To  seed local initial damage, 
we take initial conditions which are identical for each replica except on a single bond (sites $i=L/2$ and $L/2+1$), 
where all the replicas differ.\footnote{The configurations are given by
\eq{
s_{L/2}^a(t=0) &= \lfloor(a-1)/2\rfloor\,,\nonumber & 
s_{L/2+1}^a(t=0) &= (a-1)\,{\rm mod}\, 2\,.
}}
The results are shown in Fig.~\ref{fig:surv-prob}.
The data is well converged with system size and is consistent with the expected power law forms.

Our exponent estimates for $\alpha_n$ and $\delta_n$ are summarised in Table~\ref{tab:1+1D-exp}.
Interestingly, the values of the exponents $\delta_n$ are consistent with the exponents $\alpha_n$ within error bars.
This raises the question of whether there is a  symmetry of the model relating these exponents (for a given $n$). We discuss this in Sec.~\ref{sec:trs}.
The time reversal symmetry of the $n=2$ case, imposing ${\alpha_2=\delta_2}$, is well-known in the context of directed percolation. In Sec.~\ref{sec:trs} we demonstrate that there is also a time-reversal symmetry for ${n=3}$,  imposing ${\alpha_3=\delta_3}$.

Next  we show data for the densities $\varrho_n(t)$ \textit{away} from the critical point, showing explicitly that all of the damage densities $\varrho_n(t)$ act as ``order parameters'' for a common transition at ${\varepsilon=0}$ and that they have the expected finite-size scaling forms. 

 Fig.~\ref{fig:rho-dyn-sca} confirms that each $\varrho_n$ obeys the scaling ansatz in Eq.~\ref{eq:rho-n-t-scaling} as a function of $\varepsilon$ and $t$.
The bare data for different values of $\varepsilon$, around the critical point, both in the damaged and healed phases, are shown in the left column.
For ${\varepsilon<0}$, $\varrho_n(t)$ saturates to a finite constant as ${t\to\infty}$: 
this is the damaged phase with a finite fraction of damaged sites in the stationary state.
In contrast $\varrho_n(t)$ decays rapidly with $t$ for ${\varepsilon>0}$ (faster than a power-law) indicating the healing of damage.
These two behaviors are separated by the critical point, with the power-law decay discussed above. 
Upon scaling $\varrho_n(t)$ with $t^{\alpha_n}$ and $t$ with $|\varepsilon|^{\nu_\perp}$, we find that the data for different $\varepsilon$ collapses onto two common curves for $\varepsilon\gtrless 0$, as shown in the right panels of Fig.~\ref{fig:rho-dyn-sca}.
The best scaling collapse is achieved with the $\alpha_n$ obtained from Fig.~\ref{fig:rho} and with ${\nu_\parallel=1.734(1)}$ for all $n$.  This confirms the theoretical expectation that $\nu_\parallel$ takes the same value for all $n$, and is equal to the conventional directed percolation value~\cite{hinrichsen2000non}.

The data above reflected the inifinite system-size limit.
For completeness we now give data for finite-size scaling of $\varrho_n(t)$ at criticality.\footnote{The scaling ansatz in Eq.~\ref{eq:rho-n-t-scaling} is valid in the infinite system-size limit, or more generally for 
${t\ll L^z}$, where ${z = \nu_\parallel/\nu_\perp}$ is the dynamical exponent. Recall that the general scaling form for finite $L$ is 
$\varrho_n(t,L)\sim t^{-\alpha_n} f_n(\varepsilon t^{1/\nu_\parallel},tL^{-z})$,
or, at the critical point,
$\varrho_n(t,L) \sim L^{-\alpha_n}\hat f_n(tL^{-z})$.}
This yields the exponents  $z$ and  ${\nu_\perp=\nu_\parallel/z}$, which  match the directed percolation values as expected.

The left panels of Fig.~\ref{fig:fss-crit} show bare data for $\varrho_n(t,L)$ for $n=2$, $3$, and $4$ for several $L$ at the critical point, and the right panels show scaling collapses as a function of the scaling variable $t/L^z$. 
The results show that the critical power law $\varrho_n(t,L)\sim t^{-\alpha_n}$ holds up to a timescale of order ${L^z}$ and that common exponents $\nu_\parallel$ and $z$ describe all~$n$.
The best collapse is obtained for $\alpha_n$ values obtained earlier in Fig.~\ref{fig:rho} and $z=1.581$  for all $n$ consistent with the directed percolation value.

\section{Continuum descriptions: \\ Mean field case}
\label{sec:continuummeanfield}

We will argue that, in finite dimensions, a continuum description of the damage spreading transition may be given in terms of Langevin equations for the damage variables, or equivalently in terms of a Martin-Siggia-Rose-Janssen-de Dominicis field theory \cite{janssen1976lagrangean,bausch1976renormalized}.
But it is simplest  to first treat  the  ``mean-field''-like model of Sec.~\ref{sec:overviewofmodels}, 
where $N$ sites interact without any constraint from  spatial locality.
In this case we obtain Langevin equations for densities $\rho_\pi(t)$ that depend only on time.
In the thermodynamic limit of the mean field model, ${N\to \infty}$, the noise vanishes and these equations become simple ordinary differential equations specifying a deterministic decay of the damage densities, given an initial state.

Anticipating the finite-dimensional case: 
The Langevin equations that we will obtain
generalize the standard Langevin description of directed percolation \cite{hinrichsen2000non,odor2004universality}.
Let us recall this. It is a stochastic PDE for the density of the schematic form 
(here $D$, $r$, $g$ and $\nu$ are model-dependent constants) 
\be\label{eq:DPlangevin}
\partial_t \rho(x,t) = D \nabla^2 \rho(x,t) + r \rho(x,t) - g \rho(x,t)^2 + \eta(x,t)
\ee
with noise of variance
\be\label{eq:DPlangevinnoise}
\< \eta(x,t) \eta(x',t') \> = \nu \rho(x,t) \delta(x-x') \delta(t-t').
\ee
The version of this equation
for a mean field version of directed percolation (with all-to-all interactions between $N$ sites)
is similar, but  without the spatial dependence, and with a noise strength that vanishes in the thermodynamic limit (see below).

Both in mean field theory, and for field-theoretic renormalization group, we can drop terms which are higher-order in $\rho$ than the ones shown in Eqs.~\ref{eq:DPlangevin},~\ref{eq:DPlangevinnoise}.
Within mean-field theory this is justified by the smallness of $\rho$ near the critical point, while in a finite number of dimensions it is justified on the grounds of RG-irrelevance.

The crucial parameter driving the transition is $r$, whose value  is a function of the microscopic rates.
In mean field theory the critical point is at ${r=0}$  and more generally the critical point is where the renormalized value of $r$ vanishes. 

Note that, thanks to the form of the noise variance, the RHS of 
(\ref{eq:DPlangevin}) vanishes when ${\rho=0}$: 
this is inevitable since ${\rho=0}$ is an absorbing state.

Eq.~\ref{eq:DPlangevin} may be obtained in a quantitatively controlled way, for a class of models, by first examining the mean field limit, and then adding spatial structure.
Below we perform the analogous steps for the $n$-replica damage problem.
In place of the  activity density $\rho$ of directed percolation we have the damage densities $\rho_\pi$. As we will see there are some slight surprises.

\subsection{General form of mean field equations}
\label{sec:meanfieldeqnsgeneralform}

In Sec.~\ref{sec:overviewofmodels} we described a class of classical circuits  that are ``mean field'', in the sense that any site can serve as an input in the update of any other site --- there are no restrictions from spatial locality.

Recall that each physical update is of the form 
\be
 s_\text{out}  = F_t(s_\text{in}, s_\text{in}'),
\ee
taking the states of \textit{two} sites as input in order to update a single ``output'' site.
When we consider $n$ replicas of the physical dynamics, the important variables are the damage types of the sites. 
In the present class of models, the effective Markovian dynamics of the damage variables are fully determined by the conditional probability 
\be\label{eq:Ppisigmasigma}
P(\pi | \sigma, \sigma')
\ee
that the ``output'' has damage type $\pi$, \textit{given that} the two input sites have damage types $\sigma$ and $\sigma'$.
Unless otherwise specified, the number of replicas $n$ can be arbitrary in the following. 
It is not necessary to explicitly label quantities like $P(\bullet|\bullet,\bullet)$ with an $n$ value, since the value of $n$ can be read off from the inputs.

The precise form of this probability depends on the specific model we choose. 
In order to fix  the model, we must specify, first, the number $q$ of physical states per site: for example $q=2$ for a Boolean circuit acting on bits.
Second, we must specify a probability distribution $P(F)$ for the update functions. 
Each update is equivalent to a logic gate, and choosing $P(F)$ amounts to choosing the probabilities for the different kinds of logic gate in our randomly-assembled circuit.
Our formalism assumes that two logic gates have the same probability if they are related by a permutation of the state labels ${\{1,\ldots, q\}}$,\footnote{More formally: Our explicit models are chosen to possess  the statistical  invariance $P(F)=P(\mathcal{P}\circ F \circ \mathcal{P}')$, where $\mathcal{P}$ is any $S_q$
 permutation of the $q$ possible output states, and $\mathcal{P}'$ is any permutation of the $q^2$ possible input states. 
 However, this choice is only for convenience.
Retaining the former invariance (under $\mathcal{P}$) is sufficient for the basic formalism in this Section. 
We may argue that the universal behavior  stays the same even if all invariances are broken, at least if the breaking is sufficiently weak.} but $P(F)$ is otherwise arbitrary.
The specific choices do not matter for the universal behavior, so long as $P(F)$ is not fine-tuned  (App.~\ref{app:finetuning}). However,
 the case ${q=2}$ leads to some simplifications as we will see below.
Below we will give both general formulas  and a specific example.
(Full details of derivations can be found in  App.~\ref{app:meanfieldequations}.)

Let $\rho_\sigma(t)$ be the fraction of the $N$ sites that have damage type $\sigma\in \Pi_n$ at a given time $t$.
In a general model 
(i.e. for a sufficiently large value of $q$) 
a given site can have damage of any type $\pi\in \Pi_n$, 
so we begin in this Subsection with equations for all these densities.

In their initial form these equations look rather formal.
But in the following Subsections we will see that the universal description near the critical point simplifies radically, 
 because only a much smaller number of densities becomes ``massless'' at the critical point.
In Secs.~\ref{sec:MFTn2},~\ref{sec:MFTn2} we  discuss the special cases $n=2$ and $n=3$, and in  Eq.~\ref{sec:meanfieldgeneraln} we  give the reduced form of the equations for general $n$, which allows a general result for the mean field decay exponents $\alpha_n$ (cf. Sec.~\ref{sec:numerics}).

In the thermodynamic limit ${N\to \infty}$,
the mean field dynamics gives a deterministic differential equation for densities.
More generally, if 
 $N$ is large but finite, 
the dynamics is stochastic and is described by an  It\^o \cite{van1992stochastic} stochastic differential equation:\footnote{The number $n$ of replicas is left implicit. Formally we can also set $n=\infty$, see App.~\ref{app:meanfieldequations}.}
 \ba\label{eq:SDEfordensitiesgeneric}
 \dd \rho_\pi & =  \lf p_\text{out}(\pi) - \rho_\pi  \ri \dd t + \dd B_\pi.
 \end{align}
The terms proportional to $\dd t$ give the deterministic evolution that survives at ${N=\infty}$, while $\dd B_\pi$ is a noise contribution. 
In this equation,
\be\label{eq:poutdefn}
p_\text{out}(\pi) = \sum_{\sigma, \sigma'}  P(\pi|\sigma,\sigma')
\rho_\sigma \rho_{\sigma'},
\ee
which is implicitly a function of the densities $\{\rho_\sigma\}$, is the probability that an update gives an output of damage type $\pi$.
The $p_\text{out}(\pi)$ term in Eq.~\ref{eq:SDEfordensitiesgeneric} is due to updates that produce sites of type $\pi$, 
and the $-\rho_\pi$  term is due to sites of type $\pi$ getting updated (possibly reducing the number of sites of type $\pi$).

The noise $\dd B_\pi$
arises from the fact that the updates are random.
It  has mean zero and covariance
\ba  \notag
 & \< \dd B_\pi \dd B_\sigma\>  = 
 \\
 & \,\,\,\,\,\, \f{\dd t}{N}  \left[ 
\left( p_\text{out}(\pi)  + \rho_\pi \right)  \delta_{\pi \sigma}
 - p_\text{out}(\pi)  \rho_\sigma
- p_\text{out}(\sigma)  \rho_\pi
\right].\label{eq:appnoisecovariancegeneral}
 \end{align}
In the mean field model the noise strength vanishes in the thermodynamic limit. However it would be important for certain observables
(such as the survival probability) at large finite $N$. 
The above form for the noise covariance will also be important in finite dimensions.

Let us write the deterministic part of the evolution in Eq.~\ref{eq:SDEfordensitiesgeneric} in a  more explicit form 
(we will reinstate  the noise below).
We only need the equations for the nontrivial densities, i.e. for $\{ \rho_\pi\}$ with 
${\pi < \mathbbm{1}}$
(since ${\rho_{\mathbbm{1}}= 1- \sum_{\sigma<\mathbbm{1}}\rho_\sigma}$). Using a prime to denote a sum that runs only over nontrivial partitions, we have
\be\label{eq:rhoevoMK}
\f{\dd \rho_\pi}{\dd t}
= 
{\sum_{\sigma}}' \mathcal{M}_{\pi \sigma} \rho_\sigma
-
{\sum_{\sigma,\sigma'}}' \mathcal{K}^\pi_{\sigma\sigma'} \rho_\sigma \rho_{\sigma'},
\ee
with
\ba\label{eq:Mdefgeneral}
\mathcal{M}_{\pi \sigma} & = 2 P(\pi|\sigma, \mathbbm{1}) - \delta_{\pi,\sigma}
\\
\label{eq:Kdefgeneral}
\mathcal{K}^\pi_{\sigma,\sigma'} & = P(\pi|\sigma,\mathbbm{1}) + P(\pi|\mathbbm{1},\sigma')  -  P(\pi|\sigma,\sigma').
\end{align}
Recall that in the standard Langevin equation for directed percolation, the parameter ``$r$'' governing the linear term is the crucial one that vanishes at the critical point.
In Eq.~\ref{eq:rhoevoMK} we have a matrix $\mathcal{M}$ defining the linear terms. 
We will see below that the critical point is associated with the vanishing of an eigenvalue of this matrix. 
The corresponding eigenmodes are the important ``critical'' fields.

For a concrete example, let us consider the model where  each  update function $F$ is drawn as follows:
with probability $c$, 
we take $F$ to be completely irreversible, i.e. to map all input values to the same output value; while with probability ${\bar c= 1-c}$, 
we pick $F$ uniformly at random from the set of all possible update functions. 
In this case
\be\label{eq:Mdefspecific}
\mathcal{M}_{\pi \sigma} = 2\bar c \f{q!\,\mathbbm{1}_{\pi \geq \sigma}}{q^{|\sigma|} (q-|\pi|)!} - \delta_{\pi,\sigma},
\ee
where $\mathbbm{1}_{\pi \geq \sigma}$ is one if $\pi$ is either equal to $\sigma$ or coarser than $\sigma$ (and zero otherwise), and $|\sigma|$ is the number of blocks in~$\sigma$.
Note that $\mathcal{M}$ respects the partial ordering of partitions: $\mathcal{M}_{\pi\sigma}$ is nonzero only when $\pi\geq \sigma$, i.e. when $\pi$ can be obtained by merging blocks of $\sigma$.
This is consistent  with the absorbing state properties of the dynamics:
if  replicas $a$ and $b$ agree on every site, this remains true at all later times.

The kernel for the nonlinear terms in the present model  is 
\be\label{eq:Kdefspecific}
\mathcal{K}^\pi_{\sigma,\sigma'}\hspace{-0.5mm} = \hspace{-0.5mm}
\f{\bar c q!}{(q-|\pi|)!} \hspace{-0.5mm} \left[
\f{\mathbbm{1}_{\pi \geq \sigma}}{q^{|\sigma|}} \hspace{-0.5mm} + \hspace{-0.5mm} \f{\mathbbm{1}_{\pi \geq  \sigma'}}{q^{|\sigma'|}} \hspace{-0.5mm} - \hspace{-0.5mm} \f{\mathbbm{1}_{\pi \geq \sigma\wedge \sigma'}}{q^{|\sigma\wedge\sigma'|}}
\right].
\ee
In this formula the partition
\be\label{eq:wedgedefnmfsec}
{\sigma\wedge \sigma'},
\ee
also referred to as the ``meet'' of $\sigma$ and $\sigma'$, is the coarsest partition satisfying ${\sigma\wedge \sigma'\leq \sigma}$ and ${\sigma\wedge \sigma'\leq \sigma'}$ \cite{stanley_enumerative_1999}.\footnote{For example, 
\be{(1)(23)\wedge (2)(13) =(1)(2)(3)}\ee 
(at $n=3$)
and
\be 
(12)(34)\wedge (13)(24) = (1)(2)(3)(4)
\ee
(at $n=4$). Each block in $\sigma\wedge \sigma'$ is the intersection of a block in $\sigma$ and a block in $\sigma'$.}
The physical significance of $\sigma\wedge\sigma'$ is that, if two sites have damage types $\sigma$ and $\sigma'$, then $\sigma\wedge \sigma'$ is  the damage type of the pair when it is viewed as a composite system.
Again, the indicator functions in Eq.~\ref{eq:Kdefspecific} ensure that the equation of motion is consistent with the basic absorbing state properties of the dynamics.

We will describe  the simple examples of  ${n=2}$ and ${n=3}$ replicas before discussing the general form of the equations above. 
The most important model-dependent variable
is the probability $p_s$  
for local damage to propagate to an updated site in the $n=2$ problem:
\be\label{eq:psgeneraldef}
{p_s} \equiv P\big( (1)(2)  \big| (1)(2), \mathbbm{1} \big).
\ee
In the model defined just above this takes the value
\be\label{eq:psdef}
{{p_s} = \bar c (1-q^{-1})}.
\ee
The parameter $p_s$ can be read off from the 2-replica dynamics, but is also the most important parameter in the $n$-replica dynamics.

\subsection{Mean field for two replicas}
\label{sec:MFTn2}

To begin with let us consider the simplest case.
For ${n=2}$ there is only one nontrivial partition, namely $(1)(2)$.
Let us write simply ${\rho=\rho_{(1)(2)}}$. 

For the specific model defined immediately above Eq.~\ref{eq:Mdefspecific}, we find
\ba\label{eq:n2noisevariance}
\f{\dd \rho(t)}{\dd t}
& = 
r \rho(t) - g \rho(t)^2 + \eta(t)
& 
& (n=2),
\end{align}
with white noise $\eta(t)$ (now written in physicists' notation, but still interpreted in the It\^o sense)
\be
\< \eta(t) \eta(t') \> = \nu \rho(t) \delta(t-t') + O(\rho^2).
\ee
In this model the nonuniversal constants above are:
\ba\label{eq:n2nonuniversalconsts}
r & = 2 {p_s} - 1,
& 
g & = {p_s},
& 
\nu & = \f{1+2 {p_s}}{N}.
\end{align}
We have retained only the $O(\rho)$ term in the noise variance because 
$\rho$ is small near the mean-field critical point (at late times).

Taking the thermodynamic limit ${N=\infty}$ gives a mean-field  directed percolation phase transition \cite{derrida1986random} at ${r=0}$.
For negative $r$ the density decays exponentially to zero, 
while for positive $r$, 
a nontrivial state survives as ${t\to\infty}$, with a  finite density ${\rho(\infty) = r/g}$ of damaged sites. 
Precisely at the critical point, the late-time density decay is
\be\label{eq:2reprhodecay}
\rho(t) \sim \f{1}{gt}.
\ee
These formulae give the  mean-field exponents
${\beta^{\text{MF}}_2=1}$
for the 2-replica damage order parameter in the ``active''  phase, and 
${\alpha^{\text{MF}}_2=1}$
for the decay of  2-replica damage at the critical point, starting from a random initial state. (The standard directed percolation exponents were reviewed in Sec.~\ref{sec:dpreview}.)

In terms of the 
parameter $\bar c$ in the definition of the model, the
critical point ${r_*=0}$ maps (by Eqs.~\ref{eq:psdef},~\ref{eq:n2nonuniversalconsts}) to 
\be
\bar c_* = \f{q}{2(q-1)}.
\ee
For $q>2$ this lies in the interior of the parameter space.
For $q=2$, the critical point lies at the boundary ${\bar c=1}$ of the  parameter space.
This is not a fundamental difference between $q=2$ and $q>2$; the $q=2$ critical point can be moved into the interior of the parameter space by some slight change to $P(F)$.\footnote{See e.g. the discussion in Sec.~\ref{sec:choiceof1Dmodel} for the 1+1D model.} We will discuss slightly more general models below.

\subsection{Mean field for three replicas}
\label{sec:meanfieldthreereplica}

Now consider how the structure changes in the presence of one more replica.
This will introduce the idea that some of the partition types correspond to non-critical fields.

For ${n=3}$ there are in general four nontrivial partitions, 
\be
(1)(23), \quad (2)(13), \quad (3)(12), \quad (1)(2)(3).
\ee
However,  for $n=3$
the deterministic part of the dynamical equations in fact take a simpler form if we use another basis for the densities. 
We define the \textit{pairwise}
damage $\rho_{a,b}$ for each pair of replicas $a<b$.
For example, $\rho_{1,2}$ is 
the fraction of sites where replica 1 differs from replica 2, and can be seen to be given by
\be\label{eq:pairwisediffdefn}
\rho_{1, 2} \equiv \rho_{(1)(23)}
+ 
\rho_{(2)(13)} + \rho_{(1)(2)(3)}.
\ee
Examples of such pairwise damages $\rho_{ab}$ were shown in Fig.~\ref{fig:traj}, where they were denoted $\rho_{(a)(b)}$.
These pairwise damages, together with 
\be
\rho_{F} \equiv \rho_{(1)(2)(3)},
\ee
give an legitimate  basis
of four  independent densities which is an alternative to the basis labelled by ${\pi\in \Pi_3}$.
Note that for a model of bits (${q=2}$) the finest damage type cannot appear at a site, i.e. ${\rho_F=0}$ identically.

\subsubsection{Pairwise damages}

First, consider the pairwise differences:
we will return to $\rho_F$ shortly. The general formulas in Sec.~\ref{sec:meanfieldeqnsgeneralform} give
\ba\label{eq:2replicadifferenceseq}
\f{\dd \rho_{1, 2}(t)}{\dd t}
& = r \rho_{1, 2}(t) - g \rho_{1, 2}(t)^2 + \eta_{1, 2}(t),
\end{align}
and similarly for the other two symmetry-related fields.

Note that  the deterministic terms in Eq.~\ref{eq:2replicadifferenceseq} have precisely the same form as in the ${n=2}$ problem, with the same definitions for the constants.
If we start with the equations written in terms of ${\rho_\pi}$, 
which have several types of quadratic term, the reduction to Eq.~\ref{eq:2replicadifferenceseq} looks like a miraculous cancellation. 
But from another point of view, 
this reduction is  obvious. 
The key point is that 
we are always free to simply ignore the state of (say) the third replica, and if we do this we must recover the dynamics of the 2-replica problem, 
which is fully specified by $\rho_{1,2}$, which must follow the dynamics discussed in Sec.~\ref{sec:MFTn2}.
(The possibility of neglecting one replica imposes important constraints on the field theory couplings and on the RG flows which we discuss in Secs.~\ref{sec:fieldtheories},~\ref{sec:rgstructure}.)

However, nontrivial 3-replica correlations are present in the form of the noise $\eta_{a,b}$, whose covariance we will discuss below (Sec.~\ref{sec:n3meanfieldfull}). 
For example, $\eta_{1,2}$ is not independent of  $\eta_{2,3}$.
This will be especially  important in finite dimensions.
Before giving the full critical theory we describe how the fourth field, $\rho_F$, can be eliminated.

\subsubsection{Eliminating the higher damage density}

In the special case ${q=2}$, the  density $\rho_F$ for the partition $(1)(2)(3)$ vanishes,
and the mean field theory is formulated in terms of only three fields. 
For ${q>2}$, 
 $\rho_F$  does not vanish identically.
 However  at late times it effectively becomes a function of the lower damage densities.
  
The equation of motion for $\rho_F$ is 
\ba\notag
\f{\dd \rho_{F}}{\dd t}  = 
 r_F \rho_F 
 & +g_F
\begin{pmatrix}
\rho_{1,2} \\  \rho_{2,3} \\ \rho_{1,3}
\end{pmatrix}^T
\begin{pmatrix*}[r]
-1 & 1 & 1 \\
1 & -1 & 1 \\
1 & 1 & -1 
\end{pmatrix*}
\begin{pmatrix}
\rho_{1,2} \\  \rho_{2,3} \\ \rho_{1,3}
\end{pmatrix}
\\
& + 
\eta_F(t) + 
O(\rho_F \rho_{a,b}, \rho_F^2). 
\label{eq:n3rhoFdecay}
\end{align}
with the ``decay rate'' $r_F$  and 
the constant $g_F$ equal to  
\ba\label{eq:rFeq}
r_F & = -\f{2-(q-2)r}{q}.
&
g_F & = g \f{q-2}{2q}.
\end{align}
In Eq.~\ref{eq:n3rhoFdecay} we have dropped terms such as  $\rho_F\rho_{1,2}$ and $\rho_F^2$, which we now show are subleading. We also neglect the noise $\eta_F$, which vanishes in the thermodynamic limit in the mean field model.

The critical point is 
at $r=0$: near this point, the pairwise damages $\rho_{a,b}$ behave as discussed around Eq.~\ref{eq:2reprhodecay}.
The key point is that $r_F$ in 
Eq.~\ref{eq:n3rhoFdecay} remains \textit{negative}
at the critical point.

Loosely speaking, this means that $\rho_F$ does not become ``critical''.
In the finite dimensional field theory, it will be a field that retains a finite mass at the critical point, unlike $\rho_{a,b}$ which becomes massless.

Exponential decay in $\rho_F$, which would be induced by the first term in 
Eq.~\ref{eq:n3rhoFdecay},
is prevented by the nonlinear terms, which ``feed'' $\rho_F$.
For $|r|\ll 1$ and $t\gg 1$ 
we can neglect the left-hand side of Eq.~\ref{eq:n3rhoFdecay}, giving the identification
\ba\label{eq:r123identification}
\rho_F & \simeq \f{q-2}{4}
\lf
(\rho_{1,2} \rho_{2,3} + \text{2 terms}) 
- \f{1}{2} \lf \rho_{1,2}^2 + \text{2 terms}\ri
\ri.
\end{align}
Returning to our original notation, this is (after dropping the subleading term in Eq.~\ref{eq:pairwisediffdefn}) 
\be\label{eq:rho123intermsoflower}
\rho_{(1)(2)(3)} 
\simeq 
\f{q-2}{2} \lf
\rho_{(1)(23)} \rho_{(2)(13)}
+ \text{two terms}
\ri.
\ee
Therefore, within mean field theory, 
the critical exponents for damage type $(1)(2)(3)$ are  ${\beta^\text{MF}_3=2}$, ${\alpha^\text{MF}_2=2}$.
For example, 
 at the critical point,
 \be
 \rho_{(1)(2)(3)} \sim \f{6(q-2)}{t^2}.
\ee

When we move to finite dimensions, the analog of (\ref{eq:rho123intermsoflower}) will be that the 
field $\rho_{(1)(2)(3)}$ can be integrated out, so that the critical point is described by a field theory for  only three massless fields.
The three massless fields can be taken to be either 
${\{\rho_{(1)(23)}, \rho_{(2)(13)}, \rho_{(3)(12)}\}}$, or ${\{ \rho_{1,2}, \rho_{2,3}, \rho_{1,3}\}}$: 
up to higher-order corrections, 
this is simply to a change of basis. 
The  lattice observable $\rho_{(1)(2)(3)}$
can then be expressed as a composite operator, i.e. using products of the elementary fields, as in Eq.~\ref{eq:rho123intermsoflower}.
While  in mean field theory
the decay exponent for the composite operator is a simple multiple of that for  the elementary field, 
this is not expected to be the case below the upper critical dimension: 
the observable  $\rho_{(1)(2)(3)}$ will have a nontrivial scaling dimension that is not simply related to that of e.g. $\rho_{1,2}$.

\subsubsection{Full critical theory for ${n=3}$}\label{sec:n3meanfieldfull}

Above we found that there was effectively a reduction, in the critical regime,
to a theory with only three fields. 
The simplest version of this reduction is in the model with ${q=2}$, for which $\rho_{(1)(2)(3)}$ vanishes identically.\footnote{At first this may seem like significant fine-tuning, because we appear to lose an important observable.
However, this is not really the case.
Even when the damage type $(1)(2)(3)$ is not possible on a single spatial site, it is still a possible damage type for a pair of sites. 
There is no reason why we cannot define observables for pairs of sites even in the mean field model, and it becomes natural in finite dimensions.
If in the mean  field model we define $\rho^{\text{(2-site)}}_{(1)(2)(3)}$ as the fraction of site pairs that host damage type $(1)(2)(3)$, we see immediately that, for $q=2$, ${\rho^{\text{(2-site)}}_{(1)(2)(3)} = (\rho_{(1)(23)} \rho_{(2)(13)} + \text{cyclic})}$, showing that there is an analog of Eq.~\ref{eq:rho123intermsoflower} even for ${q=2}$.} 

Finally, let us return to the evolution of the pairwise damages in Eq.~\ref{eq:2replicadifferenceseq}.
We have not yet specified the statistics of the noise which is present at finite $N$ and which will become important in finite dimensions. 
We neglect 
$\rho_F$, since it is  parametrically smaller than $\rho_{a,b}$ at late times
(for the full result and derivation, see App.~\ref{app:meanfieldequations}).
Then we find that the noises in the equations
\ba\label{eq:2replicadifferenceseqagain}
\f{\dd \rho_{a, b}(t)}{\dd t}
& = r \rho_{a, b}(t) - g \rho_{a,b}(t)^2 + \eta_{a, b}(t),
\end{align}
($1\leq a<b\leq 3$) satisfy
\ba \label{eq:n3variance}
\< \eta_{1,2}(t)\eta_{1,2}(t')\> & \simeq \nu \rho_{1,2}(t)  \delta(t-t')
\\
\< \eta_{1,2}(t)  \eta_{2,3}(t') \> 
& \simeq  \f{\nu}{2}\lf
\rho_{1,2} + \rho_{2,3} - \rho_{1,3}
\ri
\delta(t-t')
\label{eq:n3covariance}
\end{align}
and symmetrically.

The first equation matches Eq.~\ref{eq:n2noisevariance}, as it must: for observables that do not involve the third replica,
the stochastic evolution must reduce to that in Sec.~\ref{sec:MFTn2}.
As a result of (\ref{eq:n3covariance}) the three fields are however nontrivially coupled.

Let us mention a  confusion that we suffered in our initial exploration of this problem, which is resolved by the above equations.
First note that the noise $\eta_{1,2}$ must vanish identically when ${\rho_{1,2}=0}$, in order to respect the absorbing state constraints.\footnote{If replicas 1 and 2 agree everywhere ($\rho_{1,2}=0$), then they remain in agreement. 
Therefore the right-hand-side (\ref{eq:2replicadifferenceseq}) must vanish when $\rho_{1,2}=0$.}
Similarly, the noise $\rho_{2,3}$ must vanish identically when ${\rho_{2,3}=0}$.
At first glance we might then conclude that the expression for the covariance
${\< \eta_{1,2} \eta_{2,3} \>}$ should  start at order $\rho_{1,2}\rho_{2,3}$, to be consistent with both requirements.
However this inference is not correct, because there is a linear term, ${\rho_{1,2} + \rho_{2,3} - \rho_{1,3}}$ which also satisfies the requirement of vanishing when either $\rho_{1,2}$ or $\rho_{2,3}$ vanishes.
For example, this combination vanishes when 
${\rho_{1,2}=0}$ because 
$\rho_{1,2}=0$ implies
that
${\rho_{2,3} = \rho_{1,3}}$.
(If replicas 1 and 2 are in the same state, we can exchange  replica index 2 for replica index 1 on any observable.)

This linear combination appears on the right-hand-side of (\ref{eq:n3covariance}). 
The relative factor of $1/2$ between 
(\ref{eq:n3covariance}) and (\ref{eq:n3variance}) is 
``universal'', i.e. independent of the details of the model.\footnote{This can be seen by using Eqs.~\ref{eq:n3variance},~\ref{eq:n3covariance} to  compute the noise $\eta_{(1)(23)}$ acting on $\rho_{(1)(23)}$.
Since $\rho_{(1)(23)}$ cannot become negative, the noise $\eta_{(1)(23)}$ must vanish when ${\rho_{(1)(23)}=0}$. This fixes the relative coefficient (in agreement with the result above from direct computation in a particular model).}

When we move to finite dimensions, the nontrivial covariance in the second line is the key feature of the $n=3$ field theory, 
preventing it from simply reducing to three copies of the $n=2$ theory (with one copy for each pairwise difference).

But  a striking feature of the $n=3$ equations is that, having dropped the subleading terms, all the numerical coefficients 
--- namely $r$, $g$ and $\nu$ which appear in Eqs.~\ref{eq:2replicadifferenceseq},~\ref{eq:n3variance},~\ref{eq:n3covariance} ---
are inherited from the $n=2$ problem.
The discussion above shows that the 
structure of 
Eqs.~\ref{eq:2replicadifferenceseqagain}--\ref{eq:n3covariance} is completely fixed on general grounds:
at the given order in $\rho$, 
the mean-field equations are \textit{independent} of the precise details of the microscopic model, except for the values of these numerical coefficients.
We will return to this point when we discuss the field theory in finite dimensions.

The equations  in the alternative basis of ${\{\rho_{\sigma_1}, \rho_{\sigma_2}, \rho_{\sigma_3}\}}$, with ${\sigma_1 = (1)(23)}$, etc., are given in the next Section. 
In this basis the noise covariance matrix is diagonal, and the nontrivial ``interactions'' between the fields instead appear in the deterministic terms.
We will discuss the interplay between these two bases in Sec.~\ref{sec:trs}.
It turns out that the ${n=3}$ theory has a time-reversal symmetry, where the symmetry transformation is closely connected with the change of basis.

\subsection{Mean field for $n$ replicas}
\label{sec:meanfieldgeneraln}

Finally we turn to the case of a general number of replicas. 

We saw in the previous Section that the number of ``elementary'' density fields --- those with the slowest power-law decay at the critical point --- is in general less than the total number of partitions $\sigma$.
We saw for ${n=3}$ that the higher density was essentially a ``composite field'', expressable in terms of the elementary densities at late times.

What is the smallest set of elementary densities for general $n$?
For  $n=3$ it was sufficient to use the set of pairwise damages $\rho_{a,b}$ for ${a<b}$, 
so a natural guess might be that the pairwise damages $\{\rho_{a,b}\}$ would form a complete basis of $n(n-1)/2$ elementary densities for general $n$.
Interestingly, this guess is incorrect,
as we can see by examining the general dynamical equation.

At linear order in the densities, Eq.~\ref{eq:rhoevoMK} reads
\be\label{eq:rhoevoMKlinear}
\f{\dd \rho_\pi}{\dd t}
= 
{\sum_{\sigma}}' \mathcal{M}_{\pi \sigma} \rho_\sigma
+ \ldots
\ee
with ${\mathcal{M}_{\pi \sigma}  = 2 P(\pi|\sigma, \mathbbm{1}) - \delta_{\pi,\sigma}}$.
The basic question is how many eigenvalues of the matrix $\mathcal{M}$ vanish at the critical point": each vanishing eigenvalue represents a ``critical'' mode.
The number of critical modes is in fact given by the number of two-block partitions.

Since $\mathcal{M}_{\pi\sigma}$
is only nonzero if $\pi\geq \sigma$, 
$\mathcal{M}$  is a triangular matrix, 
whose eigenvalues are given by the diagonal elements. These eigenvalues strictly decrease as a function of $|\pi|$, the number of blocks in the partition (if the dynamics is nontrivial).
The eigenvalue for the two-block partitions is simply $r$:
this can be checked from Eq.~\ref{eq:Mdefspecific} for our particular model, but can also be shown to be more general.
At the critical point ${r=0}$,
the eigenvalues vanish only for the 
two-block partitions $\pi$, with the other
eigenvalues of $\mathcal{M}$ remaining positive.\footnote{Eq.~\ref{eq:decayrateexplicit} below gives the explicit formula for (minus) the eigenvalue in the particular model.}
The sub-block of $\mathcal{M}$ for the two-block partitions is simply
\ba
\mathcal{M}_{\pi \sigma} & = r \delta_{\pi,\sigma}
& &(\text{for } |\pi|=|\sigma|=2).
\end{align}
There is therefore  a zero ``eigenmode''
for each two-block partition.
This mode, denoted 
\ba\label{eq:caretnotation}
\hat \rho_\pi  & = \rho_\pi + \sum_{\mu\leq \pi} a^{(\pi)}_\mu \rho_\mu,\\
& =  \rho_\pi+ (\text{densities of finer partitions})
\end{align}
 is a sum only over densities for partitions $\mu$ that are at least as fine as $\pi$. 

This structure directly generalizes the one we found for ${n=3}$. 
There we had three critical modes\footnote{In Sec.~\ref{sec:n3meanfieldfull} we focussed on the basis ${\{\rho_{12}, \rho_{23}, \rho_{13}\}}$ for the critical modes, 
but a  linear   basis change relates these to ${\{\hat \rho_{\sigma_1},\hat \rho_{\sigma_2},\hat \rho_{\sigma_3}\}}$ for the three two-block partitions [$\sigma_1 =(1)(23)$, etc.] with for example ${\hat \rho_{\sigma_1} = \rho_{\sigma_1} + \f{1}{2} \rho_{(1)(2)(3)}}\simeq \rho_{\sigma_1}$.}
together with a non-critical mode $\rho_{(1)(2)(3)}$
whose decay rate --- the eigenvalue of $\mathcal{M}$ --- remained negative at the critical point.
At late times, the non-critical mode was subleading, 
and could be expressed in terms of the critical modes (Eq.~\ref{eq:rho123intermsoflower}).
The simplest version of this reduction occured for models of bits, ${q=2}$, where the density $\rho_{(1)(2)(3)}$ vanished identically.

For general $n$,
the universal properties of the transition are captured by working only with fields labelled by two-block partitions, 
with higher partitions corresponding to composite fields, 
in parallel to  Eq.~\ref{eq:rho123intermsoflower}
(see Sec.~\ref{sec:meanfielddecayexponents} for more detail on the higher partitions).
The number of two-block partitions is 
\be
\text{Number of critical modes:} \quad
{\stirling{n}{2} = 2^{n-1}-1},
\ee
and  is strictly larger than the number of pairwise damages, except in the special cases $n=2$ and $n=3$.

As for ${n=3}$, the reduction to two-block partitions is simplest in the special case ${q=2}$, 
when these are the only possible partitions at a site.
Restricting Eqs.~\ref{eq:rhoevoMK}-\ref{eq:Kdefgeneral} to two-block partitions gives
\ba\label{eq:2blockonlygeneral}
\f{\dd \rho_{\pi}}{\dd t}
 & \simeq
 r \rho_\pi 
- \sum_{\sigma,\sigma'\in \mathcal{S}_2^n} 
\mathcal{K}^\pi_{\sigma,\sigma'}
\rho_\sigma \rho_{\sigma'},
\end{align}
where all the indices run over the set $ \mathcal{S}_2^n$ of  two-block partitions. For these partitions  Eq.~\ref{eq:Kdefgeneral} becomes
\ba\label{eq:Ktensor2blockrestriction}
\mathcal{K}^\pi_{\sigma,\sigma'} = P(\pi|\sigma,\sigma') 
- 2 {p_s} \delta_{\pi,\sigma'}.
\end{align}
For the model defined above (\ref{eq:Mdefspecific}), at ${q=2}$, this is 
\be\label{eq:KspecificMFnsec}
{\mathcal{K}}^\pi_{\sigma,\sigma'}=  g \lf
\delta_{\pi, \sigma}
+
\delta_{\pi, \sigma'}
-
\f{\mathbbm{1}_{\pi\geq \sigma\wedge \sigma'}}{2^{|\sigma\wedge\sigma|-2}}
\ri.
\ee
The constant $g$ is the same one that appeared in the ${n=2}$ and ${n=3}$ problems.
As usual the indicator function $\mathbbm{1}_{\pi\geq \sigma\wedge \sigma'}$ ensures that the dynamics respects the absorbing state constraints (see the discussion below Eq.~\ref{eq:wedgedefnmfsec}).
We will discuss the form of   $\mathcal{K}$ for a fully general model in Sec.~\ref{sec:moreoninteractiontensor} below; it is not much more complicated than (\ref{eq:KspecificMFnsec}), but involves one additional free parameter. 
For now we use the expression (\ref{eq:Ktensor2blockrestriction}), which applies to any model in this ``universality class''.

When ${r=0}$ the right-hand-side of (\ref{eq:2blockonlygeneral}) is quadratic, 
so by dimensional analysis we would expect that all the two-block  densities decay like $1/t$ as $t\to\infty$ (potentially with different prefactors); this is  justified more carefully in 
App.~\ref{app:somedetails}.

The structure above generalizes straightforwardly to larger $q$.
We noted above that we may diagonalize the linear terms in the evolution equation by a change of basis ${\hat \rho_\pi = \rho_\pi + \ldots}$, 
where, thanks to the structure of $\mathcal{M}$,
which respects the partial ordering of partitions,
the ellipses only contain densities for partitions \textit{finer} than $\pi$.
The subblock of $\mathcal{M}$ for two-block partitions gives $\stirling{n}{2}$ eigenvalues equal to $r$ 
so, for $|\pi|=2$, 
\ba\label{eq:2blockonlygeneralanyq}
\f{\dd \hat \rho_{\pi}}{\dd t}
 & \simeq
 r \hat \rho_\pi 
- \sum_{\sigma,\sigma' \in \mathcal{S}_2^n}
\mathcal{K}^\pi_{\sigma,\sigma'} \hat\rho_\sigma \hat\rho_{\sigma'} 
+ \ldots,
\end{align}
where again all indices run over two-block partitions, and where the ellipses
contains quadratic terms 
involving partitions with more than two blocks.
We will argue below that these terms are subleading and can be neglected.
Therefore we are back to the structure that we obtained for ${q=2}$.

The noise $\hat\eta_\pi$ for the two-block density modes,
which should be added 
to Eq.~\ref{eq:2blockonlygeneralanyq}
when $N$ is finite,
satisfies
\ba\label{eq:noise2blockgeneral}
\< \hat \eta_\pi (t)  \hat \eta_\sigma (t')\> 
& = \nu\hat\rho_\pi \delta_{\pi\sigma} \delta(t-t')
& & (\text{for $|\pi|=2$})
\end{align}
neglecting subleading terms. That is, the noise covariance matrix is diagonal in the basis of two-block partitions (at leading order in the densities).

We have written the interaction tensor $\mathcal{K}$ for a specific model in Eq.~\ref{eq:KspecificMFnsec}.
We discuss its general form in Sec.~\ref{sec:moreoninteractiontensor}.
We find that,  when we consider more general models 
with the same mean-field description, 
the form of $\mathcal{K}$ remains very rigid.
$\mathcal{K}$ is  specified, for all $n$, by two parameters:
the overal normalization $g$ and a single additional model-dependent constant, $\theta$.\footnote{Note that, for ${n>4}$, there are different kinds of 2-block partition: for example when $n=4$ we have those of shape $(12)(34)$ and those of shape $(1)(234)$.
We can use the mean field equation (\ref{eq:2blockonlygeneralanyq})
to find the density of each type of partition, starting from a ``generic'' initial condition in which all possible damage types occur. 
Since all two-block densities decay like $1/gt$,
ratios such as $\rho_{(12)(23)}/\rho_{(1)(234)}$
tend to constants at late time. 
These constants can depend on $\theta$.} 
For ${n\leq 3}$, the constant $\theta$ drops out, so that the only freedom in $\mathcal{K}$ is its overall scale $g$.

\subsubsection{Decay exponents for higher damages}
\label{sec:meanfielddecayexponents}

So far, we have found that all of the two-block densities yield ``critical'' densities $\hat \rho_\pi$ that decay like $1/t$ at the critial point  ${r=0}$.
Now let's determine the decay exponents for more general densities.
We make the assumption --- to be confirmed self-consistently below --- that increasing the number of blocks can only increase the decay exponent, not decrease it, and use this to simplify the equations.

For an arbitrary partition with $|\pi|>2$, we find from (\ref{eq:rhoevoMK}) that
\ba\label{eq:higherpartitionsdecay}
\f{\dd \rho_\pi}{\dd t} \simeq - \lambda_\pi \rho_\pi 
+ {\sum_{\sigma,\sigma'}}' P(\pi|\sigma,\sigma') \rho_\sigma \rho_{\sigma'}
\end{align}
(the sums run over all nontrivial partitions) with 
\be\label{eq:decayrateexplicit}
\lambda_\pi = 1 - 2 P(\pi|\pi,\mathbbm{1}).
\ee
When ${|\pi|>2}$, $\lambda_\pi$ remains positive at the critical point.
For example, in the specific model we have discussed, 
\be
\lambda_\pi = 
1- q^{2-|\pi|}\f{(q-2)!}{(q-|\pi|)!}
\ee
when $r=0$.
At late times, $\dd \rho_\pi/\dd t$ is much smaller than $\rho_\pi$,  so 
the nonzero value of $\lambda_\pi$ means that we can neglect the left-hand side of Eq.~\ref{eq:higherpartitionsdecay}. This gives:
\ba\label{eq:steadystatevaluehigherpartitions}
\rho_\pi & \simeq \f{1}{\lambda_\pi} \,
{\sum_{\sigma,\sigma'}}' P(\pi|\sigma,\sigma') \rho_\sigma \rho_{\sigma'}
& 
& (\text{for $|\pi|>2$}).
\end{align}

This equation determines the scaling of $\rho_\pi$ at late times, in the  mean-field model, to be
\be\label{eq:meanfielddimensionsansatz}
\rho_\pi \sim t^{ - 
\mathcal{N}_2(\pi) } 
\ee
where 
$\mathcal{N}_2(\pi)$  is the smallest number of two-block partitions from which $\pi$ can be formed by the ``meet'' operation  (see Sec.~\ref{sec:meanfieldeqnsgeneralform} and Ref.~\cite{stanley_enumerative_1999}), i.e. as 
\be\label{eq:2blockdecomp}
 \pi = \sigma_1 \wedge \sigma_2 \wedge \cdots \wedge \sigma_{\mathcal{N}_2(\pi)},
 \qquad 
 |\sigma_i| = 2.
\ee 
This quantity depends only on the number of blocks in $\pi$, and not the sizes of the blocks, so we will also write $\mathcal{N}_2(|\pi|)$.
We have ${\mathcal{N}_2(k) =\lceil \log_2(k) \rceil}$, i.e.\footnote{To see that a partition $|\pi|=2^p$  can be formed using $p$ two-block partitions, we can identify the blocks of $\pi$ with points of the $p$-dimensional hypercube, and define the two-block partition $\sigma_i$ in Eq.~\ref{eq:2blockdecomp} using the  sign of the $i$th coordinate of the hypercube.} 
\ba
\mathcal{N}_2(2) &  = 1,
\\
\mathcal{N}_2(3) &  = \mathcal{N}_2(4)  = 2 , \label{eq:meanfieldexponent34}
\\
\mathcal{N}_2(5) & = \mathcal{N}_2(6) = \ldots = \mathcal{N}_2(8) = 3 ,
\\
 \mathcal{N}_2(9) & =  \mathcal{N}_2(10) = \ldots = \mathcal{N}_2(16) = 4,
\end{align}
etcetera.
One may check that these exponents are
are consistent with Eq.~\ref{eq:steadystatevaluehigherpartitions} by noting that
\be
\min_{\substack{\sigma, \sigma' \\ \sigma\wedge\sigma' \leq \pi}}
 \left[ 
\mathcal{N}_2(\sigma) + \mathcal{N}_2(\sigma') 
\right]
= \mathcal{N}_2(\pi).
\ee

Eq.~\ref{eq:meanfielddimensionsansatz} defines the mean-field values of the density decay exponents, ${\alpha^\text{MF}_\pi = \mathcal{N}_2(|\pi|)}$.
The exponents dictating the scaling of the steady state densities just inside the active phase are also ${\beta^\text{MF}_\pi= 
\mathcal{N}_2(|\pi|)}$.

We see that, within mean field theory, the number of distinct decay exponents (for a given number of replicas) is not only much smaller than the number of partitions $\pi$, but also much smaller than the number of possible values of $|\pi|$.

The RG discussion in Sec.~\ref{sec:rgstructure} and the numerical results in Sec.~\ref{sec:numerics} show that only the first of these properties remains true below the upper critical dimension.
Below the UCD, the  decay exponent $\alpha_{|\pi|}$
 of $\rho_\pi$
depends only on $|\pi|$, 
but different values of $|\pi|$ lead in general to distinct dimensions.
For example the simulations  in Sec.~\ref{sec:numerics} show that ${\alpha_3 \neq \alpha_4}$ in 1+1D, in contrast to the mean field result in Eq.~\ref{eq:meanfieldexponent34}.

\subsubsection{General form of the $n$-replica equations}
\label{sec:moreoninteractiontensor}

We have seen that the critical theory may be reduced to the form
\ba\label{eq:2blockonlygeneralanyqrepeated}
\f{\dd  \hat\rho_{\pi}}{\dd t}
 & \simeq
 r  \rho_\pi 
- g \sum_{\sigma,\sigma' \in \mathcal{S}_2^n}
\hat{\mathcal{K}}^\pi_{\sigma,\sigma'} \hat\rho_\sigma \hat\rho_{\sigma'} 
+ \eta_\pi,
\end{align}
with
\ba\label{eq:noise2blockgeneralrepeated}
\< \hat \eta_\pi (t)  \hat \eta_\sigma (t')\> 
& = \nu\hat\rho_\pi \delta_{\pi\sigma} \delta(t-t')
& & (\text{for $|\pi|=2$}).
\end{align}
In these equations  $\hat \rho_\pi$ represents either the ``bare''  damage field for the two-block partition $\pi$ if ${q=2}$,
or the mode ${\hat \rho_\pi = \rho_\pi + (\ldots)}$, which diagonalizes the linear terms, if ${q>2}$. 
The distinction will not be crucial in what follows, so we now drop the caret on $\hat \rho_\pi$.
We have defined a ``normalized'' interaction tensor $\hat{\mathcal{K}}$ by extracting the numerical constant   $g$ that already appeared in the two and three-replica problems. 

Above  we have computed $\hat{\mathcal{K}}$ for a particular microscopic model with ${q=2}$:\footnote{The definition of this model is given  above Eq.~\ref{eq:Mdefspecific}. Eq.~\ref{eq:Kdefspecificq2} is obtained by specifying Eq.~\ref{eq:Kdefspecific} to the case ${q=2}$ 
(where $g=\bar c/2$) 
and using the fact that $\mathbbm{1}_{\pi\geq \sigma}= \delta_{\pi,\sigma}$ for two-block partitions.}
\be\label{eq:Kdefspecificq2}
\hat {\mathcal{K}}^\pi_{\sigma,\sigma'}= 
\delta_{\pi, \sigma}
+
\delta_{\pi, \sigma'}
-
\f{\mathbbm{1}_{\pi\geq \sigma\wedge \sigma'}}{2^{|\sigma\wedge\sigma|-2}} .
\ee
In previous Sections we saw that  there is no freedom at all in 
this tensor when  ${n\leq 3}$:
this followed from the fact that the  ${n=3}$ equations must reduce to the ${n=2}$ equations when we ``ignore'' one replica.\footnote{See App.~\ref{app:Ktensordetails} for a more explicit discussion. For ${n=3}$, 
Eq.~\ref{eq:2blockonlygeneralanyqrepeated} is equivalent to our previous formulation in 
Eq.~\ref{eq:2replicadifferenceseqagain}, 
but expressed in a different basis.} 
However, since the size of $\hat{\mathcal{K}}$ grows with $n$, we could worry that 
a large or even infinite number of  model-dependent parameters might appear when we consider general $n$.
Fortunately this is not the case ---
$\hat{\mathcal{K}}$ is fully specified, for all $n$, by a single model-dependent parameter $\theta$.

Since we deal only with two-block partitions,
${\sigma\wedge\sigma'}$
contains at most four blocks.
This has the consequence that any nonzero\footnote{$\hat{\mathcal{K}}^\pi_{\sigma,\sigma'}$ can be nonzero only if $\pi\leq \sigma\wedge \sigma'$}
element   $\hat{\mathcal{K}}^\pi_{\sigma,\sigma'}$
 reduces, regardless of the value of $n$, 
to an equivalent tensor element in the $n=4$ theory.
This can be seen in a specific example:
\be\label{eq:Kelementidentity}
\hat{\mathcal{K}}^{({\blue 1 5}4)(23)}_{({\blue 1 5}2)(34), ({\blue 15}13)(24)}
= 
\hat{\mathcal{K}}^{({\blue 1}4)(23)}_{({\blue 1}2)(34), ({\blue 1}3)(24)}
\ee
The left hand side is an element of the tensor for ${n=5}$. 
However, the only partitions that appear are ones where 
replicas 1 and 5 are always in the same block:
loosely speaking, this means that they can be treated as a single replica,
giving the ${n=4}$ element on the RHS.\footnote{In more detail: 
the $n=5$ dynamical equations must reduce to the $n=4$ equations if we restrict to configurations in which the states of replica 1 and 5 are identical. This requires (\ref{eq:Kelementidentity}) to hold.}

Therefore it is sufficient to characterize $\hat{\mathcal{K}}$ for ${n=4}$.
Replica symmetry and consistency with the ${n=3}$ equations then leave only one free parameter, denoted $\theta$ (Appendix.~\ref{app:Ktensordetails}).
For general $n$,
\ba\label{eq:Kdefgeneral2}
\hat {\mathcal{K}}^\pi_{\sigma,\sigma'}& = 
\delta_{\pi, \sigma}
+
\delta_{\pi, \sigma'}
-
\mathbbm{1}_{\pi\geq \sigma\wedge \sigma'}
\,\,
\chi^\pi_{\sigma\wedge\sigma'},
\end{align}
with        
\ba\label{eq:Kdefgeneral3}
\chi^\pi_{\sigma\wedge\sigma'} & = 
\left\{
\begin{array}{lll}
1 & \text{if } & |\sigma\wedge\sigma'|=2
\\
{1}/{2}
& \text{if } 
& |\sigma\wedge\sigma'|=3
\\
\theta
& \text{if } 
& |\sigma\wedge\sigma'|=4, \,\, \text{ Case A }
\\
\f{1}{2} - \theta 
& \text{if } 
& |\sigma\wedge\sigma'|=4, \, \, \text{ Case B }
\end{array}\right. ,
\end{align}
where Case A is where one of the blocks of $\sigma\wedge\sigma'$ coincides with a block of $\pi$ and Case B is where this is not the case.
For example,
\ba\notag
\hat{\mathcal{K}}^{(1)(234)}_{(12)(34),(13)(24)} & =  -  \theta,
& 
\hat{\mathcal{K}}^{(14)(23)}_{(12)(34),(13)(24)} & =   \theta - \f{1}{2}.
\end{align}
The result (\ref{eq:Kdefspecificq2}) corresponds to the case ${\theta=1/4}$.
App.~\ref{app:Ktensordetails} describes a slightly more general microscopic model, related to that in Sec.~\ref{sec:choiceof1Dmodel},
in which $\theta$ can be varied.

For ${n\leq 3}$ we always have $|\sigma\wedge\sigma'|\leq 3$, so that in this case the new constant $\theta$ does not play a role.

(When we consider field theory below the upper critical dimension for ${n\geq 4}$, we should take the constant $\theta$ into account in the renormalization group treatment of the model.
This  means that there is \textit{one} additional coupling constant compared to the ${n=2}$ case, i.e. compared to directed percolation.)

\section{Continuum descriptions: 
\\  Finite dimensions}
\label{sec:fieldtheories}

It is straightforward to go from the mean field results in the previous section to stochastic differential equations  for spatially local systems.
Loosely speaking, we need only take the mean-field equations with a finite noise strength, and add spatial derivatives: we give a little more detail below.
The stochastic differential equations can then be related to field theory by the  Martin-Siggia-Rose-Janssen-de Dominicis approach \cite{janssen1976lagrangean,bausch1976renormalized}  (see e.g. \cite{cardy1996scaling} for a review).

For damage spreading, we can take two points of view. 
The simplest one, which we will focus on here,
is to think in terms of a hierarchy of field theories labelled by $n$ (with Lagrangians $\mathcal{L}_n$),
 each member of the hierarchy containing the observables of the lower ones as subsectors of the theory.
In Sec.~\ref{sec:infinitefields} we will touch on the alternative viewpoint, in which there is a ``complete'' theory  
$\mathcal{L}_\infty$, with  an infinite number of fields, 
that  contains the observables for all $n$.

These field theories may be used to compute critical exponents by an epsilon expansion around four spatial dimensions, although we will not perform this here.
(Above 4D, the exponents take the mean-field values given in the previous Section.)
The field theories are also useful for analysing symmetries of the damage-spreading fixed point (Sec.~\ref{sec:trs}).

In more detail: we may use a standard construction to obtain  a spatially local model whose equations of motion can be obtained in a controlled way.
We start with a spatial lattice of ``islands'',  
each hosting a mean-field model with a finite but large number $N$ of sites. 
Each island is then governed by the stochastic equations for the densities  derived in the previous Section.
We  then incorporate ``swap'' updates into the model that exchange spin states between 
pairs of sites belonging to nearby islands. 
These updates  allow for diffusion of the densities between spatial sites, adding  spatial gradients to the stochastic equations.

\subsection{Field theory for $n=3$}
\label{sec:fieldtheoryfinitedn3}

We start with ${n=3}$, where 
the continuum theory is slightly  simpler than in the general case.
In the basis of pairwise damages, equations in Sec.~\ref{sec:meanfieldthreereplica} give
\ba\label{eq:3replicalocaleqn}
\partial_t \rho_{1, 2}
& = 
D \nabla^2 \rho_{1,2}
+
r \rho_{1, 2} - g \rho_{1, 2}^2 + \eta_{1, 2},
\\
\partial_t \rho_{2, 3}
& = 
D \nabla^2 \rho_{2,3}
+
r \rho_{2,3} - g \rho_{2,3}^2 + \eta_{2,3},
\\
\partial_t \rho_{1, 3}
& = 
D \nabla^2 \rho_{1,3}
+
r \rho_{1, 3} - g \rho_{1, 3}^2 + \eta_{1, 3},
\label{eq:3replicalocaleqnlast}
\end{align}
 for the critical fields,
with noise covariances\footnote{In the lattice construction mentioned above, the swap operations are another source of noise. Since this noise is conserving, it leads to a term with an additional derivative which is less relevant.}
\ba
\< \eta_{1,2}\eta_{1,2}\> 
& \simeq \nu \rho_{1,2}\delta^{(d+1)}
\\
\< \eta_{1,2}  \eta_{2,3} \> 
& \simeq  \f{\nu}{2}\lf
\rho_{1,2} + \rho_{2,3} - \rho_{1,3}
\ri
\delta^{(d+1)},
\end{align}
and the expressions related by cycling the replicas, where the spacetime arguments have been suppressed to avoid clutter and  ${\delta^{(d+1)}= \delta^{(d)}(x-x')\delta(t-t')}$.
For convenience we have used the basis of 2-replica damages introduced above Eq.~\ref{eq:pairwisediffdefn}.

The fact that there is only a \textit{single} independent quadratic coefficient $g$ in Eqs.~\ref{eq:3replicalocaleqn}--\ref{eq:3replicalocaleqnlast}
is evident when we use the basis of 2-replica damages
(as already mentioned below Eq.~\ref{eq:2replicadifferenceseq}).
On the other hand, if we write the equations in terms of the partition densities $\{\rho_\pi\}$ for two-block partitions, 
several types of quadratic term appear that are not related by replica symmetry.
The fact that their couplings are all determined by a single constant $g$ is physically a result of the fact that all the replicas evolve independently, see the discussion below Eq.~\ref{eq:Kdefgeneralrepeat}.

The MSR approach forms an integral over trajectories,
using additional ``response fields'' $\widetilde \rho_{a,b}\in i \mathbb{R}$ to enforce the  equations of motion \cite{cardy1996scaling}. The path integral is weighted by $\exp({-\int \dd t\dd^d x \mathcal{L}_3})$ with: 
\ba \notag
\mathcal{L}_3 = & \sum_{a<b} \left[ 
\widetilde\rho_{ab}
\lf \partial_t - D \nabla^2 - r \ri \rho_{ab}
+ g \widetilde \rho_{ab}^{\phantom{1}}\rho_{ab}^2
- \f{\nu}{2}\widetilde\rho_{ab}^2 \rho_{ab}^{\phantom{1}}
\right]
\\
& - \f{\nu}{2} \left[
\widetilde \rho_{12} \widetilde \rho_{23} \lf \rho_{12}+\rho_{23}-\rho_{13} \ri
+ \text{cyclic} \right] .
\label{eq:n3lagrangian}
\end{align}
At the end of this Section we will rewrite this Lagrangian in a more compact way after a change of basis.
Despite its apparent complexity, Eq.~\ref{eq:n3lagrangian} 
really only 
contains  a  \textit{single} nontrivial cubic coupling constant, which is ${\lambda = \sqrt{g\nu/2}}$. 
This is because  the ratio of $g$ to $\nu$ can be scaled out of the problem by rescaling the fields  (see Sec.~\ref{sec:trs}).

The Lagrangian above has an interesting relation to directed percolation.
If we retained only the first line of (\ref{eq:n3lagrangian}), we would have three decoupled copies of the critical theory for directed percolation \cite{janssen1976lagrangean,
cardy1980directed,
bronzan1974higher,
janssen1981nonequilibrium} (reviews: \cite{hinrichsen2000non,odor2004universality,cardylecturenotesnoneq}).
However they are coupled by the second line.

This coupling is important if we examine observables that involve all three replicas.
On the other hand, if we  only look at correlation functions of  $\rho_{12}$ and $\widetilde \rho_{12}$,
then the theory must necessarily reduce to the \textit{two}-replica theory. The latter  is governed by the Lagrangian  ${\mathcal{L}_2 = \widetilde\rho_{12}
\lf \partial_t - D \nabla^2 - r \ri \rho_{12}
+ g \widetilde \rho_{12}^{\phantom{1}}\rho_{12}^2
- \f{\nu}{2}\widetilde\rho_{12}^2 \rho_{12}^{\phantom{1}}}$ that follows from the results in Sec.~\ref{sec:MFTn2}.

It is remarkable that the additional nontrivial interaction in the second line of
(\ref{eq:n3lagrangian})
does not introduce any new independent \textit{coupling constant} beyond those in the two-replica theory. 
This is guaranteed by an argument mentioned in Sec.~\ref{sec:n3meanfieldfull}, which carries over to the spatially local theory, and  implies that the noise terms in the three replica theory are determined by those in the two-replica theory.

Since all of the couplings in $\mathcal{L}_3$ already appear in $\mathcal{L}_2$, the RG flows of Eq.~\ref{eq:n3lagrangian} follow immediately from those of  standard directed percolation, assuming that we restrict to terms of at most cubic order in $\rho,\widetilde\rho$ as above.
The upper critical
(spatial) dimensionality for both $\mathcal{L}_2$ and $\mathcal{L}_3$ is ${d=4}$.
In $d<4$ the massless theory ($r=0$) flows to a nontrivial fixed point, where only one of the couplings, namely $r$, is relevant.
The critical point
 can be studied in a $d=4-\epsilon$ expansion, which at two-loop order is fairly accurate for $d\geq 2$ \cite{hinrichsen2000non,odor2004universality, jensen1999low}.

Although $\mathcal{L}_3$ does not include new independent \textit{coupling constants}, it permits new \textit{operators}. Let us discuss the damage densities.

The scaling dimension of the pairwise damage densities $\rho_{ab}$ 
follows from the 2-replica theory: in ``time'' units, 
the  dimension of these fields is the standard directed percolation decay exponent $\alpha_2$ (Sec.~\ref{sec:dpreview}).

There is also a scaling operator associated with the density  $\rho_{(1)(2)(3)}$.
In Sec.~\ref{sec:meanfieldthreereplica} we argued that this density could be regarded as a composite operator, quadratic in the $\{\rho_{ab}\}$ (Eq.~\ref{eq:r123identification}).\footnote{The identification of $\rho_{(1)(2)(3)}$ with a composite operator can be obtained more explicitly by taking into account the equation of motion for $\rho_{(1)(2)(3)}$ in Sec.~\ref{sec:meanfieldthreereplica}.
Promoting this to the level of the field theory, we find that the quadratic terms for $\rho_{(1)(2)(3)}$ and its response field $\widetilde \rho_{(1)(2)(3)}$
include a positive squared mass 
(given by $r_F$ in Eq.~\ref{eq:rFeq}) even at the critical point.
As a result of this mass, this field can be integrated out. 
Its equation of motion gives an approximate identification between the bare field $\rho_{(1)(2)(3)}$ and a sum of bilinears.}

Therefore, above the upper critical dimension, the decay exponent $\alpha_3$ for $\rho_{(1)(2)(3)}$  is ${\alpha_3 = 2\alpha_2}$.
However, we expect both operators to acquire nontrival anomalous dimensions below the upper critical dimension (this is consistent with our simulations in 1+1D).

The two-block damage operators such as $\rho_{(1)(23)}$ are related to the operators discussed above by a change of basis  that was discussed in Sec.~\ref{sec:meanfieldthreereplica} (and for general $n$ in Sec.~\ref{sec:meanfieldgeneraln}).
Defining the alternative field basis  
\ba\label{eq:changeofbasisyetagain}
\hat \rho_{(1)(23)} & = \f{1}{2} \lf \rho_{12} + \rho_{13} - \rho_{12}  \ri, 
\end{align}
and cyclically 
(this is the ${n=3}$ versions of the 
basis $\{\hat \rho_\sigma\}$ defined in Eq.~\ref{eq:caretnotation}), we have
\be
\rho_{(1)(23)}  = \hat \rho_{(1)(23)} - \f{1}{2} \rho_{(1)(2)(3)},
\ee
etcetera
(this is a rewriting of Eq.~\ref{eq:pairwisediffdefn}).
 This shows for example that the two-block damage densities decay like $t^{-\alpha_2}$ with a subleading term of order $t^{-\alpha_3}$.
This relation between a damage operator (on the left-hand side of Eq.~\ref{eq:changeofbasisyetagain}) and a superposition of scaling operators (on the right) exemplifies a more general structure that will be explained in Sec.~\ref{sec:rgstructure}.

For completeness, let us rewrite the Lagrangian in the alternative basis of fields  $\{\hat \rho_{\sigma}\}$  labelled by  two-block damages (and their conjugate fields). To avoid clutter, we again drop  caret symbols: 
\ba 
\mathcal{L}_3 = & 
\widetilde\rho_\pi
\lf \partial_t - D \nabla^2 - r \ri \rho_\pi
+ 
g \, \hat{\mathcal{K}}^{\pi}_{\sigma\sigma'}
\widetilde \rho_\pi \rho_\sigma \rho_{\sigma'}
- \f{\nu}{2} \widetilde \rho_\pi^2 \rho_\pi,
\label{eq:L32blockbasis}
\end{align}
where ${\pi, \sigma,\sigma'}$ are summed over the three partitions 
\ba
\sigma_1 &= (1)(23),& 
\sigma_2&=(2)(13), &
\sigma_3&=(3)(12),
\end{align}
and with  the tensor values
\ba\notag
\hat{\mathcal{K}}^{\sigma_1}_{\sigma_1,\sigma_1} & = 1,
& 
\hat{\mathcal{K}}^{\sigma_1}_{\sigma_1,\sigma_2} & = \f{1}{2},
& 
\hat{\mathcal{K}}^{\sigma_1}_{\sigma_2,\sigma_2} & =0,
&
\hat{\mathcal{K}}^{\sigma_1}_{\sigma_2,\sigma_3} & =-\f{1}{2}
\end{align}
 (and  cyclic permutations).
After a rescaling of $\rho$ and $\widetilde \rho$ we can write the nonlinear terms as $\lambda (\hat{\mathcal{K}}^{\pi}_{\sigma\sigma'}
\widetilde \rho_\pi \rho_\sigma \rho_{\sigma'}
- \widetilde \rho_\pi^2 \rho_\pi)$,
with a single coupling constant ${\lambda = \sqrt{g\nu /2}}$. 
This structure is preserved under RG (Sec.~\ref{sec:trs}).

\subsection{Field theory for $n>3$}
\label{sec:fieldtheoryfinitedgeneraln}

Finally, let's discuss the continuum theory for an arbitrary number of replicas. 
The derivations in Sec.~\ref{sec:meanfieldgeneraln}
show that, close to (or above) the upper critical dimension we can work with  
$\stirling{n}{2}$ massless fields, 
which are in correspondence with the two-block partitions.
Since $\stirling{n}{2}$ is larger than $n(n-1)/2$,
this means that 
it is no longer possible to express the theory only in terms of pairwise replica differences.
(For example, for $n=4$ replicas there are $6$ pairwise differences $\rho_{ab}$, but there are $7$ two-block partitions.)

To simplify the derivation of the field theory,
we assume for the remainder of this section that the microscopic model has ${q=2}$ (i.e. the underlying circuit is a Boolean circuit for bits). 
Then the two-block 
partitions give a complete basis for the local densities, and we can work with only these $\stirling{n}{2}$ fields. We expect that the ${q>2}$ case can effectively be reduced to this case, at large lengthscales,  by carefully integrating out the massive fields corresponding to $\rho_\pi$ with $|\pi|>2$.

\subsubsection{Stochastic equations}

This simplification gives the stochastic equations
(see Eqs.~\ref{eq:2blockonlygeneralanyq}; all partitions $\pi,\sigma,\sigma'$ are in the set of two-block partitions, which we denote by $\mathcal{S}_2^n$)
\ba \label{eq:langevingeneral}
\partial_t \rho_\pi
& = 
D \nabla^2 \rho_\pi
+ r \rho_\pi 
 -
g
\sum_{\sigma,\sigma'\in\mathcal{S}_2^n}
\hat{\mathcal{K}}_{\sigma,\sigma'}^\pi
\rho_\sigma\rho_{\sigma'}
+ \eta_\pi.
\end{align}
The noise correlator  is diagonal in the basis  of two-block partitions (Eq.~\ref{eq:noise2blockgeneral}),
\ba\label{eq:noisecorrelatorgeneraln}
\<\eta_\pi(t)\eta_\sigma(t')\> = \nu 
\rho_\pi(t)
\delta_{\pi\sigma} 
\delta(t-t') \delta^{d}(x-x').
\end{align}
The form of $\hat{\mathcal{K}}$ was given above in Eqs.~\ref{eq:Kdefgeneral2},~\ref{eq:Kdefgeneral3}:
we have extracted the constant $g$
that  appeared in the ${n=2}$ and ${n=3}$ equations, so that $\hat{\mathcal{K}}$ obeys the normalization $\hat{\mathcal{K}}_{\sigma\sigma}^\sigma=1$ for any  $\sigma\in \mathcal{S}_2^n$.
$\hat{\mathcal{K}}$ contains a single model-dependent dimensionless parameter $\theta$.

We have derived these equations in a controlled way for a particular family of microscopic models, but we expect them to capture the universal behavior of the damage spreading transition more broadly (in the spirit of Landau theory for equilibrium transitions, or the stochastic equations of the Hohenberg-Halperin classification for dynamical critical phenomena \cite{halperin2019theory}).
In this more general context, the neglect of higher order terms --- of order $\rho^4$ in Eq.~\ref{eq:langevingeneral} or of order $\rho^2$ in Eq.~\ref{eq:noisecorrelatorgeneraln} ---  is justified on grounds of universality, e.g.  on the basis that they are irrelevant within the ${4-\epsilon}$ expansion.

A key point is that the form of the resulting equations is extremely rigid. 
The constraint that the equations for any $n$ must be consistent with the two-replica equations fixes the index structure in 
Eqs.~\ref{eq:langevingeneral},~\ref{eq:noisecorrelatorgeneraln}.\footnote{The linear terms in both equations can be fixed by considering the case where only a single nontrivial damage type  $\pi\in \mathcal{S}_2^n$ is present. In this setting  no other nontrivial damage type can be generated, and the dynamics of the damage reduces to the $n=2$ problem.}
For similar reasons, the constants $D$, $r$, $g$, $\nu$  
and $\theta$ 
(which depend on the microscopic model)  are independent of $n$.

\subsubsection{Lagrangians for general $n$}\label{subsec:lagrangiangeneraln}

We conjecture that (for a given $n$ and a given  ${d<4}$) all generic critical models with ${\theta>0}$ flow under RG to the same single fixed point. 
At this fixed point, 
the only relevant perturbation 
that is allowed by the replica structure 
is then the perturbation that corresponds to changing $r$ in Eq.~\ref{eq:langevingeneral}.

The RG flows for a sector of the couplings
(the sector that is visible for $n\leq 3$)
 are simply those of directed percolation: 
 the massless theory  contains a single cubic coupling constant, 
$\lambda = \sqrt{g \nu/2}$ (see below),
which flows to a nonzero fixed point value when ${d<4}$.
For ${n \geq 4}$ we must also consider the flow of the dimensionless constant $\theta$ (which can be viewed as the ratio of two cubic coupling constants). We expect that, for ${d<4}$,
 $\theta$ flows to some fixed point value and that perturbations to this value are RG--irrelevant.

The starting point for studying these flows is to again promote the Langevin equations to a field theory. 
After a rescaling of fields, this takes the form
\ba 
\mathcal{L}_n = & 
\widetilde\rho_\pi
\lf \partial_t - D \nabla^2 - r \ri \rho_\pi
+ 
\lambda \lf  \hat{\mathcal{K}}^{\pi}_{\sigma\sigma'}
\widetilde \rho_\pi \rho_\sigma \rho_{\sigma'}
-  \widetilde \rho_\pi^2 \rho_\pi \ri,
\label{eq:ngenerallagrangian}
\end{align}
where repeated indices are summed over 
the ${\stirling{n}{2}=2^{n-1}-1}$ different
two-block partitions. 
The time-reversal symmetries discussed in Sec.~\ref{sec:trs} ensure that the Lagrangian retains this structure under RG, up to renormalization of the couplings shown and of $\theta$.

The scaling operators of the fixed point can be labelled by partitions
 $\pi \in \Pi_n$ (not necessarily restricted to two blocks).
  This structure will be analyzed in Sec.~\ref{sec:rgstructure}.
We expect a set of scaling operators
 $\mathcal{O}_\pi$ built from damage densities.
 On the lattice, $\mathcal{O}_\pi$ would correspond to the lattice damage density $\rho_\pi$, plus subleading terms given by finer damage densities.
In the context of the $4-\epsilon$ expansion, 
  $\mathcal{O}_\pi$ for a 2-block partition $\pi$ is essentially the corresponding elementary field $\rho_\pi$.
  For higher partitions $\mathcal{O}_\pi$ is  a composite field, expressed using products of the elementary fields: 
  an example for ${n=3}$ is the operator 
  ${\mathcal{O}_{(1)(2)(3)} =\rho_{(1)(2)(3)}}$ that we discussed in the previous Section.
A key simplification is that the scaling dimension of $\mathcal{O}_\pi$ depends only on the number $|\pi|$ of blocks in $\pi$, 
and not on the number of replicas in each block, 
or directly on the value of $n$. 
It would be very interesting to compute the  spectrum of scaling dimensions in ${4-\epsilon}$ dimensions.
In addition to the operators $\mathcal{O}_\pi$ which measure damage, there are corresponding ``dual'' operators $\widetilde{\mathcal{O}}_\pi$ which \textit{create} damage (Sec.~\ref{sec:rgstructure}).

Note that one is free to use alternative bases in field space for writing $\mathcal{L}_n$ in Eq.~\ref{eq:ngenerallagrangian}. 
We saw that for ${n=3}$ the deterministic terms in the Langevin equations simplified when we used the basis of two-replica overlaps. 
For $n=4$ the basis of two-replica overlaps is not complete, but we obtain a complete basis if we supplement it with one additional field. 

\subsubsection{Lagrangian with an infinite number of fields}
\label{sec:infinitefields}

Finally we mention an alternative perspective on the hierarchy of field theories that is slightly more abstract.
So far we have thought in terms of a distinct field theory for each value of $n$
(albeit with very close relations between these theories, which for example involve the same coupling constants).
An alternative formalism is to define a field theory $\mathcal{L}_\infty$ for ``$n=\infty$''. 
Formally the Lagrangian still resembles the right-hand-side of Eq.~\ref{eq:ngenerallagrangian}.
However, we now wish to include an  infinite number of fields $\rho_\pi$ 
that are labelled by two-block partitions of the infinite set of replicas $\{1,2,3,4,\ldots\}$.
To make sense of this 
we can restrict to the case where one of the two blocks of $\pi$ is finite.
Summing over $\pi$ is then equivalent to summing over finite subsets of $\{1,2,3,4,\ldots\}$, of which there are a countable infinity.

From a  pragmatic point of view, any observable involving a finite number of replicas can be calculated using $\mathcal{L}_n$ for finite $n$, so we are not forced to invoke the more abstract object $\mathcal{L}_\infty$. 
However the existence of $\mathcal{L}_\infty$ allows us to think of a single field theory and a single RG fixed point that controls all possible observables for any number of replicas.

\section{Renormalization group structure}
\label{sec:rgstructure}

We turn to the renormalization group structure of the damage spreading critical point, with the aim of understanding the spectrum of scaling dimensions for the key local observables in the theory.
The analysis in this Section does not rely on a continuum field theory formulation or on any specific dimensionality, so it will be simpler now to switch back to  the language of lattice models and real space RG.
The analysis should hold in any number of dimensions.

One tool for classifying observables is replica symmetry. 
But in fact, in the present problem, the replica structure  leads to   constraints that are much stronger than those imposed by symmetry.  
These are associated with the way that the partial order property of partitions (reviewed below) interacts with the absorbing state properties of the damage evolution.

We have argued that the densities $\rho_\pi$,
which detect the presence of a given pattern $\pi$ of difference and agreement between the replicas, 
form a natural set of local observables for damage spreading.
In order to understand the spectrum of scaling dimensions at the critical RG fixed point,
we must understand how an RG transformation acts on these densities.
As we discuss below, this RG action
mixes the densities in a nontrivial way.
In order to obtain the spectrum of scaling dimensions, we must construct the RG scaling operators\footnote{Loosely speaking these are operators that are invariant (up to rescaling) under RG and so have definite scaling dimensions \cite{cardy1996scaling}.}
that diagonalize the RG transformation.
Loosely speaking, these scaling operators are linear combinations of $\rho_\pi$ for different $\pi$.
Fortunately, the partial order structure for  partitions mentioned above means that the RG action is extremely constrained.\footnote{See Ref.~\cite{dai2020quantum} for a different setting where the RG transformation must respect a partial ordering.} 
As a result the spectrum of scaling operators/dimensions is simpler than might have been expected.

A quick caveat: 
We will simplify the discussion 
by restricting to the ``simplest'' operator of a given damage type. 
We neglect subleading operators of a given damage type, or operators 
carrying additional symmetry charges (see comments at the end of Sec.~\ref{sec:moregeneraloperators}). 
This neglect simplifies the relation between lattice operators and scaling operators.\footnote{That is, 
when we identify some lattice operator with a scaling operator,  the identity really holds only up to subleading terms given by  scaling operators with higher dimensions.}
We believe that this is sufficient to see the key constraints on the spectrum, but a fuller analysis would be worthwhile
and could be carried out in the epsilon expansion.
We focus first on the density operators $\rho$ that measure damage, deferring a briefer discussion of the ``dual'' operators $\widetilde\rho$ that \textit{create} damage to Sec.~\ref{sec:moregeneraloperators}.

We find below that for each partition ${\pi\in \Pi_n}$ 
(for a given choice of $n$)
there is a leading scaling operator $\mathcal{O}_{\pi}({\bf x})$.
However, rather 
than being equal to $\rho_\pi({\bf x})$,
the scaling operator involves a sum of partitions including both $\pi$ and all the partitions ${\sigma< \pi}$ that are \textit{finer} than $\pi$ (see below):
\be\label{eq:formofscalingop}
\mathcal{O}_{\pi}({\bf x}) = \sum_{\sigma\leq\pi} v_{\pi,\sigma} \rho_\sigma ({\bf x}).
\ee
For example,
\begin{align}
\notag
\mathcal{O}_{(1)(2)} & = \rho_{(1)(2)}, 
\\
\mathcal{O}_{(1)(2)(3)}  & = \rho_{(1)(2)(3)},
\label{eq:scalingopexamples}
\\\notag
\mathcal{O}_{(12)(3)} &  = \rho_{(12)(3)} + \f{\rho_{(1)(2)(3)}}{2}. 
\end{align}
(The factor of 1/2 is explained in Sec.~\ref{subsec:RG} around Eq.~\ref{eq:forgettingrho}.)
The scaling dimension of $\mathcal{O}_\pi$ only depends on the number $|\pi|$ of blocks in the partition, so we will  write the scaling dimension (inverse length dimension) as $\Delta_{|\pi|}$ instead of ${\Delta_\pi}$.

Inverting Eq.~\ref{eq:formofscalingop}, the 
density $\rho_{\pi}$ is a  combination of the scaling operator for $\pi$ and the scaling operators for $\sigma <\pi$:
\be\label{eq:Ofiner}
\rho_{\pi}({\mathbf x}) = \sum_{\sigma\leq \pi} (v^{-1})_{\pi,\sigma}\mathcal{O}_\sigma ({\mathbf x}).
\ee

More concretely, if for example we initialize the replicas in completely random initial states, so that there is  damage throughout the system, then for $n=2$ the local damage density decays as (suppressing the spatial argument) 
\be
\big\langle\rho_{(1)(2)}(t)\big\rangle \sim C_2 t^{-\Delta_2/z}.
\ee
where $C_2$ is a constant.
Note that the factor of $z$ is because of the convention that the scaling dimension $\Delta$ gives the inverse \textit{length} dimension of the operator; 
the decay exponent $\alpha_2=\Delta_2/z$ 
(more generally, ${\alpha_n = \Delta_n/z}$)
can be viewed as the scaling dimension measured in time units.\footnote{More precisely, this is true below the upper critical dimension ${d=4}$. Above ${d=4}$ the decay exponent is no longer given by the scaling dimension, because of the dangerous irrelevance of the interaction terms.
This is analogous to the breakdown of hyperscaling for the Ising model above $d=4$ \cite{cardy1996scaling}.}
For $n=3$ we have
\ba\label{eq:rgsec123decay}
\big\langle\rho_{(1)(2)(3)}(t)\big\rangle & \sim C_3 t^{-\Delta_3/z},
\end{align}
 but 
\ba
\big\langle \rho_{(1)(23)}(t)\big\rangle & \sim \f{C_2}{2} t^{-\Delta_2/z} -\f{C_3}{2} t^{-\Delta_3/z},\label{eq:n32blockdensity}
\end{align}
where the coefficients are fixed by
the mapping between operators at different $n$ that is discussed 
around Eq.~\ref{eq:forgettingrho}.
Here we are ignoring  terms associated with subleading operators in a given sector (Sec.~\ref{sec:moregeneraloperators});
it is plausible that these contributions are smaller at large time than the subleading correction shown in Eq.~\ref{eq:n32blockdensity}.
Similarly, for ${n=4}$
\ba\label{eq:subleadingopdecay}
\big\langle\rho_{(12)(34)}(t)\big\rangle & \sim D t^{-\Delta_2/z} + O(t^{-\Delta_3/z}),
\\ \label{eq:subleadingopdecay2}
\big\langle\rho_{(1)(234)}(t)\big\rangle & \sim D' t^{-\Delta_2/z} + O(t^{-\Delta_3/z}),
\\\label{eq:subleadingopdecay3}
\big\langle \rho_{(12)(3)(4)}(t)\big\rangle & \sim D'' t^{-\Delta_3/z} + O(t^{-\Delta_4/z}),
\\\label{eq:subleadingopdecay4}
\big\langle\rho_{(1)(2)(3)(4)}(t)\big\rangle & \sim D''' t^{-\Delta_4/z},
\end{align}
and so on. The reasoning in Sec.~\ref{subsec:RG}
shows that, above   ${D+D'=C_2/2}$ and ${D''=C_3/3}$; we have checked that the numerica data in Sec.~\ref{sec:generalpartitionssimulations} are consistent with these relations.

Note that there can be 
  inequivalent  operators
with the same scaling dimension, even after accounting for equivalence under replica symmetry:
for example  $\rho_{(12)(34)}$ and $\rho_{(123)(4)}$.
This richer structure
would become apparent if we considered multi-point correlation functions, 
universal amplitudes associated with power law decays, etc.

\subsection{RG transformation and poset structure}\label{subsec:RG}

We use a heuristic picture in terms of real-space renormalization group in order to constrain the structure of the scaling operators in the fixed-point theory.
While this simplified RG picture would not be sufficient for  accurate computations of exponents, it should be enough to fix exactly the discrete structure that we discuss, which arises from basic combinatorial properties of partitions.

Recall that one writes $\tau' \leq \tau$ if the partition ${\tau'\in \Pi_n}$ is finer than the partition ${\tau\in \Pi_n}$, i.e if $\tau'$ can be obtained from $\tau$ by splitting groups (this includes the case ${\tau'=\tau}$). 
This structure makes $\partitions{n}$ into a partially ordered set \cite{stanley_enumerative_1999} (only partially ordered because  some pairs $\sigma$, $\tau$ obey neither $\sigma\leq \tau$ nor $\tau\leq \sigma$).
The ordering relations between partitions are illustrated by the Hasse diagrams, reproduced in  Fig.~\ref{fig:hassediagrams} for the cases $n=3$ and $n=4$.

Now consider the  effective $n$--replica dynamics
for some value of $n$
(for now we will leave the $n$-dependence of various objects implicit)
in a  model on a spatial lattice.
We let ${\sigma({\mathbf x})\in \partitions{n}}$ represent the partition associated with spacetime point ${{\mathbf x} = (x,t)}$. As discussed below, it is convenient to take the model to be in continuous time
(see e.g. the continuous time update in Eq.~\ref{eq:MFupdate}), so ${x\in \mathbb{Z}}$ but ${t\in \mathbb{R}}$. 
These choices should not affect the universal behavior.
In addition, let us assume that  $q$, the  number of local physical  states on a site in the single-replica problem, 
is large enough to allow any partition  ${\tau\in\partitions{n}}$ to occur at a single site. 
(This is not a crucial restriction, since if $q$ is smaller we can consider operators on multiple sites, as we do in our simulations.)
For each partition type $\tau$, let  $\rho_\tau({\mathbf x})$ be the corresponding density operator, 
which is 1 if ${\sigma({\mathbf x}) = \tau}$ and zero otherwise:
\be
\rho_\tau({\mathbf x}) = \delta_{\tau, \sigma({\mathbf x})}.
\ee

For models like those we have discussed in this paper, the spreading problem is described by effective Markovian dynamics for the damage variables.
Imagine that we perform a real-space RG transformation 
for these dynamics,
by forming block degrees of freedom made out of $b$ spatial sites.
We imagine that  the critical point is described by a fixed point of this RG transformation.\footnote{That is, at the fixed point, the local weights associated with spacetime trajectories are left invariant under the RG transformation.}
As usual, in order to find a fixed point, we must rescale time by a factor $b^z$, where $z$ is the dynamical exponent (of directed percolation).
It is because $b^z$ is not an integer that it is convenient to consider continuous-time dynamics.

The blocked degrees of freedom are partitions $\sigma'({\mathbf{X}})$, where ${\mathbf{X} = (X, T)}$,  ${X\in \mathbb{Z}}$ labels a spatial block, and ${T\in \mathbb{R}}$ is the rescaled time. 
 The crucial point is that,
for consistency with the absorbing state properties of the dynamics, 
$\sigma'({\mathbf{X}})$ should be defined as
the greatest lower bound of the partitions for the sites $x^1, \ldots, x^b$ within the block: 
\be\label{eq:coarsegrainingrelation}
 \sigma'\lf X,T \ri 
 =
 \bigwedge\bigg\{
 \sigma( x, \,b^{-z} T )
\, : \,
x \in X
  \bigg\}.
\ee 
That is, $\sigma'({\mathbf{X}})$
is  the coarsest partition which is nevertheless a refinement of all the partitions within the block.
This is equivalent to saying that two replicas $a$ and $b$ disagree in the coarse-grained variable $\sigma'(\mathbf{X})$ if and only if they disagree on at least one of the microscopic  sites. 
This definition ensures that the coarse-grained dynamics inherits the key absorbing-state property of the microscopic dynamics: if two replicas agree everywhere in space at some time $t$, then they continue to agree at all later times.

The RG transformation also acts on operators. 
Let $\mathcal{O}({\mathbf{x}})$ be an operator
(for example a density) 
at site ${x}$ prior to coarse-graining.
Let $X$ denote the block containing the ``microscopic'' site ${x}$.
After coarse-graining, $\mathcal{O}({\mathbf x})$ 
is replaced with an operator in the coarse-grained theory that is 
quasi-localized around ${\bf  X}$. For simplicity we make the  approximation
(standard in textbook treatments of the real-space RG \cite{cardy1996scaling})
that the coarse-grained operator also lives strictly at  ${\bf  X}$.
We believe this is sufficient to reveal the basic constraints on the RG transformation.

The observables at a microscopic site are the densities $\rho_\tau(\mathbf{x})$. 
In the above  approximation, coarse-graining maps such an operator to a sum of density operators $\rho'_{\tau'}$ in the coarse-grained theory, with some coefficients $A_{\tau'\tau}$:\footnote{This equation means that correlators in the initial theory involving the operator on the left hand side are equal to correlators in the coarse-grained theory involving the operator sum on the right-hand side \cite{cardy1996scaling}.}
\be\label{eq:RGtransformationRho}
\rho_\tau({\mathbf{x}}) 
\quad 
\overset{{\text{RG step}}}{\longrightarrow}
\quad
 \sum_{\tau'\in \partitions{n}}\rho'_{\tau'}({\mathbf X}) A_{\tau' \tau}.
\ee
Heuristically,  
we can think of 
$A_{\tau'\tau}$ as the probability that the microscopic spin is $\tau$, conditional on the block spin being $\tau'$ --- see App.~\ref{app:blockspindetails} for more detail of this picture. 

Within this  approximation of restricting  to single-site operators,
 the scaling operators 
 are determined by  eigenvectors $v$ of the matrix $A$. If ${A v = b^{- \Delta } v}$,
then the operator  $ {\mathcal{O}({\mathbf{x}}) = \sum_\tau v_\tau  \rho_\tau({\mathbf x})}$ 
is transformed under RG into 
$ {b^{-\Delta} \mathcal{O}'({\mathbf{X}})}$
(again the prime indicates that this is an operator in the coarse-grained theory),
telling us that the combination $\mathcal{O}$ is a scaling operator with scaling dimension $\Delta$ \cite{cardy1996scaling}.

Now let's consider the constraints on the RG transformation matrix $A$. We now write $A^{(n)}$ to make explicit the fact that this matrix depends on the number of replicas we are considering.

\smallskip

\textit{(1) Preserving the identity operator.}
The trivial operator 1, which can also be written ${1=\sum_{\tau\in \Pi_n}\rho_\tau^{(n)}}$, 
is invariant under RG transformations. 
In our approximation of restricting to single-site operators, 
this invariance requires the rows of $A^{(n)}$ to sum to unity. 

\smallskip

\textit{(2) Respecting the partial order on $\Pi_n$.}
The RG transformation matrix  element $A^{(n)}_{\tau',\tau}$  
can be nonzero only when ${\tau'\leq\tau}$.
To see this, note that the operator 
$\rho_\tau({\mathbf x})$ 
is nonzero only when the microscopic variable $\sigma({\mathbf x})$ is equal to $\tau$.
Therefore, by Eq.~\ref{eq:coarsegrainingrelation},
$\rho_\tau({\mathbf x})$  takes a nonzero value only when the block spin is ${\leq \tau}$.
Therefore when mapped to the coarse-grained theory (Eq.~\ref{eq:RGtransformationRho}),  $\rho_\tau({\mathbf x})$ includes a sum only over partitions ${\tau'\leq \tau}$. This is equivalent to the stated constraint on $A^{(n)}$.

$A^{(n)}$ is therefore not only a  triangular matrix (for an appropriate ordering of the indices), 
but for $n>2$ has more zeroes than are required by triangularity. 
Such matrices form the ``incidence algebra'' associated with $\partitions{n}$ \cite{stanley_enumerative_1999}. The inverse of $A$ has the same property.

This structure means that for each $\pi\in \Pi_n$
 there is an eigenvector $v^{(\pi)}$, 
with $v^{(\pi)}_\tau$ being nonzero only for ${\tau\leq \pi}$. This gives the result for the scaling operators stated above in Eq.~\ref{eq:formofscalingop}.
Within the present approximation, 
the diagonal element $A_{\pi\pi}^{(n)} \equiv
 b^{-\Delta_\pi}$
sets the scaling dimension $\Delta_\pi$ of the corresponding scaling operator. 
In a more complete treatment, the diagonal element $A_{\pi\pi}^{(n)}$ would be replaced by a sub-block 
associated with multiple operators in the ``$\pi$ sector'', and we would diagonalize this sub-block to find the leading scaling dimension in this sector.

The above constraint arises from the absorbing state properties of the dynamics, which forced us to use an RG transformation like (\ref{eq:coarsegrainingrelation}).
The following constraints rely  on the additional fact that the physical replicas  do not interact with each other (prior to averaging over the random circuit).

\smallskip

\textit{(3) ``Merging'' relation.}
Let us say that we choose initial conditions such that replica ${n-1}$ and replica $n$ have exactly the same state on every site. They may then be treated as a single replica (``merged''), 
and the problem reduces to the ${[n-1]}$-replica problem.

Consistency with this reduction imposes a constraint on the RG transformation matrices $A^{(n)}$.
Let ${\tau,\tau'\in \Pi_n^{(n-1)\leftrightarrow n}}$ be
chosen from the subset ${{\Pi_n^{(n-1)\leftrightarrow n}}\subset \Pi_n}$ of partitions for which   replicas ${n-1}$ and $n$ are in the same block, and let $\widetilde \tau$, $\widetilde \tau'$ be the corresponding partitions in $\Pi_{n-1}$ obtained by merging the two replicas. Then we require
${A_{\tau, \tau'}^{(n)}  = A_{\widetilde\tau, \widetilde\tau'}^{(n-1)}}$.

This relation implies that the diagonal elements $A^{(n)}_{\tau,\tau}$, which determine the scaling dimensions $\Delta_\tau$, depend only on the number of blocks in $\tau$.\footnote{E.g. merging implies that  ${A^{(3)}_{(1)(23), (1)(23)} =
A^{(2)}_{(1)(2), (1)(2)}}$, giving $\Delta_{(1)(23)}  =\Delta_{(1)(2)}$.}
Therefore we write ${\Delta_\tau = \Delta(|\tau|)}$.

In fact, the merging constraint on $A$ can also be obtained as a special case of the following.

\smallskip

\textit{(4) ``Forgetting'' relation.} 
Within the $n$-replica theory, we may write down
observables that  
depend on only ${n-1}$ of the replicas, say replicas ${\{1,2,\ldots n-1\}}$. We must obtain the same results regardless of whether we view these observables as living in the $n$-replica theory or in the ${(n-1)}$-replica theory.

Let us think of $\rho^{(n-1)}_{\widetilde\tau}$ as an observable on the first ${n-1}$ replicas, in a system of $n$ replicas. 
This observable may be written in terms of the densities $\rho^{(n)}$ of the $n$-replica theory by summing over all of the $n$-replica partitions that are consistent with $\widetilde\tau$:
\be\label{eq:forgettingrho}
\rho_{\widetilde\tau}^{(n-1)}
=
\sum_{\substack{
\pi \, :
\\
f(\tau) = \widetilde \tau
}}
\rho_{\tau}^{(n)} .
\ee
Here ${f:\Pi_n\to \Pi_{n-1}}$ maps partions of ${\{1,\ldots,n\}}$ to partitions of ${\{1,\ldots,n-1\}}$ by forgetting about the last replica.

The above relation allows us to ``promote'' operators/scaling operators in the ${(n-1)}$-replica theory to 
operators/scaling operators in the $n$-replica theory, as discussed below.
It also constrains the RG transformation:
when we coarse-grain the above operator, we must get the same result regardless of whether we apply the RG transformation $A^{(n-1)}$ to the left-hand side of (\ref{eq:forgettingrho}), or the transformation $A^{(n)}$ to the right-hand side.
This requires
\ba
\sum_{\substack{
\tau\in \Pi_n \, :
\\
f(\tau) = \widetilde \tau
}}
A^{(n)}_{\tau'\tau}
& = 
A^{(n-1)}_{f(\tau'), \widetilde \tau}
 \, ,
& 
& \text{for all $\tau' \in \Pi_n$, \, 
$\widetilde \tau \in \Pi_{n-1}$}.
\end{align}

\smallskip

The above constraints suffice to fix the scaling operators of the type under discussion for $n=2$, $n=3$.
The incidence algebra structure implies   that
$\rho_{(1)(2)}$ 
is a scaling operator
(within our approximation of neglecting subleading operators within the same 
topological sector)
in the $n=2$ theory and 
$\rho_{(1)(2)(3)}$
is a scaling operator in the ${n=3}$ theory. 
(We omit the superscript $n$ since its value can be inferred from the partition label.)
The other nontrivial scaling operators for ${n=3}$ follow either from examining the structure of $A^{(3)}$, 
or more simply just by using 
Eq.~\ref{eq:forgettingrho} to ``promote'' the scaling operator $\rho_{(1)(2)}$ to a scaling operator in the ${n=3}$ theory. Doing this gives the operator
\ba\notag
\rho_{(1)(2)}
\rightarrow &
\rho_{(13)(2)} + \rho_{(1)(23)}+\rho_{(1)(2)(3)}
\\
 = &  
\lf 
\rho_{(13)(2)} + \f{\rho_{(1)(2)(3)}}{2} 
\ri 
+
\lf 
\rho_{(1)(23)} + \f{\rho_{(1)(2)(3)}}{2} 
\ri.\label{eq:throwinthirdreplica}
\end{align}
We have split the second line into contributions that take the form in Eq.~\ref{eq:formofscalingop}. We see that we have three scaling operators of the form 
\be
\mathcal{O}_{(1)(23)} = 
\rho_{(1)(23)} + \, \f{1}{2}
\rho_{(1)(2)(3)}
\ee
and cyclically.\footnote{Despite the fact that we have worked with a simplified RG transformation,
the coefficient $1/2$ here is exact, because a version of the map (\ref{eq:throwinthirdreplica}) between operators in the $n-$ and $(n-1)$-replica theories remains valid if we extend the RG treatment to  account for mixing with multi-site operators  etc.}

\subsection{Comparison with simulations in 1+1D}
\label{sec:generalpartitionssimulations}

\begin{figure}[t]
\includegraphics[width=\linewidth]{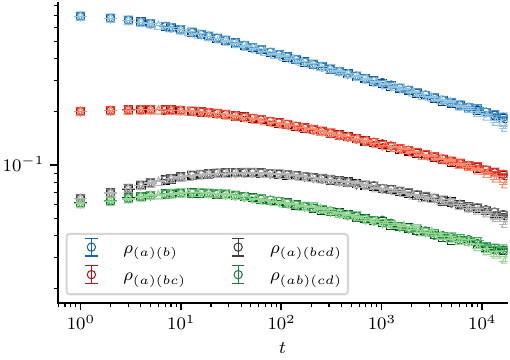}
\caption{Decay of damage densities for two-block partitions:
the operator $\rho_{(1)(2)}$ of the $n=2$ theory; 
the operators $\rho_{(a)(bc)}$ of the $n=3$ theory; 
the operators $\rho_{(a)(bcd)}$ and $\rho_{(ab)(cd)}$ of the $n=4$ theory.
All these have leading exponent $\alpha_2$, but all except the first have subleading corrections from scaling operators asssociated with finer partitions (see Eqs.~\ref{eq:Ofiner}--\ref{eq:subleadingopdecay2}).
Different colours denote different different operators (see legend) whereas different markers with different colour intensities denote different $L=$ 512, 1024, 2048 and 4096.}
\label{fig:all-two-blocks}
\end{figure}

We now show numerical data for some of the more general densities we have just discussed, before returning briefly to the RG in the next Subsection  to discuss operators that are not damage densities.

In Sec.~\ref{subsec:numresults} we showed simulation results for the decay of 
$\rho_{(1)(2)}$, $\rho_{(1)(2)(3)}$, and 
$\rho_{(1)(2)(3)(4)}$, for $n=2$, $3$, and $4$ respectively. 
We have just argued that these densities
differ from scaling operators only by subleading terms in the same partition ``sector''.
We have  argued that a more general density 
$\rho_\pi$ is a sum of scaling operators 
in multiple sectors, 
corresponding to partitions at least as fine as $\pi$. 
As a result, $\<\rho_\pi(t)\>$ has a leading decay  given by $t^{-\alpha_k}$ with ${k=|\pi|}$, but has subleading corrections from other sectors that start at order $t^{-\alpha_{k+1}}$.
We now show that this general picture is consistent with simulations.

In practise we simulate four replicas, 
collecting the damage densities $\rho_\pi$ for all  ${\pi\in\Pi_4}$, and 
reconstructing the densitites for fewer replicas as suitable linear combinations of these ${n=4}$ densities.
This is the  ``forgetting a replica'' relation in Eq.~\ref{eq:forgettingrho}.

The permutation symmetry of the replicas\footnote{Which is respected by our  initial condition, after averaging}
implies ${\< \rho_{(1)(234)}\> = \< \rho_{(2)(134)}\>=\cdots}$, and we write 
$\rho_{(a)(bcd)}$ to denote the average value for any of these equivalent partitions (and similarly for other partition shapes). 
Numerically, we average over the equivalent partitions, via
\begin{align}
\rho_{(a)(bcd)}& =\frac{1}{4}  \left\langle\rho_{(1)(234)}+\rho_{(2)(134)}+ \text{2 terms}\right\rangle\,,
\\
\rho_{(ab)(cd)}& = \frac{1}{3}\Braket{\rho_{(12)(34)}+\rho_{(13)(24)}+\rho_{(14)(23)}}\,,
\\
\rho_{(a)(b)(cd)}& = \frac{1}{6}\left\langle\rho_{(1)(2)(34)}+\text{5 terms} \right\rangle\,,
\\
\rho_{(a)(b)(c)(d)}& = \braket{\rho_{(1)(2)(3)(4)}}\,.
\label{eq:rho-a-b-c-d}
\end{align}
The three-replica damages can be expressed as 
\eq{
\rho_{(a)(bc)}&=\rho_{(a)(bcd)}+\rho_{(ab)(cd)}+\rho_{(a)(b)(cd)}\,,\\
\rho_{(a)(b)(c)}&=3\rho_{(a)(b)(cd)}+ \rho_{(a)(b)(c)(d)}\,,\label{eq:rho-a-b-c}
}
and the two-replica damage as 
\eq{
\begin{split}
\rho_{(a)(b)} =2&\rho_{(a)(bcd)}+2\rho_{(ab)(cd)}\\ +&5\rho_{(a)(b)(cd)}+\rho_{(a)(b)(c)(d)}\,.\label{eq:rho-a-b}
\end{split}
}

\begin{figure}[t]
\includegraphics[width=\linewidth]{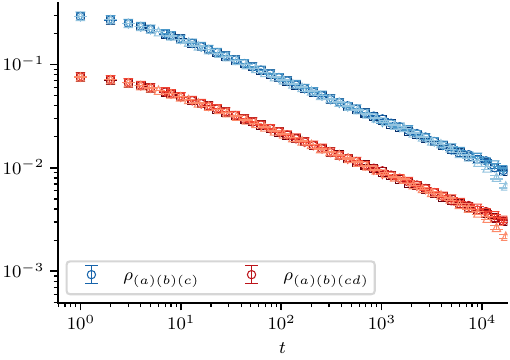}
\caption{Decay of damage densities for three-block partitions:
the operator $\rho_{(1)(2)(3)}$ of the $n=3$ theory, and the operators $\rho_{(a)(b)(cd)}$ of the $n=4$ theory. Both have leading exponent $\alpha_3$, but the latter has a subleading corrections with exponent $\alpha_4$ that is absent in the former (see Eqs.~\ref{eq:Ofiner}-\ref{eq:subleadingopdecay4}).
}
\label{fig:all-three-blocks}
\end{figure}

Fig.~\ref{fig:all-two-blocks} shows 
all the densities for
\textit{two-block} partitions in $\Pi_2$, $\Pi_3$ or $\Pi_4$. All these have the same asymptotic decay exponent ${\alpha_2= \Delta_2/z}$.
However we expect a subleading correction 
proportional to $t^{-\alpha_3}$
in all except $\rho_{(a)(b)}$ (Eqs.~\ref{eq:n32blockdensity}--\ref{eq:subleadingopdecay2}).
Indeed, $\rho_{(a)(b)}$ presents a cleaner power law than the others.

Analogously, Fig.~\ref{fig:all-three-blocks} shows the three-block density ${\rho_{(a)(b)(c)}\simeq C_3 t^{\alpha_3}}$ of the ${n=3}$ theory
(Eq.~\ref{eq:rgsec123decay})
and the three-block density ${\rho_{(a)(b)(cd)}= D'' t^{-\alpha_3} + O(t^{-\alpha_4})}$ of the ${n=4}$ theory (Eq.~\ref{eq:subleadingopdecay4}).

The data is consistent with the two relations between prefactors mentioned below Eq.~\ref{eq:subleadingopdecay4}, which come from the   forgetting relation in Eq.~\ref{eq:forgettingrho}.

\subsection{More general operators}
\label{sec:moregeneraloperators}

So far we have considered the operators spanned by the damage densities $\rho_{\pi}$:
let us  comment briefly on some more general operators, starting with  the operators that \textit{create} damage and whose scaling dimensions determine the decay of the survival probabilities $\mathcal{S}_n$ (Sec.~\ref{sec:trs}). 

In a transfer matrix treatment of the sum over trajectories\footnote{I.e. if we think of the    probability distribution $P\lf \sigma(1), \sigma(2),\ldots, \sigma(L) \ri$ as a 
as a ket 
$\ket{P}= \sum_{\sigma(1),\ldots,\sigma(L)} 
P\lf \sigma(1), \sigma(2),\ldots, \sigma(L) \ri
\ket{\sigma(1),\ldots,\sigma(L)}
$ that evolves in time by the action of the transfer matrix/Markov transition matrix.} (see e.g. \cite{grassberger1979reggeon}) 
the densities $\rho_\tau$ are diagonal observables
of the schematic form $\ket{\tau}\bra{\tau}$.
We may also define off-diagonal operators. For example, inserting the (single-site) operator
$\ket{\tau}\bra{\mathbbm{1}}$ at position $x$ and at timeslice $t$ modifies the sum over trajectories by creating a damaged site of type $\tau$ at the spacetime point $(x,t)$.
Let us denote this lattice operator $\widetilde\rho_\tau$. Its role in correlation functions is similar to that of the response field  $\widetilde \rho_\tau$ in the Martin-Siggia-Rose-Janssen-de Dominicis approach.

To properly treat such operators we would have to refine our crude real-space RG picture above.
However, one may argue more heuristically that a
the physically sensible
RG transformation of such operators should have the schematic form (see Ref.~\cite{dai2020quantum} for other problems with an analogous structure)
\be\label{eq:creationopRG}
\widetilde\rho_\tau ({\mathbf x}) \overset{\text{RG step}} \longrightarrow  \sum_{\tau' \geq \tau}
\widetilde\rho_{\tau^\prime}^{\,'}({\mathbf X}) B_{\tau',\tau}.
\ee
The constraint ${\tau'\geq \tau}$ imposed by the poset structure here is \textit{opposite} to that for the density operators $\rho_\tau$.
To see this,
consider applying $\widetilde\rho_\tau$ at the origin in a damage-free state. This creates damage of type $\tau$. 
However, one possibility is that this damage survives only for an O(1) time, leaving behind either the undamaged state, or more generally a state with a coarser damage type ${\tau'>\tau}$.
So a sensible coarse-graining scheme should allow $\widetilde\rho_\tau$ to ``descend'' to $\rho_{\tau'}$ for ${\tau'\geq \tau}$ under temporal coarse-graining.
As a result of Eq.~\ref{eq:creationopRG}, we then expect scaling operators of the schematic form
\be\label{eq:formofscalingcreationoperators}
\widetilde{\mathcal{O}}_\pi =  \widetilde\rho_\pi
+ \sum_{\pi'> \pi} w^{(\pi)}_{\pi'} \widetilde{\rho}_{\pi'},
\ee
with scaling dimensions $\widetilde\Delta_\pi$.
The structure of these operators could be further constrained by generalizing the arguments in the previous Subsection. We do not enter into this here. 
However note that the ability to ``merge'' replicas that are in the same state implies that the scaling dimensions $\widetilde{\Delta}_\pi=\widetilde{\Delta}_{|\pi|}$
will depend only on the number of blocks in $\pi$.

In the next section we  discuss time-reversal symmetries that exist at least for ${n=2}$ and ${n=3}$. These symmetries impose further constraints on the RG structure, enforcing $\Delta_2=\widetilde{\Delta}_2$ and $\Delta_3=\widetilde{\Delta}_3$.

In the remainder of this Section we discuss some finer points.

When we treated the damage densities, we made the simplification of considering only one density operator $\rho_\pi$ in each damage sector $\pi\in \Pi_n$. 
For completeness, we  note  that in reality multiple operators exist in a given sector
(even if they may be less relevant in the RG sense).
For simplicity we restrict to diagonal operators in the language above, corresponding in the  continuum field theory to operators built from the fields $\rho$ and not from the fields $\widetilde \rho$.

We say that such an operator is ``in the sector $\pi$'' if, like $\rho_\pi$,
it (a) is only nonzero when the damage state of the system is equal to $\pi$ or finer, (b) can be nonzero when the state of the system is $\pi$.

As in any field theory, there are certainly subleading operators with the  same ``quantum numbers'' as any given operator. For example, $\nabla^2 \rho_\pi$ has the same symmetries and is in the same sector as $\rho_\pi$, but has a scaling dimension larger by 2.
There are also damage operators with different spatial symmetry properties, such as  $\nabla \rho_\pi$, which transforms under spatial reflections/rotations.

Slightly less trivially, in some cases we can define operators that are in the same sector as $\rho_\pi$ but which transform differently under global replica symmetry.
In App.~\ref{app:nonsymmetricoperators} we illustrate this for the sector ${(1)(2)(3)}$, where
in addition to the ``symmetric'' operator $\rho_{(1)(2)(3)}$ we can define operators in a two-dimensional irreducible representation of $S_3$ symmetry. We suggest that, below the upper critical dimension, these operators are less relevant than the symmetric operator $\rho_{(1)(2)(3)}$ in the same sector (App.~\ref{app:nonsymmetricoperators}).

\section{Time-reversal symmetries}
\label{sec:trs}

The directed percolation universality class has a  ``duality'' or ``time-reversal'' or ``rapidity reversal'' symmetry \cite{grassberger1979reggeon,hinrichsen2000non} 
that  is apparent both in  the Martin-Siggia-Rose-Janssen-de Dominicis path integral and on the lattice. 
Surprisingly, we find that the $n=3$ theory also has a   time-reversal symmetry. This imposes  nontrivial relations between scaling dimensions.
We will show that such a symmetry is present both in the continuum and on the lattice.

We first review the time-reversal symmetry of DP. Recall the MSR Lagrangian \cite{hinrichsen2000non}. After a rescaling of the fields we have
\be
\mathcal{L}_2  = 
\widetilde\rho
\lf \partial_t - D \nabla^2 - r \ri \rho
+ 
\lambda \lf
\widetilde \rho \rho^2
- \widetilde \rho^2 \rho
\ri,
\ee
with ${\lambda = \sqrt{g\nu/2}}$.
Integrating by parts shows that this Lagrangian is formally invariant under \cite{hinrichsen2000non}
\ba
\rho & \longleftrightarrow - \widetilde \rho,
& 
t & \longrightarrow - t.
\end{align}
This symmetry ensures that the physical field $\rho$ and the response field $\widetilde\rho$ have the same scaling dimension.\footnote{To be more precise,
this formal symmetry establishes the equality of scaling dimensions to all orders in perturbation theory. As we have formulated the field theory, $\rho$ and $\widetilde\rho$ have different integration contours, and the path integral is also divergent! 
These issues do not affect perturbation theory,
but a nonperturbative result would require a more careful formulation of the path integral. 
Fortunately, the validity of the symmetry relations beyond perturbation theory is guaranteed in the examples we will discuss by lattice versions of the time-reversal symmetry.}
As a result, the exponent $\alpha_2$,
governing the decay $\rho(t)\sim t^{-\alpha_2}$ 
of the density starting from 
a fully active state,
is equal to the exponent governing the probability
$P(t)\sim t^{-\delta_2}$ that activity survives up to time $t$ if we start with a single active site \cite{hinrichsen2000non}.

Now consider ${n=3}$.
In the basis of two-block partitions (and 
again after rescaling fields) the Lagrangian is  (\ref{eq:ngenerallagrangian})
\ba 
\mathcal{L}_3 = & 
\widetilde\rho_\pi
\lf \partial_t - D \nabla^2 - r \ri \rho_\pi
+ 
\lambda \left( 
{\hat{\mathcal{K}}}^{\pi}_{\sigma\sigma'}
\widetilde \rho_\pi \rho_\sigma \rho_{\sigma'}
- \widetilde \rho_\pi^2 \rho_\pi
\right),\label{eq:L3partitionbasis}
\end{align}
where ${\pi, \sigma,\sigma'}$ are summed over the three partitions 
\ba
\sigma_1 &= (1)(23),& 
\sigma_2&=(2)(13), &
\sigma_3&=(3)(12),
\end{align}
with  the tensor values (and their cyclic permutations)
\ba\notag
\hat{\mathcal{K}}^{\sigma_1}_{\sigma_1,\sigma_1} & = 1,
& 
\hat{\mathcal{K}}^{\sigma_1}_{\sigma_1,\sigma_2} & = \f{1}{2},
& 
\hat{\mathcal{K}}^{\sigma_1}_{\sigma_2,\sigma_2} & =0,
&
\hat{\mathcal{K}}^{\sigma_1}_{\sigma_2,\sigma_3} & =-\f{1}{2}.
\end{align}
For simplicity we will write $\rho_a\equiv \rho_{\sigma_a}$. 

Recall also that for ${n=3}$ there 
is another natural choice of basis for the density fields, which is given by the  two-replica differences 
${\rho_{ab} = \rho_a + \rho_b}$ for ${a<b}$.
This basis yields the Lagrangian in Eq.~\ref{eq:n3lagrangian} (with ${g\to \lambda}$, ${\nu/2\to \lambda}$).
In full, the change of basis which relates these two formulations of the field theory is
${\rho_{ab} = \rho_a + \rho_b}$ together with 
${\widetilde\rho_{12} = \f{1}{2} \lf  \widetilde \rho_{1}+ \widetilde \rho_2 - \widetilde \rho_3  \ri}$ and cyclically.

The nontrivial tensor  $\hat{\mathcal{K}}$ in the cubic term means that the Lagrangian $\mathcal{L}_3$ in Eq.~\ref{eq:L3partitionbasis} is not invariant under  
a ``diagonal'' rapidity reversal operation
${(\rho_\pi \leftrightarrow -\widetilde \rho_\pi,\, t\rightarrow - t)}$
that acts separately on fields with   different indices $\pi$.
Remarkably,  though,  
$\mathcal{L}_3$ \textit{is} invariant 
under a 
non-diagonal time-reversal symmetry
 under which  $t\to - t$, and
\ba\label{eq:trsn3transformations}
\widetilde \rho_1 & \longrightarrow - \lf 
\rho_2 + \rho_3
\ri ,
& 
\rho_1 & \rightarrow - \f{1}{2} \lf 
 \widetilde\rho_2 +\widetilde\rho_3  - \widetilde\rho_1
\ri, 
\end{align}
etc. Interestingly, this is the combination of the diagonal rapidity reversal transformation  with the basis change described in the previous paragraph.
Indeed, switching between the two bases moves the nontrivial tensor structure from one of the cubic terms to the other.

This time-reversal symmetry 
ensures that $\rho_\pi$ and $\widetilde \rho_\pi$ have the same scaling dimension, at least within perturbation theory around the upper critical dimension.
In fact, as we discuss next, it is also possible to find a lattice analog of this time-reversal symmetry, showing that the  nontrivial constraints on correlators and exponents hold exactly in all dimensions.

One ``microscopic'' way to understand the time-reversal symmetry of directed percolation is via the Markov transition matrix \cite{grassberger1979reggeon}.
For the standard directed percolation model reviewed in Sec.~\ref{sec:tworeplicareview},
the transition matrix $\mathcal{T}$, which advances the probability distribution over activity configurations by two timesteps,
satisfies $\mathcal{T}=\mathcal{S}.\mathcal{T}^T.\mathcal{S}^{-1}$,
where $\mathcal{S}$ represents a local change of basis. 
Since $\mathcal{T}$ is transposed on the right-hand side, this 
symmetry exchanges initial and final states (also modifying them via the action of $\mathcal{S}$). It yields nontrivial relations between transition probabilities.

In App.~\ref{app:latticetrs} we show that the effective Markov process for the 1+1D model with ${n=3}$
(defined in Sec.~\ref{sec:effectivedynamicsn3})
also has a time-reversal symmetry
\be\label{eq:latticetimereversalmaintext}
S.\mathcal{T}^T.
S^{-1}
=\mathcal{T},
\ee
for an appropriate local basis change $\mathcal{S}$.

We may argue that the time reversal symmetry imposes the relation $\Delta_3 = \widetilde \Delta_3$ for the scaling dimensions, and $\alpha_3 = \delta_3$ for the decay exponents (details in App.~\ref{app:latticetrs}).

It remains to be checked whether there are analogous time-reversal symmetries for larger values of $n$.

\section{Conclusions and future directions}
\label{sec:conclusions}

This paper has presented a theory for the damage spreading phase transition in (spatiotemporally random) classical cellular automata. The transition is described by a critical theory with an infinite hierarchy of sectors; for any $n$, truncating a the sector with at most $n$-replica observables gives a well-defined sub-theory. 
A continuum description is provided by stochastic partial differential equations for  damage densities $\rho_\pi$ (for the two-block partitions $\pi$) or by the equivalent Lagrangian. Note that, although we use replicas, the results make sense without the need for a replica limit.

The RG structure of the damage-spreading critical point is also interesting. The 
 fact that the underlying replicas evolve independently, with identical deterministic dynamics, leads to an unusual classification of operators related to the partial order on the lattice of partitions.
A side-product of the analysis is a hierarchy of reaction-diffusion-like models with absorbing state transitions, with the same nontrivial RG structure. 

One of the applications of the results in this paper is to the  question of how  irreversible dynamics concentrates an initially uniform measure on states. 
If the dynamics was reversible this measure would be trivially preserved, but here the dynamics is contracting in the discrete ``phase space'' (see comments  under the ``Entropies and probabilities'' heading below).

We now discuss some broader lessons and directions for the future. 
We begin with topics closest to the current setting (for example, concrete calculations using the continuum descriptions above) and end with applications of the present results to superficially quite different models. 

{\bf Mean field.} One task is a fuller  analysis of the mean field equations for all $n$, including for example analysis of survival probabilities for local initial damage (using the mean field equations at large finite $N$, which include the stochastic terms).

{\bf Critical scaling (simulations).} On the numerical front, it would be interesting to push the ${1+1D}$ simulations to larger $n$. For example, how do the critical exponents $\alpha_n$ and $\delta_n$ scale with $n$ at large $n$? (They grow  logarithmically with $n$ within mean field theory.)
It would also be interesting to  investigate higher dimensions $d$, where it may well be  possible to match with  field-theory results (see below).\footnote{One way  to accelerate simulations at the critical point may be to only update spins $s_i^a$ in spatial regions where damage is present. This takes advantage of the  statistical symmetry of the update rules, which mean that  updates in ``passive'' regions can be omitted without changing the statistics of the damage.} 

{\bf Critical scaling ($\epsilon$ expansion).}  Exponents could be computed in the $\epsilon$ expansion around four spatial dimensions using the field theories in Sec.~\ref{sec:fieldtheories}. It may be an interesting combinatorial problem to compute the exponents  $\alpha_n$ and $\delta_n$ for general $n$.  In addition, whereas for ${n\leq 3}$ the RG flows of the field theory are known from directed percolation, they remain to be computed for general $n$, as this requires  one additional coupling ($\theta$ in Eq.~\ref{eq:Kdefgeneral3}) to be taken into account. 
It is encouraging that, for directed percolation, the epsilon expansion seems to work well above 1+1D.\footnote{At two-loop order, errors in the standard exponents are on the scale of a percent in 3+1D, and of a few percent in 2+1D \cite{odor2004universality}.} 

The $\epsilon$ expansion is also a setting where the operator classification can be studied in a concrete way, applying ideas from Sec.~\ref{sec:rgstructure}. 
This includes not only damage densities but also  off-diagonal operators.

Sec.~\ref{sec:choiceof1Dmodel} noted that, in addition to the generic damage-spreading problem, it is also possible to define a fine-tuned version, by restricting the allowed update functions. The exponents of this problem could also be studied numerically, or with the field theory.

{\bf Entropies and probabilities.} One application of the multi-replica observables in this paper is to characterize the dynamical systems  information-theoretically --- 
for example, to understand how irreversibility 
leads to loss of entropy if the initial ``state'' is indefinite --- and in terms of the heterogeneity between circuits.
For a given a realization $C$ of the classical circuit, 
an initially unifom measure on bit strings will be generally be transformed by the dynamics into a more concentrated measure, with a smaller entropy, at later times $t$.

For a simple example of the relation between partitions and probabilities, consider just a single bit $s_i$ at site $i$. 
The measure at time $t$ dictates probabilities ${\{p^C, 1-p^C\}}$ for the two states of the bit at time $t$.  (The probability $p^C$ is associated with the average over initial states, for a fixed realization $C$ of the random circuit.)
A more symmetric parameterization is ${\widetilde p^C = p^C(1-p^C)}$. It is easy  to show (App.~\ref{app:partitionsandprobabilities}) that 
\ba
\mathbb{E}\, \widetilde p^C & = \f{1}{2} \< \rho_{(1)(2)}\>,
& 
\mathbb{E}\,  (\widetilde p^C)^2 & = \f{1}{2} \< \rho_{(12)(34)}\>
\end{align}
(where $\mathbb{E}$ is the average over circuits).
At criticality, both of these decay with the same exponent $\alpha_2$. This allows us to separate the average over initial states from the average over circuits --- it shows that
  even after averaging over pairs of initial states, different circuits yield very different averaged damage profiles (App.~\ref{app:partitionsandprobabilities}).

We have emphasized that the critical theories in this paper, for example the hierarchy of Lagrangians $\mathcal{L}_n$,
make sense for any fixed value of $n$: the computation of damage densities does not require a replica limit.  
However some quantities, such as entropies, can be expressed using replica limits of damage observables~(App.~\ref{app:partitionsandprobabilities}). 

Inside the damage-spreading phase, the decay of the entire systems's entropy, starting from an initially uniform distribution, 
 takes a time exponentially large in the system volume.
(This is reminiscent of entropy decay 
in irreversible quantum dynamics \cite{gullans2020dynamical,fidkowski2021dynamical,li2021statistical,nahum2021measurement,bulchandani2024random,de2025universality,gerbino2024dyson,gerbino2026universal} and connects to results on random maps \cite{derrida1987random}.) We will discuss the entropy decay elsewhere.

{\bf Lessons about RG structure.} A basic lesson of this study is that operators (observables) in the theory of damage spreading are labelled by discrete data which go beyond the   ``quantum numbers'' associated with the global symmetry of the theory.
This classification is reminiscent of the  ``topological'' classification of scaling operators found in Ref.~\cite{dai2020quantum} for a quantum Hamiltonian and an associated stochastic process. 
What is the best way to think about  and perhaps  unify such models?\footnote{An algebraic perspective might be useful. The absorbing state properties of directed percolation and of the models here give rise to noninvertible operators
that commute with the Markov transition matrix. (In the directed percolation case this is just $\otimes_j \mathcal{R}$, where $\mathcal{R}$ resets a site to empty. For $n>2$ we can write multiple ``resetting'' operators.)
This is reminiscent of ``noninvertible symmetries'' \cite{shao2023s,schafer2024ictp} but in the present case  the operators are not topological line operators.}

{\bf Population dynamics.} The effective Markovian lattice models  for ${n>2}$, where the degrees of freedom are partitions,  may be reinterpreted as population dynamics models with multiple species 
\cite{janssen1997spontaneous,janssen1999coupled,tauber1998multicritical,janssen2001directed,cardy2023cut,bartels2024emergent} and may be interesting in that context.
They are simplest in the case $q=2$ when the  nontrivial local states are labelled by 2-block partitions. In Sec.~\ref{sec:effectivedynamicsn3} we have given rules for the $n=3$ case, where a given site can be  empty, or occupied by one of three colors of particles (for $n>3$ there are inequivalent particle types). 
The rates for these processes are highly constrained by the underlying replica structure, 
though when the models are taken as Markov processes in their own right we are free to break these constraints. It would be interesting to explore the models' place in broader phase diagrams.

An interesting feature of the processes is that they provide nontrivial probabilistic ``couplings'' \cite{den2012probability} for several copies of  directed percolation. For example, as illustrated in Fig.~\ref{fig:traj}, a trajectory of the $n=3$ model determines three separate (correlated) directed percolation trajectories, each obtained by deleting all particles of one color and neglecting the colors of the other particles.

{\bf Applications of results to more general models.} The models considered here were classical circuits with  randomness that is ``quenched'' in the sense that it is held fixed when we consider the dynamics, but which is random both in space and time.

We can ask to what extent the results are applicable to models that are ``less random'' in various ways 
(either in finite $d$ or in the all-to-all setting).
These could be models that  are spatiotemporally inhomogeneous,   but with a  structure that is complex rather than random;
or  models that are random only in space but translation-invariant in time; or models whose update rules are translationally invariant both in space and time.

For reasons discussed in Ref.~\cite{derrida1986random}, mean-field results for random-in-time models can have applications  to time-translation-invariant Kauffman models, in a certain time regime. Similarly, for some  spatially local circuits with a large number $q$ of states at each site, or a large interaction range, there can be a long timescale before the distinction between random-in-space and random-in-spacetime circuits becomes manifest.\footnote{Depending on the dynamics, the typical timescale for revisiting a  given   local spin configuration may be very long, and over this timescale the system is effectively exploring ``new" randomness associated with distinct  inputs to the update function, even if the update function is fixed in time.} This can extend even to completely translationally invariant models, with the randomness being supplied by the initial condition  (again, up to a large timescale, and for models in an appropriate regime).

So far we have discussed cellular automata.  It is also interesting to ask about transitions  to chaotic 
\cite{pomeau1986front,sipos2011directed,hof2023directed,das2018light,liu2021butterfly,deger2022constrained,klamser2025directed} or damage-spreading phases in systems with continuous degrees of freedom. We will describe applications of the present ideas to continuous-variable systems elsewhere.

\acknowledgments 
We are grateful to Austen Lamacraft for discussions that were important in motivating this project. We also thank Bernard Derrida, Guy Bunin, Denis Bernard and Guilhem Semerjian for useful discussions. AN is supported by the European Union  (ERC, STAQQ, 101171399). Views and opinions expressed are however those of the authors only and do not necessarily reflect those of the European Union or the European Research Council Executive Agency. Neither the European Union nor the granting authority can be held responsible for them.
SR acknowledges support from the Department of Atomic Energy, Government of India, under Project Nos. RTI4019 and RTI4013, from SERB-DST, Government of India, under Grant No. SRG/2023/000858, and from a Max Planck Partner Group grant between ICTS-TIFR, Bengaluru and MPIPKS, Dresden.

\appendix

\section{Fine-tuning in the model with $p_1=0$}
\label{app:finetuning}

In Sec.~\ref{sec:choiceof1Dmodel} we discussed a two-parameter space of models, with update functions of three types.  We stated that the phase diagram boundary with $p_1=0$ should be regarded as fine-tuned. 
On this boundary, ``type-1'' update functions (which take the same value for 3 choices of the argument and a distinct value for the 4th choice of the argument) are forbidden.

To see that this phase diagram boundary is fine-tuned, consider a system of four replicas, and restrict to the case where the initial state satisfies
\be\label{eq:app4replicaconstraint}
s^1_i + s^2_i + s^3_i + s^4_i = 0\qquad (\operatorname{mod} 2)
\ee
for every physical site $i$.
By checking cases one finds that this property is preserved if type-1 updates are absent, but not in the more general case where type-1 updates are allowed.

In other words, the states in the 4-replica problem that obey (\ref{eq:app4replicaconstraint}) form an invariant subspace for circuits in the $p_1=0$ ensembles, but not in more general circuits.  This is a form of fine-tuning. 

This fine-tuning also affects correlation functions for damage operators. 
For example, consider an initial state for four replicas in which the only nontrivial damage is of types
$(12)(34)$, $(13)(24)$, and $(14)(23)$.
Such a state lies in the subspace defined above. 
As a result, damage of type $(123)(4)$ (for example) can never be produced. 
In field theory language, this 
can be translated into a statement about the vanishing of correlation functions which involve the operator $\rho_{(123)(4)}$ together with operators that insert damage of the $(ab)(cd)$ types. 
The fact that these correlation functions vanish in the fine-tuned models, but are nonzero in more general models, strongly suggests that the endpoint of the critical line with $p_1=0$ is a multicritical point with distinct universal properties (when $n>3$ replicas are considered).

Relatedly, within the general family of 1+1D circuits considered, one can see (using the universality of NAND gates, which are type-1 gates) that it is possible to construct a circuit that implements any Boolean function. When type-1 gates are excluded, this is no longer the case. 
We expect that this is the more fundamental reason why the $p_1=0$ boundary of the phase diagram is fine tuned. 

Our simulations in the main text are 
made at the point on the critical line with maximal $p_1$. 
This in turn imposes ${p_0=0}$.
However, this does not represent an important constraint on the set of allowed circuits. For example,  even if type-0 updates are not allowed microscopically, the equivalent effect can be achieved by combining  layers of type-2 and type-1 updates. This is unlike the case where type-1 updates are forbidden microscopically, where  constraint is imposed which is still present even after coarse-graining.

\section{Appendix: Mean-field theory for $n$ replicas}
\label{app:meanfieldequations}

We consider the evolution of the multi-replica damage variables $\rho_\pi$ (Sec.~\ref{sec:introducengt2damage}) for  the mean field model described in Sec.~\ref{eq:MFupdate}. (More precisely, this is a family of models, distinguished by the choice of the probability distribution $P(F)$ from which  the update functions are drawn.)

Let us review the construction of the all-to-all-coupled circuit.
Let us allow the number of states $q$ at each site to be general, so $s_i = 1, \ldots, q$, for sites $i=1,\ldots,N$.
The dynamics is as follows:

 $\bullet$ In each infinitesimal time interval $\dd t$, an update occurs with probability $N\dd t$.

 $\bullet$ If an update occurs, 
the site $k$ to be updated is chosen uniformly at random. 
Two ``input'' sites $i$ and $j$ are also chosen  indendently at random.\footnote{This allows for the possibility that $i=j$. It is convenient to allow this because it avoids factors like $(1-1/N)^{-1}$ in the probabilities for the state of the inputs.}

 $\bullet$ The previous state $s_k(t)$ of the site $k$ is overwritten with the output state $s_k(t + \dd t)$, which is given by a random function of the inputs:
\be
 s_\text{out}  = F_t(s_\text{in}, s_\text{in}'),
\ee
where $s_\text{out}=s_k(t+\dd t)$, $s_\text{in}=s_i(t)$, $s_\text{in}'=s_j(t)$, 
where $F_t$ is a function chosen at random from a probability distribution $P(F)$. If updates occur at times $t_1, \ldots, t_T$, then there are corresponding independently drawn update functions $F_{t_\alpha}$ for ${\alpha = 1,\ldots, T}$.

$\bullet$ A given instance of the random circuit is defined by the set of update times $\{t_1, \ldots, t_T\}$
(where the number $T$ of updates can vary between instances),
by the spins which are involved in each of the updates, and by the choice of functions $F_{t_\alpha}$. When we consider multiple replicas, all replicas are evolved with the same circuit instance.)

 $\bullet$ Some  of the formulas below will be given for more general choices of $P(F)$, but for our principle example, we define $P(F)$ by drawing $F$ as follows:

 --- With probability $c$, we pick a uniformly random value in $\{1,\ldots, q\}$, and   choose the $F$ whose output is equal to this value regardless of its inputs.

 --- With probability $\bar c = 1-c$, we choose a uniformly random $F$: for each of the $q^2$ possible inputs $(s_\text{in}, s_\text{in}')$, we assign a uniformly random output $s_\text{out}$.

This choice of distribution for the functions is statistically invariant under $S_{q^2}$ permutations of the $q^2$ input states and under $S_q$ permutations of the output states.
Many other choices could be made that also respect this convenient invariance property.
We expect that (barring fine-tuning: see App.~\ref{app:finetuning}) the universal properties at the critical point will be independent of the precise distribution.
However, as we will see below, the \text{bare} coupling constant values in the field theory depend nontrivially on the choice once we consider $n>3$ replicas.

Now we consider the effect of the updates on the damage variables.
We are free to consider any number $n$ of replicas. 
In principle we have a separate set of mean field equations for each $n$, 
but in fact the equations for a given $n$ determine those for all smaller $n$ in a trivial way.\footnote{An alternative to thinking of a separate set of equations for each $n$ is to set $n=\infty$, but to restrict to partitions with a finite number of blocks.}

At a given time $t$, each site $i=1, \ldots, N$ can be assigned to a damage type (partition) $\pi_i\in\Pi_n$ (which may be the trivial partition $\pi_i= \mathbbm{1} $).
The number of sites of type $\pi$ is $N\times \rho_\pi(t)$.

Since all replicas of the system are evolved with the same circuit, the dynamics defined above also defines the dynamics of the $n$-replica system, starting from some initial state $\{s_i^a\}$.
 Because of the statistical invariance of the update rules under permutations of the local spin states $\{1,\ldots, q\}$ and --- in this mean field model --- also under permutations of the lattice sites,
 this leads to an autonomous Markovian dynamics of the variables $\{\rho_\pi\}$. In other words, it is not necessary to keep track of the specific spin state: it is sufficient to keep track only of the fraction of sites with a given type of damage.

 The change to $\rho_\pi$ in $\dd t$ is
 \ba\label{eq:appdeltarhopi}
\dd \rho_\pi = 
\left\{
\begin{array}{ll}
0 &   \,\, \text{with probability  $1-N \dd t$} \\
  \Delta_\pi/N &   \,\, \text{with probability  $N \dd t$} 
\end{array}
\right.,
\end{align}
 where the first line accounts for the probability that no update takes place in $\dd t$, and  
   $\Delta_\pi$ is the change to the number of sites of type $\pi$ in an update, given that one occurs.
 ($\Delta_\pi$ should not be confused with the scaling dimensions discussed in Sec.~\ref{sec:rgstructure}!)
   $\Delta_\pi$ may be written as the difference of two binary quantities $C_\pi=0,1$
    and $A_\pi =0,1$ ($C$ for Creation and $A$ for Annihilation)
 \be
 \Delta_\pi = C_\pi - A_\pi,
 \ee 
 where $C_\pi$ is 1 if the  site $k$ that is updated (see above) is of type $\pi$ after the update (i.e. if a state of type $\pi$ is ``cretated'' at site $k$),
and $A_\pi$ is 1 if site $k$ was of type  $\pi$ before the update (i.e. if a state of type $\pi$ is ``annihilated'' at site $k$). 

Since both $A_\pi$ and $C_\pi$ are either 1 or 0, and since each of them is nonzero only for a single value of $\pi$ (because only one site's state is updated), we have 
      \ba
 \< C_\pi\> & = p_\text{out}(\pi),
 &  
    \<C_\pi C_\sigma\> & = \delta_{\pi \sigma} \<C_\pi\>,
   \\
 \< A_\pi\> & = \rho_\pi, &
            \<A_\pi A_\sigma\>  & = \delta_{\pi \sigma} \<A_\pi\>.
   \end{align}
Also, once we condition on the state instantaneously before the update, $A$ and $C$ are independent, so
  \be
  \< C_\pi A_\sigma\> = \< C_\pi \> \<A_\sigma\> = p_\text{out}(\pi) \rho_\sigma.
  \ee
Using these identities,
 \ba\label{eq:appdeltapiaverages}
\< \Delta_\pi \> 
& = p_\text{out}(\pi) - \rho_\pi,
\\  \notag
\langle  \Delta_\pi \Delta_{\sigma} \rangle 
& = 
\left( p_\text{out}(\pi)  + \rho_\pi \right)  \delta_{\pi \sigma}
- p_\text{out}(\pi)  \rho_\sigma
- p_\text{out}(\sigma)  \rho_\pi.
\end{align}
 Here $p_\text{out}(\pi)$ is the probability that, using two random sites as input, you get $\pi$ as output. 
 It is (implicitly) a function of the instantaneous densities $\{\rho_\sigma \}$: this dependence has been suppressed to avoid clutter. 
The form of $p_\text{out}(\pi)$, as a function of the densities, is determined by the choice of the probability distribution $P(F)$ for the update functions. This is discussed below.\footnote{We always assume (in order to have an exact reduction to a Markov process for the $\rho_\pi$ variables) that $P(F)$ is invariant under permutations of the site labels $\{1,\ldots,Q\}$. 
Usually, for convenience, we will impose the stronger property of being  invariant under $S_q\times S_{q^2}$, where the $S_q$ factor represents permutations of the output and the $S_{q^2}$ represents permutations acting on the  $q^2$ possible input states for the two input variables.}

From Eqs.~\ref{eq:appdeltarhopi},~\ref{eq:appdeltapiaverages} we have
\ba
 \< \dd \rho_\pi \>  & = \lf p_\text{out}(\pi) - \rho_\pi  \ri \dd t,
\\
 \< \dd \rho_\pi  \dd \rho_\sigma\>  & =
\f{\dd t}{N}
 \big[ 
 \left( p_\text{out}(\pi)    + \rho_\pi \right)  \delta_{\pi \sigma}\notag
 \\
& \qquad \qquad \qquad - p_\text{out}(\pi)  \rho_\sigma
- p_\text{out}(\sigma)  \rho_\pi
\big].
\end{align}
These equations are exact even for finite $N$.  When $N$ is large we can neglect higher cumulants of the fluctuations and we obtain an It\^o stochastic differential equation for the density,
 \ba\label{eq:appSDEfordensitiesgeneric}
 \dd \rho_\pi & =  \lf p_\text{out}(\pi) - \rho_\pi  \ri \dd t + \dd B_\pi,
 \end{align}
with a nontrivial covariance for the noise,
\ba  \notag
 & \< \dd B_\pi \dd B_\sigma\>  = 
 \\
 & \,\,\,\,\,\, \f{\dd t}{N}  \left[ 
\left( p_\text{out}(\pi)  + \rho_\pi \right)  \delta_{\pi \sigma}
 - p_\text{out}(\pi)  \rho_\sigma
- p_\text{out}(\sigma)  \rho_\pi
\right].\label{eq:appnoisecovariancegeneral}
 \end{align}
It is sufficient to retain the equations for the nontrivial densities ($\pi\neq \mathbbm{1}$, or equivalently $\pi<\mathbbm{1}$), since ${\rho_{\mathbbm{1}}= 1 -\sum_{\pi < \mathbbm{1}} \rho_\pi}$.
 
For the mean-field model in the thermodynamic limit ${N\to \infty}$, 
  we are left with a deterministic evolution of the densities:
 \be\label{eq:appmeanfielddeterministicgeneral}
\f{\dd \rho_\pi}{\dd t} =  p_\text{out}(\pi) - \rho_\pi .
 \ee
The stochastic equation at finite but large $N$ (Eq.~\ref{eq:appSDEfordensitiesgeneric}) will be useful for deriving a finite-dimensional description (Sec.~\ref{sec:fieldtheories}).

It remains to fix the form of $p_\text{out}(\pi)$ as a function of the densities.
Since the two sites that are used as input for a given update are sampled independently and uniformly from the population, we have
\be\label{eq:apppoutintermsofA}
p_\text{out}(\pi) = \sum_{\sigma, \sigma'}  P(\pi|\sigma,\sigma')
\rho_\sigma \rho_{\sigma'},
\ee
where the sum is over the damage states $\sigma$ and $\sigma'$ of the first and second inputs, respectively, and $P(\pi|\sigma,\sigma')$ is the probability that the output is of type $\pi$, \textit{given that} the inputs are of types $\sigma$, $\sigma'$.

The formula above makes sense for any number $n$ of replicas, with the partitions lying in $\Pi_n$. 
Rather than saying that there is a separate function $P(\bullet|\bullet,\bullet)$ for each $n$, it is notationally more convenient to view 
 $P(\bullet|\bullet,\bullet)$  as a single function 
 that allows inputs  for any $n$ (i.e. that takes arguments  in $\cup_n \Pi_n^3$).
 This is why we do not decorate $P$ with a subscript ``$n$''.\footnote{Similarly $\rho_\pi$ does not need a subscript $n$ because the value of $n$ can be read off from $\pi$.}

The sums run over all partitions, including the trivial one. Using $\rho_{\mathbbm{1}}=1-\sum_{\sigma<\mathbbm{1}}\rho_\sigma$, 
we can also write
\ba \notag
& p_\text{out}(\pi) = 
\delta_{\pi,\mathbbm{1}} + \sum_{\sigma<\mathbbm{1}} \rho_\sigma \left[
P(\pi|\sigma,1) + P(\pi|1,\sigma) - 2\delta_{\pi,\mathbbm{1}}
\right]
\\
& +  \hspace{-1.5pt} \sum_{\sigma,\sigma'<\mathbbm{1}} \rho_\sigma \rho_{\sigma'} \left[ 
P(\pi|\sigma,\sigma')  \hspace{-1.5pt} - \hspace{-1.5pt} P(\pi|\sigma,\mathbbm{1})  \hspace{-1.5pt} -  \hspace{-1.5pt} P(\pi|\mathbbm{1},\sigma')  \hspace{-1.5pt}  +  \hspace{-1.5pt} \delta_{\pi,1}
\right].\notag
\end{align}
We may simplify this further because our update procedure ensures $P(\pi|\sigma,\sigma')=P(\pi|\sigma',\sigma)$ and because we only need the equations of motion for nontrivial $\pi$.
For $\pi<\mathbbm{1}$,
\ba \notag
 p_\text{out}(\pi) & = 
2 \sum_{\sigma<\mathbbm{1}} \rho_\sigma 
 P(\pi|\sigma,1)  
 \\
  & + \hspace{-1.5pt}\sum_{\sigma,\sigma'<\mathbbm{1}}\hspace{-1.5pt} \rho_\sigma \rho_{\sigma'} \hspace{-1.5pt}\left[ 
  P(\pi|\sigma,\sigma')  \hspace{-1.5pt} - \hspace{-1.5pt} P(\pi|\sigma,\mathbbm{1})  \hspace{-1.5pt} -  \hspace{-1.5pt} P(\pi|\mathbbm{1},\sigma')  
  \right].
\end{align}
Plugging this into the equation for the deterministic mean field dynamics (\ref{eq:appmeanfielddeterministicgeneral}) gives, for nontrivial $\pi$, 
\be
\f{\dd \rho_\pi}{\dd t}
= 
\sum_{\sigma<\mathbbm{1}} \mathcal{M}_{\pi \sigma} \rho_\sigma
-
\sum_{\sigma,\sigma'<\mathbbm{1}} \mathcal{K}^\pi_{\sigma\sigma'} \rho_\sigma \rho_{\sigma'},
\ee
with
\ba\label{eq:appMdefgeneral}
\mathcal{M}_{\pi \sigma} & = 2 P(\pi|\sigma, \mathbbm{1}) - \delta_{\pi,\sigma}
\\
\label{eq:appKdefgeneral}
\mathcal{K}^\pi_{\sigma,\sigma'} & = P(\pi|\sigma,\mathbbm{1}) + P(\pi|\mathbbm{1},\sigma')  -  P(\pi|\sigma,\sigma').
\end{align}

As discussed in the main text, the quantity
\be
{p_s} \equiv P\left( (1)(2) | (1)(2), \mathbbm{1} \ri,
\ee
which is the probability for isolated damage to survive one update in the two-replica case, 
is the key quantity determining the mean-field transition point.

If $\pi$ and $\sigma$ are two-block partitions (which is the only possible type of nontrivial partition when ${q=2}$, or when ${n=2}$)
then   $P(\pi|\sigma,\mathbbm{1})$ is nonzero only for $\pi=\sigma$.\footnote{Since, for two-block partitions, $\pi\geq\sigma$ implies $\pi=\sigma$.}
In this case, the value reduces to the $n=2$ result,
\ba
P(\pi|\pi,\mathbbm{1}) & = {p_s}  
& &(\text{for } |\pi|=2),
\end{align}
because all the replicas in a given block have the same state, and behave like a single replica.
Therefore the sub-block of the matrix $\mathcal{M}$ that refers to two-block partitions has a simple form,
\ba
\mathcal{M}_{\pi \sigma} & = (2 {p_s} - 1) \delta_{\pi,\sigma}
& &(\text{for } |\pi|=|\sigma|=2).
\end{align}
This fact is relevant to the number of ``massless'' fields at the critical point, as  discussed in the main text.

In some cases we only need the noise covariance (\ref{eq:appnoisecovariancegeneral})  to linear order in the nontrivial densities.
At this order, the covariance matrix is diagonal: 
\ba \label{eq:appnoisecovariancegenerallinearorder}
 \< \dd B_\pi \dd B_\sigma\>  = 
 \f{ \delta_{\pi\sigma}  \dd t}{N}  
 \lf \rho_\pi + 2\sum_{\mu\leq \pi} P(\pi|\mu, \mathbbm{1}) \rho_\mu   \ri  + O(\rho^2\dd t).
 \end{align}
 If we restrict to two-block partitions (which are the the only possible nontrivial partitions when $q=2$) then this simplifies to 
\ba 
 \< \dd B_\pi \dd B_\sigma\> &  \simeq 
 \f{  \dd t}{N}   \delta_{\pi\sigma} \rho_\pi ( 1 + 2 {p_s}) 
 &
& (\text{keeping only 2-block}). 
 \end{align}

\subsubsection{A specific model for $P(F)$}

For concreteness, let us fix on the choice of $P(F)$ described just above Eq.~\ref{eq:Mdefspecific}.
This $P(F)$ corresponds to drawing functions $F$ as follows:
with probability $c$, we decide to pick
(at random) one of the $q$ possible completely irreversible functions, i.e. one of the functions whose output  is independent of its inputs. 
With probability $\bar c=1-c$, we instead decide to choose a function uniformly at random from all possible functions [of which there are $q^{(q^2)}$].\footnote{Formally, this is $P(F)=\f{1-c}{q^{q^2}} + \f{c}{q} \chi_\text{irr}(F)$, where $\chi_\text{irr}(F)=1$ if $F$ takes the same value on all inputs, and $\chi_\text{irr}(F)=0$ otherwise.}

To write down the kernel $P(\pi|\sigma,\sigma')$ in Eq.~\ref{eq:apppoutintermsofA}
it is convenient to use some standard notation  \cite{stanley_enumerative_1999}. 
The number of blocks in $\sigma$ is denoted $|\sigma|$.
The ``meet'' of $\sigma$ and $\sigma'$  (or their greatest lower bound) is denoted $\sigma \wedge \sigma'$: this is the coarsest partition such that $(\sigma \wedge \sigma') \leq \sigma$ and $(\sigma \wedge \sigma') \leq \sigma'$.
Each block of  $\sigma \wedge \sigma'$ is the intersection of some block from $\sigma$ with some block from $\sigma'$.

Physically, ${\sigma\wedge \sigma'}$ is the  damage type for the pair of sites, viewed as a composite system.
Each block of ${\sigma\wedge\sigma'}$ represents a set of replicas that share the same spin state (for the pair of input sites), 
while replicas in different blocks have \textit{different} spin states.
In the update, each block behaves like a single replica: since all the replicas in the block have the same input state, they have the same output state.
Because of the $S_{q^2}$ invariance of $P(F)$  under permutations of the input states, 
 $P(\pi | \sigma,\sigma')$ only depends on $\sigma\wedge \sigma'$, and not on $\sigma$ and $\sigma'$ separately.
 
First observe that for $P(\pi| \sigma, \sigma')$ to be nonzero, 
we must have $\pi \geq (\sigma\wedge \sigma')$. 
That is, the output must be at least as coarse as the (2-site) input, because replicas that agree on the input states must also agree on the output state.
Let $\mathbbm{1}(\pi \geq \sigma\wedge \sigma')$ denote the indicator function which is  1 if this condition holds and 0 otherwise.

What is the probability of getting from $\sigma\wedge \sigma'$ to $\pi$?
With probability $c$, $F$ is a fully irreversible function, implying that $\pi=\mathbbm{1}$.
On the other hand, with probability $\bar c = 1-c$, 
$F$ is chosen uniformly at random from all functions.
This means that, for each possible input of $F$, we make a random choice of the output from ${\{1,\ldots,q\}}$.
This amounts to assigning each of the $k\equiv |\sigma\wedge \sigma'|$ blocks of replicas to one of the possible output spin states  $\{1,\ldots, q\}$ 
(randomly and independently).
This can lead to merging of blocks, if the function $F$ assigns them to the same output.

We are interested in the probability that such a random assignment
leads to  a spin state of type $\pi$,  which has $l= |\pi|$ blocks.
(Each block of $\pi$ is the union of some number of the blocks of $\sigma\wedge\sigma'$.)
There are $q^k$ possible assignments in total, of which 
\be\label{eq:appnumberofcompatibleassignments}
q (q-1) (q-2)\ldots (q-l+1) =  \f{q!}{(q-l)!}
\ee
lead to $\pi$.  
(To obtain $\pi$, we have to assign distinct spin states to each of its $l$ blocks: Eq.~\ref{eq:appnumberofcompatibleassignments} is the number of ways of doing this.)
Therefore, in an update with a uniformly random function, the probability of obtaining $\pi$ from ${\sigma\wedge \sigma'}$ is 
\be
P_{\text{uniform F}}(\pi | \sigma, \sigma') = \f{q! \times \mathbbm{1}(\pi \geq \sigma\wedge \sigma')}{q^{|\sigma\wedge \sigma'|} (q-|\pi|)!}
\ee
Altogether,
\be
P(\pi | \sigma,\sigma')
= c \delta_{\pi, \mathbbm{1}} 
+ \bar c 
\f{q!}{q^{|\sigma\wedge \sigma'|} (q-|\pi|)!} \mathbbm{1}(\pi \geq \sigma\wedge \sigma'),
\ee
and by Eq.~\ref{eq:apppoutintermsofA} (using $\sum_\sigma \rho_\sigma=1$)
\be
 p_\text{out}(\pi)   =
  c \delta_{\pi , \mathbbm{1}} 
 + \bar c
\sum_{\substack{{\sigma,\sigma'}\\{\pi \geq \sigma\wedge \sigma'}}} 
 \rho_\sigma \rho_{\sigma'} 
 \f{q!}{q^{|\sigma\wedge \sigma'|} (q-|\pi|)!}.
\ee
The kernels $\mathcal{M}$ and $\mathcal{K}$ in Eqs.~\ref{eq:appMdefgeneral},~\ref{eq:appKdefgeneral} are (for $\pi,\sigma<\mathbbm{1}$)
\be\label{eq:appMdefspecific}
\mathcal{M}_{\pi \sigma} = 2\bar c \f{q!\mathbbm{1}(\pi \geq \sigma)}{q^{|\sigma|} (q-|\pi|)!} - \delta_{\pi,\sigma}
\ee
and ($\pi,\sigma,\sigma'<\mathbbm{1}$)
\be\label{eq:appKdefspecific}
\mathcal{K}^\pi_{\sigma,\sigma'}\hspace{-0.5pt} \hspace{-0.5pt}\hspace{-0.5pt}= \hspace{-0.5pt}\hspace{-0.5pt}
\f{\bar c q!}{(q\hspace{-0.5pt}-\hspace{-0.5pt}|\pi|)!}\hspace{-0.5pt} \hspace{-0.5pt}\left[\hspace{-0.5pt}\hspace{-0.5pt}
\f{\mathbbm{1}(\pi\hspace{-0.5pt} \geq\hspace{-0.5pt} \sigma)}{q^{|\sigma|}}\hspace{-0.5pt}\hspace{-0.5pt} + \hspace{-0.5pt}\hspace{-0.5pt}\f{\mathbbm{1}(\pi \hspace{-0.5pt}\geq \hspace{-0.5pt} \sigma')}{q^{|\sigma'|}} \hspace{-0.5pt}\hspace{-0.5pt}-\hspace{-0.5pt} \hspace{-0.5pt}\f{\mathbbm{1}(\pi\hspace{-0.5pt} \geq\hspace{-0.5pt} \sigma\wedge \sigma')}{q^{|\sigma\wedge\sigma'|}}\hspace{-0.5pt}\hspace{-0.5pt}
\right].
\ee
If we set $q=2$, the only nontrivial partitions $\sigma$ that can arise are two-block partitions with $|\sigma|=2$. 
(For $n$ replicas, there are ${\stirling{5}{2}=2^{n-1}-1}$ such partitions.)
Then 
\ba
 \mathcal{M}_{\pi \sigma} & =
(2 {p_s}-1) \delta_{\pi \sigma},
& 
{p_s} 
& = \f{\bar c}{2} 
& 
& (q=2)
\end{align}
and ($\pi,\sigma,\sigma'<\mathbbm{1}$)
\be
\mathcal{K}^\pi_{\sigma,\sigma'} = \f{\bar c}{2} \lf
\mathbbm{1}(\pi \geq \sigma)
+
\mathbbm{1}(\pi \geq  \sigma')
-
\f{\mathbbm{1}(\pi \geq \sigma\wedge \sigma')}{2^{|\sigma\wedge\sigma'|-2}} 
\ri.
\ee

\subsubsection{Details of some statements in main text}
\label{app:somedetails}

In Sec.~\ref{sec:meanfieldthreereplica}, Eqs.~\ref{eq:n3variance},~\ref{eq:n3covariance} we stated results for the noise covariance for $n=3$,
for the specific model above,
in terms of the pairwise damage variables
${\rho_{1,2} = \rho_{(1)(23)} + \rho_{(2)(13)} + \rho_{(1)(2)(3)}}$ and ${\rho_F = \rho_{(1)(2)(3)}}$.
Let us give the more detailed formulas.
In the original basis of partitions, the noise covariance matrix is \textit{diagonal} at order $\rho$ (Eq.~\ref{eq:appnoisecovariancegenerallinearorder}), and we easily find that at this order (writing 
$\rho_1 = \rho_{(1)(23)}$,
$\dd B_1 = \dd B_{(1)(23)}$, etc., and $\dd B_F = \dd B_{(1)(2)(3)}$)
\ba
\< \dd B_1 \dd B_1\> & = \f{\dd t}{N} \left[
(1+2{p_s}) \rho_1 + \f{2 p_s}{q} \rho_F
\right]
\\
\< \dd B_F \dd B_F \> & = 
\f{\dd t}{N}
\lf 1+2 {p_s} - \f{4 {p_s}}{q} \ri \rho_F
\end{align}
(and symmetrically). Changing basis using (by Eq.~\ref{eq:pairwisediffdefn}) 
\be
\rho_{1} = \f{1}{2} \lf \rho_{1,2} + \rho_{1,3} - \rho_{2,3} - \rho_F \ri 
\ee
(and symmetrically) gives
\ba
\< \dd B_{1,2}\dd B_{1,2}\>
& = \nu \rho_{1,2},
\\
\< \dd B_{1,2}\dd B_{2,3}\>
& = 
\f{\nu}{2} (\rho_{1,2}+ \rho_{2,3} - \rho_{1,3}) 
+ \lf \f{\nu}{2} - \nu' \ri \rho_F
\\
\< \dd B_{1,2} \dd B_F\> & = (\nu - 2\nu') \rho_F,
\end{align}
where $\nu=(1+2{p_s})/N$ is defined in Eq.~\ref{eq:n2nonuniversalconsts} and ${\nu' = 2 {p_s}/(qN)}$.

Around Eq.~\ref{eq:2blockonlygeneral} in the main text
we stated that, for a general model with $q=2$, all the two-block densities decay like $1/t$ (up to constant prefactors) when starting from a fully random initial state. Let us confirm this. 

Recall the result $\rho_{a,b}\sim 1/gt$ 
for the pairwise damages 
(from Sec.~\ref{sec:MFTn2} or Sec.~\ref{sec:meanfieldthreereplica}, but applicable for any $n$).
Since $\rho_{a\neq b}$ is a sum of two-block densities when $q=2$,
and using replica symmetry, 
this result implies that there is some ${k\in \{1,\ldots, n-1\}}$, 
such that $\rho_\sigma\sim \text{const.}/t$
for partitions with the structure  ${(1\cdots k)(k+1\cdots n)}$.

Next consider a partition $\pi$ with blocks of sizes ${(k+1, n-k)}$.
It is possible to find $\sigma$ and $\sigma'$, both of type ${(k,n-k)}$,
such that ${\pi\geq \sigma\wedge \sigma'}$.
Therefore,
(barring fine-tuning of the model)
partitions of type $\pi$ are generated at a nonzero 
rate
$P(\pi|\sigma,\sigma')$
by updates acting on inputs of type $\sigma$, $\sigma'$.
The final term in Eq.~\ref{eq:2blockonlygeneral}
then shows that $\rho_\pi$ is also of order $1/t$ at late times.
A similar argument applies to partitions with blocks of size ${(k-1,n-k+1)}$.
Continuing this way, we see that all two-block partitions have densities $\rho= O(1/t)$ as $t\to\infty$.

\section{The interaction tensor $\mathcal{K}$}
\label{app:Ktensordetails}

We  add to the discussion of the   interaction tensor $\mathcal{K}^\pi_{\sigma,\sigma'}$ in Sec.~\ref{sec:fieldtheoryfinitedgeneraln}.

We have already seen that for 
$n=2$ and $n=3$ this tensor 
reduces to a single constant, which we denoted $g$.
Here we rederive this fact in a slightly different way. 
We then observe that, 
once ${n>4}$, the interaction tensor
contains more free parameters, at least in the  ultraviolet.
(At the critical point --- and within the framework of the $4-\epsilon$ expansion ---  we conjecture that $\mathcal{K}$ flows to a universal form in the infrared.)

In the family of models we are considering, 
$\mathcal{K}^\pi_{\sigma,\sigma'}$ is defined (Eq.~\ref{eq:Kdefgeneral}) by the output probabilities for a single update:
\be\label{eq:Kdefgeneralrepeat}
\mathcal{K}^\pi_{\sigma,\sigma'} = 
\mathcal{K}^\pi_{\sigma',\sigma} = 
P(\pi|\sigma,\mathbbm{1}) + P(\pi|\sigma',\mathbbm{1})  -  P(\pi|\sigma,\sigma').
\ee
Recall that we are presently restricting to the case $q=2$,
where all these partitions are two-block.

For ${n=3}$,
replica symmetry immediately reduces  $\mathcal{K}$   to three distinct nonzero constants that are \textit{not} related to each other by replica  permutations.
Temporarily using the shorthand ${\sigma_1=(1)(23)}$, ${\sigma_2 = (2)(13)}$, ${\sigma_3=(3)(12)}$, these are
\ba
\mathcal{K}^{\sigma_1}_{\sigma_1,\sigma_1} &  = g,
&
\mathcal{K}^{\sigma_1}_{\sigma_1,\sigma_2} & = 
\f{g}{2},
&
\mathcal{K}^{\sigma_1}_{\sigma_2,\sigma_3} & =-\f{g}{2}.
\end{align}
Despite the fact that these are not related to each other by replica permutations, they all reduce to a single constant $g$, 
which is a quantity in the 2-replica theory,
\be
g = 2 P\big( (1)(2)\big|  (1)(2), \mathbbm{1} \big)  - P\big( (1)(2)\big|  (1)(2),  (1)(2) \big).
\ee
This reduction is a consequence of the fact that the 3 physical replicas do not interact with each other (the replicas are correlated only as a result of experiencing the same randomness\footnote{It may be interesting to explore how this fact constrains the coupling constants in replica field theories for disordered systems more generally.}).
Since the replicas do not interact,
any probability that  only refers to two out of the three replicas must reduce to the corresponding probability in the two-replica theory. 
In the present setting, with ${q=2}$, the resulting constraints  fully determine the update probabilities in the three-replica theory 
(see footnote\footnote{To see the logic, consider 
 the sum
\be\label{eq:simplesum}
P\big( \sigma_1 \big| \sigma_1, \mathbbm{1} \big) + 
P\big( \sigma_2 \big| \sigma_1, \mathbbm{1} \big).
\ee
This
is the probability that the output is \textit{either} $(1)(23)$ \textit{or} $(2)(13)$, gven the specified inputs.
Since we have restricted to $q=2$ models, where the state $(1)(2)(3)$ cannot occur, 
this is simply the probability that replicas 1 and 2 differ in the output.
This probability is  independent of what the initial state of the third replica was, so  (\ref{eq:simplesum}) reduces to a probability in the two-replica theory, 
\be
P\big( \sigma_1 \big| \sigma_1, \mathbbm{1} \big) + 
P\big( \sigma_2 \big| \sigma_1, \mathbbm{1} \big)
=
P\big( (1)(2) \big| (1)(2), \mathbbm{1} \big),
\ee
which we denoted ${p_s}$ in Eq.~\ref{eq:psgeneraldef}.
Since $P(\sigma_2|\sigma_1, \mathbbm{1})=0$,
this gives us 
$P(\sigma_1|\sigma_1, \mathbbm{1})={p_s}$.
A straightforward generalization gives
 ${P(\sigma_1|\sigma_1,\sigma_1)=\gamma}$,
${P(\sigma_1|\sigma_1,\sigma_2)=\gamma/2}$,
 ${P(\sigma_1|\sigma_2,\sigma_3)=(p_s - \gamma/2)}$ 
 (and symmetrically), where ${\gamma \equiv P\big( (1)(2)\big|(1)(2),(1)(2)\big)}$.} for more detail).
 
The more general arguments of Sec.~\ref{sec:n3meanfieldfull},~\ref{sec:fieldtheoryfinitedn3} above showed that even in models with $q>2$, the constants in the three-replica Lagrangian (or three-replica mean field equation) reduce to those in the two-replica theory once we discard both massive fields and perturbations that are irrelevant at the upper critical dimension.

By contrast, when ${n\geq 4}$  the cubic terms contain one new constant, $\theta$, that is not fixed by the two-replica theory (Sec.~\ref{sec:moreoninteractiontensor}).

For an explicit demonstration of this additional degree of freedom in the continuum description,  let us consider a more general microscopic model than the one in Eq.~\ref{eq:Kdefspecificq2}, again for bits~(${q=2}$).

The most general  probability distribution for the local updates
that is consistent with our basic structure
(Sec.~\ref{sec:meanfieldeqnsgeneralform})
is described in Sec.~\ref{sec:choiceof1Dmodel}.\footnote{In Sec.~\ref{sec:choiceof1Dmodel} the update functions are  discussed in the context of 1+1D models, but the  possible local update functions themselves are the same as in the mean field model and its finite-dimensional extension.}
For the models of bits that we are presently considering, 
update functions $F$ can be 
classed as  ``Type 0'',
``Type 1'', or ``Type 2'' in order of increasing reversibility. 
(For example, a Type 0 function sends all inputs to the same output.)
The general model  is then parameterized by probabilities $(p_0, p_1, p_2)$ for these types with ${p_0+p_1+p_2=1}$.
We recover the previous model, defined above Eq.~\ref{eq:Mdefspecific}, by setting ${p_1 = \bar c/2}$, ${p_2 = 3\bar c/8}$.\footnote{Recall that the latter is the case where, with probability $\bar c$, we choose a uniformly random update function, and with probability ${1-\bar c}$, we choose a completely irreversible function that sends all inputs to the same output.}

We easily check that ${p_s = p_1/2 + 2 p_2/3}$ in such a model.
More generally, for any $n$, the nontrivial update probabilities for  $|\pi|=|\sigma|=|\sigma'|=2$ are 
\ba
P(\pi|\sigma, \sigma') = 
p_s \, R^\pi_{\sigma,\sigma'},
\end{align}
where
\ba \label{eq:Rtensorresult}
R^\pi_{\sigma,\sigma'}
=\mathbbm{1}_{\pi\geq \sigma\wedge \sigma'}\times
\left\{
\begin{array}{lll}
1 & \text{if } & |\sigma\wedge\sigma'|=2
\\
{1}/{2}
& \text{if } 
& |\sigma\wedge\sigma'|=3
\\
\theta 
& \text{if } 
& |\sigma\wedge\sigma'|=4, \text{ Case A }
\\
\f{1}{2} - \theta 
& \text{if } 
& |\sigma\wedge\sigma'|=4, \text{ Case B. }
\end{array}\right.
\end{align}
The constant is  
\ba
\theta & = \f{3 p_1}{2(3p_1 + 4 p_2)},
\end{align}
and in Eq.~\ref{eq:Rtensorresult} we have distinguished between two cases in which $|\sigma\wedge\sigma'|$ has four blocks.
In  Case A, 
one of the blocks of $\sigma\wedge\sigma'$ coincides with a block of $\pi$
(and the the other three blocks of 
$\sigma\wedge\sigma'$ live inside the other block of $\pi$).
In Case B,  every block of $\sigma\wedge\sigma'$ is strictly smaller than one of the blocks of  $\pi$.

For the present models we also have $P{(\pi|\sigma, \mathbbm{1})= p_s \delta_{\pi,\sigma} = g \delta_{\pi,\sigma}}$, 
so the tensor appearing in the Lagrangian may be written
\ba
\mathcal{K}^\pi_{\sigma,\sigma'}&  
 = g \lf \delta_{\sigma,\pi}
 + \delta_{\sigma', \pi} - R^\pi_{\sigma,\sigma'}\ri.
\end{align}
We may check that this reduces to Eq.~\ref{eq:Kdefspecificq2} for the appropriate parameter values ${p_1=\bar c/2}$, ${p_2 = 3\bar c/8}$. 

For ${n\leq 3}$ it is impossible for ${|\sigma\wedge \sigma'|}$ to be greater than three,  
so in these cases the constant $\theta$   in Eq.~\ref{eq:Rtensorresult} drops out, and   the tensor $\mathcal{K}$ reduces to a single constant $g$
as we found above.
However the explicit example above shows that there is additional freedom in the form of the tensor  once ${n>3}$. 

We show in Sec.~\ref{sec:moreoninteractiontensor} 
(by noting that all the constants in $\mathcal{K}$ are determined by the four-replica theory,  regardless of the value of $n$) that $\theta$ is the only additional parameter that appears in the $\mathcal{K}$ tensor when we go above ${n=3}$.

\section{Schematic block spin transformation}

\label{app:blockspindetails}

Continuing from the discussion below Eq.~\ref{eq:RGtransformationRho}, let us  recall the usual RSRG setup. 
In the present case, the ``partition function'' $Z$ is a sum over spacetime configurations (trajectories), with some boundary conditions. Schematically we write $Z = \int_{\{\sigma\}} W(\sigma)$, where 
$\{\sigma\}$ represents the configuration of partitions in spacetime and $W$ is the weight of a trajectory. 
(Although the degrees of freedom are discrete, we are imagining continuous time, hence the integral symbol.)

The block spin configuration $\{\sigma'\}$ is defined deterministically in terms of $\{\sigma\}$ as described in the text: at a given time, $\sigma'({\mathbf{X}})$ is defined to be greatest lower bound of the partitions inside the block at  that time.
We write $\int_{\{\sigma\}} =\int_{\{\sigma'\}}  \int_{\{\sigma\} | \{\sigma'\}}$, where the subscript in the second integral means we integrate over all microscopic spin configurations consistent with the block spin configuration.
Then $Z = \int_{\{\sigma'\}} W'(\{\sigma'\})$ where $W'(\{\sigma'\})=  \int_{\{\sigma\} | \{\sigma'\}}W(\{\sigma\})$.

Now consider an operator insertion $\rho_\tau(x)$. We easily check that in the coarse-grained expression this operator is replaced with 
\be
\<\rho_\tau({\bf x})\>_{\{\sigma'\}} \equiv  \f{1}{W'(\{\sigma'\})}  \int_{\{\sigma\} | \{\sigma'\}} W(\sigma) \, \rho_\tau({\bf x}).
\ee
On the right-hand-side, we must compute an expectation value conditioned on the full configuration of block spins (together with the microscopic boundary conditions). 
But if the RSRG procedure is to work, such expectation values must have a quasilocality property, i.e. they must depend only weakly on data far from ${\bf x}$.
The crudest approximation is to treat the average as dependent only on the value of $\sigma'({\bf X})$ for the ${\bf X}$ that contains ${\bf x}$. Within this approximation, 
\be
A_{\tau' \tau}=
\<\rho_\tau\>_{\tau'},
\ee
where the subscript on the average is the value of the block spin that we condition on. That is, within this approximation $A_{\tau'\tau}$ is the probability that the microscopic spin is $\tau$, given that the block spin is $\tau'$.

\section{Symmetry classification in  a sector}
\label{app:nonsymmetricoperators}

Following on from the discussion at the end of Sec.~\ref{sec:moregeneraloperators},
consider  the sector ${(1)(2)(3)}$ in the ${n=3}$ theory.
We have discussed the scaling operator corresponding to $\rho_{(1)(2)(3)}$.
Note that this operator is a singlet (invariant) under the global $S_3$ symmetry associated with permutations of the three replicas. We can also define operators that belong to the same sector, 
but that transform in a nontrivial representation of $S_3$. 

To see this, it is useful to recall from Sec.~\ref{sec:fieldtheoryfinitedn3}
that, in the continuum field theory, $\rho_{(1)(2)(3)}$ is essentially the composite operator
\ba\label{eq:rho123again} \notag
\rho_{(1)(2)(3)} & \sim \hat \rho_{\sigma_1} \hat \rho_{\sigma_2}
+ \hat \rho_{\sigma_2} \hat \rho_{\sigma_3}
+ \hat \rho_{\sigma_3} \hat \rho_{\sigma_1}
\\
& =  \hat \rho_{\sigma_1} \hat \rho_{\sigma_2} + \text{cyclic permutations},
\end{align}
where on the left $\hat \rho_\sigma$ denotes one of the elementary fields of the field theory,
with ${\sigma_1 = (1)(23)}$,
 ${\sigma_2 = (2)(13)}$,
  ${\sigma_3 = (3)(12)}$.
The expression on the right hand side of (\ref{eq:rho123again}) can also be given a meaning in the lattice. 
Recall that, in a model of bits, 
we can define nontrivial density operators 
$\rho_\sigma(j)$  at a \textit{site}  $j$ only for the two block partitions,
and the three-block  density has to be defined on a \textit{bond}  $x=(j, j+1)$. 
Explicitly,
\ba\notag
\rho_{(1)(2)(3)}(x)  =  & 
\notag
\Big(  \rho_{\sigma_1}(j) \hat \rho_{\sigma_2}(j+1)  + \text{cyclic perms} \Big) 
\\ & +
\Big( 
j \leftrightarrow j+1
\Big).\label{eq:rho123againlatt}
\end{align}
There are 6 terms here:
$\rho_{(1)(2)(3)}(x)$ is nonzero iff 
the two sites are assigned to two distinct two-block partitions (two distinct colours in the notation of Sec.~\ref{sec:effectivedynamicsn3})
and there a 6 such assignments. 
A naive continuum limit of (\ref{eq:rho123againlatt}) gives 
(\ref{eq:rho123again}).

It is straightforward to adapt
Eq.~\ref{eq:rho123againlatt} to give operators that transform nontrivially under $S_3$. 
Let
\ba\label{eq:rho123again} 
\notag
\rho_{(1)(2)(3)}^+ & 
\sim \hat \rho_{\sigma_1} \hat \rho_{\sigma_2}
+ 
e^{2\pi  i /3} \hat \rho_{\sigma_2} \hat \rho_{\sigma_3}
+ e^{4\pi  i /3}  \hat \rho_{\sigma_3} \hat \rho_{\sigma_1}.
\end{align}
The real and imaginary parts of $\rho^+_{(1)(2)(3)}$ form a two-dimensional irreducible representation of $S_3$.

In four dimensions operators retain their engineering dimensions, showing that $\rho_{(1)(2)(3)}$ and $\rho^+_{(1)(2)(3)}$ have the same scaling dimension for ${d=4}$.
For ${d<4}$ we would naively expect that their scaling dimensions differ. 
We expect that $\Delta^+_{(1)(2)(3)}>\Delta_{(1)(2)(3)}$ for ${d<4}$.
We may argue for this (details omitted) by considering sums $A(t)$ and $B(t)$ of lattice probabilities that map to correlation functions of the schematic form 
$A(t)\sim {\<\rho_{(1)(2)(3)}(t)
\widetilde \rho_{(1)(2)(3)}(0)\>\sim t^{-(\Delta_3 + \widetilde \Delta_3)/z}}\sim t^{-2 \Delta_3/z}$
and 
$ B(t) \sim {\<\rho^+_{(1)(2)(3)}(t)
\widetilde \rho^+_{(1)(2)(3)}(0)\>\sim t^{-(\Delta_3^+ + \widetilde \Delta_3^+)/z}}\sim t^{-2 \Delta_3^+/z}$ respectively.
(Here we make an assumption about how time-reversal symmetry acts on $\rho^+_{(1)(2)(3)}$.)
We then note that $A(t)\geq B(t)$, because the former is a sum of probabilities with coefficient 1 while the latter is the same sum with signed coefficients.

We leave the generalization to more complex partitions 
to the future.

\section{Time-reversal symmetry}
\label{app:latticetrs}

First we show the existence of a time reversal symmetry for the ${n=3}$ model on the lattice. 
Then we argue, using the continuum formulation, that time-reversal symmetry imposes the exponent relation ${\alpha_3=\delta_3}$.

The effective Markovian dynamics
for the 1+1D $n=3$ model 
whose rules are given in Sec.~\ref{sec:effectivedynamicsn3} can be formalized using a ${4\times 4}$ transition matrix $\mathcal{T}_\text{2-site}$ 
encoding the transition probabilities 
for the update on a pair of sites.
For example,  Eq.~\ref{eq:3siteupdatepictures} gives (using Dirac notation for the matrix elements)
\ba
\bra{\ballC,\emptysite}\mathcal{T}_\text{2-site} \ket{\ballA, \ballB}
& =\operatorname{Prob}\lf
\ballA, \ballB \rightarrow \ballC,\emptysite
\ri
\\
& = \f{p_s}{2} \times \lf 1- \f{3 p_s}{2} \ri,
\end{align}
where the product form of the RHS arises because the left and right sites of the pair are updated independently.

For ${n=3}$ one can find a ${2\times 2}$ matrix, representing a change of basis on a single site, such that 
\be
(S_1\otimes S_1).(\mathcal{T}_\text{2-site})^T.(S_1\otimes S_1)^{-1}=\mathcal{T}_\text{2-site}.
\ee
Explicitly, in the basis ${(\ballA, \ballB, \ballC, \emptysite)}$ for the four states on a single site, we may take:
\be
S_1=
\lf
\begin{array}{rrrc}
1 & - 1 & - 1 & 1 \\
-1 & 1 & -1 & 1 \\
-1 & - 1 & 1 & 1 \\
1 & 1 & 1& -3+2/ps
\end{array}
\ri.
\ee

The full Markov transition matrix $\mathcal{T}$ for the entire system 
(because of the alternating even-odd structure of the 1+1D updates, 
$\mathcal{T}$ corresponds to two timesteps) can be written using tensor products of local $\mathcal{T}_\text{2-site}$ operators, so we also have 
\be\label{eq:latticetimereversal}
S.\mathcal{T}^T.
S^{-1}
=\mathcal{T},
\ee
where $S=\otimes_{i=1}^L S_1$.

The relation in Eq.~\ref{eq:latticetimereversal} is a time-reversal symmetry. 
The probability to transition from a basis  state 
$\ket{\text{in}}$ 
to a basis state $\bra{\text{out}}$ is 
$\bra{\text{out}} \mathcal{T}_\text{sys}^{t/2}\ket{\text{in}}$.
Using the time-reversal symmetry, 
\be
\bra{\text{out}} S^{-1} \mathcal{T}^{t/2}\ket{\text{in}}
= 
\bra{\text{in}}S^{-1} 
\mathcal{T}^{t/2}
 \ket{\text{out}}.
\ee
This formula also holds if the states are sums of basis states with real coefficients, and may be used to obtain nontrivial relations between different probabilities. 

However, for simplicity we will instead argue in a more heuristic way  that $\alpha_3=\delta_3$.

First recall that $z \alpha_n=\Delta_n$  is the scaling dimension of the density operator  $\rho_{(1)(2)\cdots(n)}$.
On the other hand $z \delta_n=\widetilde \Delta_n$ is the scaling   dimension of the scaling operator $\widetilde{\mathcal{O}}_\pi$ (see Sec.~\ref{subsec:RG}) which \textit{creates} damage of type ${\pi = (1)(2)\cdots (n)}$;
this is in analogy with standard directed percolation, and is explained at the end of this appendix.

For $n=3$, the first of these scaling operators has a simple expression in the field-theoretic epsilon expansion,\footnote{The RHS is the lowest-order expression in the fields that is nonzero only in configurations that exhibit damage type $(1)(2)(3)$ and which is symmetric under $S_3$ permutations.}
\be
\rho_{(1)(2)(3)}\sim \hat\rho_1\hat \rho_2 + \hat\rho_2 \hat\rho_3 + \hat\rho_3 \hat\rho_1  .
\ee
Under the duality transformation in Eq.~\ref{eq:trsn3transformations} this is transformed into a distinct scaling operator made from $\tilde \rho$ fields. 
We  argue that this should be identified with $\widetilde{\mathcal{O}}_{(1)(2)(3)}$.\footnote{
Writing out the duality-transformed 
version of $\rho_{(1)(2)(3)}$ using
Eq.~\ref{eq:trsn3transformations}, 
we see that it matches the expected form of $\widetilde{\mathcal{O}}_{(1)(2)(3)}$
in  Eq.~\ref{eq:formofscalingcreationoperators}: it is a sum of (composite) operators that create damage of type $(1)(2)(3)$ together with operators that create lower damage types.}
It follows that ${\alpha_3 = \delta_3}$.

Finally let us we  explain why $z \delta_n$ is the scaling   dimension of the  operator that creates damage of type ${\pi = (1)(2)\cdots (n)}$.

The survival probability, and therefore $\delta_n$, can be extracted from a correlation function of the form
\be
{\mathcal{S}_n(t) = \<\tilde\rho_{(1)(2)\cdots(n)}(0,0) \, \chi_{(1)(2)\cdots(n)}(t) \>}.
\ee
Here the boundary conditions are such that system is initiated in the vacuum in the far past. The operator
$\tilde\rho_{(1)(2)\cdots(n)}(0,0)$ then seeds local damage of type $(1)(2)\cdots(n)$ at the spacetime origin. 
Then, the nonlocal operator $\chi_{(1)(2)\cdots(n)}(t)$, which is equal either to zero or to one, 
enforces the constraint that all the global states of the $n$ replicas are all different at time $t$. 

In the correlator above, $\tilde\rho_{(1)(2)\cdots(n)}$ is not quite a scaling operator; instead, as discussed at the very end of Sec.~\ref{subsec:RG}, the corresponding scaling operator $\widetilde{\mathcal{O}}_{(1)(2)\cdots(n)}$ is of the form
\ba\notag
\widetilde{\mathcal{O}}_{(1)(2)\cdots(n)}
& = 
\widetilde{\rho}_{(1)(2)\cdots(n)} \\
& + (\text{creation operators for coarser partitions}).\label{eq:creationopexpagain}
\end{align}
However the indicator function $\chi_{(1)(2)\cdots(n)}$ ensures that the terms in parentheses in Eq.~\ref{eq:creationopexpagain} do not contribute to ${\<\widetilde{\mathcal{O}}_{(1)(2)\cdots(n)}(0,0) \, \chi_{(1)(2)\cdots(n)}(t) \>}$.
So we can write
${\mathcal{S}_n(t) = \<\widetilde{\mathcal{O}}_{(1)(2)\cdots(n)}(0,0) \, \chi_{(1)(2)\cdots(n)}(t) \>}$.
It may also be argued (see below) that the scaling dimension of $\chi_{(1)(2)\cdots (n)}(t)$ is zero.
Therefore
$\mathcal{S}_n(t) = \<\widetilde{\mathcal{O}}{(1)(2)\cdots(n)}(0,0) \chi_{(1)(2)\cdots(n)}(t) \>$ scales as $t^{-\delta_n}$, where  $\delta_n = \widetilde\Delta_n/z$ is given simply by the scaling dimension in time units of $\mathcal{O}_{(1)(2)\cdots(n)}$.

For completeness let us show explicitly that the scaling dimension of the operator $\chi_{(1)(2)\cdots(n)}(t)$ above is equal to zero.
See Ref.~\cite{janssen2005survival} for a discussion of how to compute survival probabilities in field theory. 
To make the point needed here in a handwaving way, it is sufficient to note that we may write 
\be
\< \chi_{(1)(2)}(t) \, (\cdots) \> = 
\lim_{\ell \to 0} 
\< \lf\int \dd^d x \rho_{12}(x,t) \ri^\ell \, (\cdots) \>,
\ee
where $\rho_{ab}(x,t)$ is the density of sites where replicas $a$ and $b$ differ.
The $\ell\to 0$ limit gives the desired indicator function which is  1 if replicas $a$ and $b$ differ somewhere, and zero if they differ nowhere. 
For a given $\ell$, dimension counting shows that $\lf\int \dd^d x \rho_{12}(x,t) \ri^\ell$ has time dimension ${\ell(d/z - \alpha_2)}$. This vanishes as $\ell\to 0$, showing that $\chi_{(1)(2)}$ has scaling dimension zero. This idea extends to larger $n$ by using a product of integrals for each $a<b$ (compare the product of $\mu^{ab}$ factors in  Eq.~\ref{eq:surv-prob}).

\section{Partitions and probabilities}
\label{app:partitionsandprobabilities}

For a simple example of the relation between partition densities $\<\rho_\pi\>$ and probabilities, consider the physical state $s_i(t)$ of a single site (say, at the origin $i=0$), in a model of bits, at time $t$.

For a \textit{fixed realization} $C$ of the classical circuit,  averaging over all possible \textit{initial states} of the system induces a probability distribution $\{p_0^C, p_1^C\} = \{ p^C, 1-p^C \}$ on the two possible states $s_0(t)=0,1$ of the bit at time $t$.
The value of $p^C$ depends on the circuit. 
For some instances of the circuit,
all initial states will give the same state of the bit, so that $\{ p^C, 1-p^C \}=\{ 0,1 \}$. For other  circuit instances, different initial conditions will give different states for the bit, so that $p$ takes a nontrivial value. 
Since we are free to rename $p^C\leftrightarrow 1-p^C$, let's write $\widetilde p^C = p^C(1-p^C)$, which is approximately equal to the smaller of the two probabilities when that is small.
We give an example of how the partition densities contain information about the distribution of $\widetilde p$ in the circuit ensemble. We will discuss more nontrivial examples elsewhere.

Let $\mathbb{E}$ denote the circuit average.
Note that  ${\mathbb{E} \sum_{s=0,1} (p^C_s)^n}$ 
is the probability that $n$ replicas, with independently random initial conditions, have the same state for the bit at time $t$. So, in terms of partitions,
\ba\label{eq:pssumpartitions}
\mathbb{E} \sum_{s=0,1} (p^C_s)^n & = \< \rho_{\mathbbm{1}} \> 
 = 1 - \sum_{\substack{\pi \neq \mathbbm{1}  \\  \pi \in \Pi_n}} \< \rho_\pi\>.
\end{align}
We have suppressed the spacetime arguments on $\rho_{\pi}(i,t)$ to avoid clutter.
The equation above generalizes to any region, if $s$ is taken to run over all states of the region, and $\rho_\pi$ is defined the corresponding multi-site damage observable, as defined in 
 Sec.~\ref{sec:simulationsnreplicaobservables}.
The left-hand side can also be written in terms of the R\'enyi entropy $S_n(t)$ of the region at time $t$, as ${\mathbb{E} \exp [-(n-1) S_n(t)]}$. We will discuss entropies elsewhere.

For the single bit, we have ${\mathbb{E} \sum_s  p_s^C = 1- 2 \mathbb{E}\,\widetilde p^C}$
and ${\mathbb{E} \sum_{s=0,1} (p_s^C)^4 = 1 - 4 \mathbb{E}\, \widetilde p^C + 2\mathbb{E}\, \widetilde (p^C)^2}$.
Eq.~\ref{eq:pssumpartitions} gives 
\ba
\mathbb{E} {\sum}_{s} (p_s^C)^2 & = 1 - \< \rho_{(1)(2)}\> 
\\ \label{eq:ps2rho}
& = 1 - 2\lf \rho_{(12)(34)} + \rho_{(1)(234)} \ri
\\
\label{eq:ps4rho}
\mathbb{E} {\sum}_{s} (p_s^C)^4 &  =  1 - 3 \<\rho_{(12)(34)}\> 
- 4\<\rho_{(1)(234)}\>
\end{align}
In (\ref{eq:ps2rho}) we used the symmetry between different partitions with the same shape that holds for the present choice of initial conditions, and the relation between two- and four-replica observables in Eq.~\ref{eq:rho-a-b}. For (\ref{eq:ps4rho}) we also used the fact that, for a single bit, only the two-block densities contribute to the final term of  (\ref{eq:pssumpartitions}).
The equations above give
\ba\label{eq:ptildevaluesapp}
\mathbb{E}\, \widetilde p^C & = \f{1}{2} \< \rho_{(1)(2)}\>,
& 
\mathbb{E}\, ( \widetilde p^C)^2 & = \f{1}{2} \< \rho_{(12)(34)}\>.
\end{align}
We have shown that, at the critical point, these scale with the same power of $t$ at late time.  

One way to interpret this is in terms of hetereogeneity between distinct circuits $C$. Consider the average damage density \textit{in a fixed circuit }$C$, which we temporarily denote by $\rho_C$, and which is simply related to $\widetilde p$ defined above:
\be
\rho^C(i,t) \equiv \<  1- \delta_{s_i^1(t), s_i^2(t)} \>^C = 2 \widetilde p^C.
\ee
In the central expression, the averaging $\<\cdots\>^C$ is only over the initial conditions $\{s_k^1(0)\}_{k=1}^L$ and $\{s^2_k(0)\}_{k=1}^L$, with the circuit $C$ held fixed.
Of course if we restore the circuit average we get the usual $n=2$ result, $\mathbb{E} \rho^C(i,t)= \< \rho_{(1)(2)}(i,t)\>$.
(Again $\mathbb{E}$ here denotes the average only over circuits $C$.)

Even when $C$ is held fixed, different choices of the pair of initial conditions will give different spacetime damage trajectories. 
A natural question is whether averaging over these trajectories, at fixed $C$, gives a similar result to averaging over $C$ also. 
The result above shows that it does not: Eq.~\ref{eq:ptildevaluesapp} shows that the fluctuations in $\rho^C(i,t)$, between different circuits,
are large compared to the mean value.

\bibliography{bibliodamagespreading}
\end{document}